\documentclass[useAMS,usenatbib,usegraphicx,uselongtable]{mn2e}
\usepackage{lscape}

\title[Disc Frequencies for Brown Dwarfs]{Disc Frequencies for Brown Dwarfs in the Upper Scorpius OB Association: Implications for Brown Dwarf Formation Theories}
\author[Riaz et al.]
{B. Riaz,$^{1}$ N. Lodieu,$^{2,3}$
 S. Goodwin,$^{4}$ D. Stamatellos,$^{5}$, 
 M. Thompson$^{1}$ \\
$^{1}$Centre for Astrophysics Research, Science \& Technology Research Institute, University of Hertfordshire, Hatfield, AL10 9AB, UK\\
$^{2}$Instituto de Astrofisica de Canarias, E-38200 La Laguna, Tenerife, Spain \\
$^{3}$Departamento de Astrof\'isica, Universidad de La Laguna, E-38205 La Laguna, Tenerife, Spain \\
$^{4}$Department of Physics \& Astronomy, University of Sheffield, Hicks Building, Hounsfield Road, Sheffield, S3 7RH, UK \\
$^{5}$School of Physics \& Astronomy, Cardiff University, Cardiff, CF24 3AA, UK }

\begin{document}

\date{}

\pagerange{\pageref{firstpage}--\pageref{lastpage}} \pubyear{2002}

\maketitle

\label{firstpage}

\begin{abstract}

We have investigated the brown dwarf (BD) and stellar disc fractions in the Upper Scorpius OB Association (USco) and compared them with several other young regions.  We have compiled the most complete sample of of all spectroscopically confirmed BDs in USco, and have made use of the Wide Field Survey Explorer (WISE) catalog to identify the disc candidates. We report on the discovery of 12 new BD discs in USco, with SpT between M6 and M8.5. The WISE colors for the new discs are similar to the primordial (transition) discs earlier detected in USco. Combining with previous surveys, we find the lowest inner disc fractions ($\sim$20-25\%) for a wide range in stellar masses ($\sim$0.01-4.0$M_{\sun}$) in the USco association. The low disc fractions for high-mass stars in USco (and the other clusters) are consistent with an evolutionary decline in inner disc frequency with age. However, BD disc fractions are higher than those for the stars in 1-3 Myr clusters, but very low in the $\sim$5 Myr old USco. Also, primordial BD discs are still visible in the $\sim$10 Myr old TW Hydrae association, whereas the higher mass stars have all transitioned to the debris stage by this age. The disc frequencies for BDs do not show any dependence on the stellar density or the BD/star number ratio in a cluster. We also find no convincing evidence that any of the well-known disc dispersal mechanisms for stars are active in BD discs. We suggest that the large differences in the observed BD disc fractions between regions may well be due to different BD formation mechanisms and therefore different initial disc fractions/properties.

We also present a WISE SED classification scheme, based on the $K_{s}$ and WISE bands of 3.4-12$\micron$. We have determined certain thresholds in the WISE spectral slope versus spectral type diagrams to distinguish between the red population of Class I/II systems and the Class III sequence. We have found the WISE [3.4]-[12] color to provide the best distinction between the photospheric and the disc population. Our work includes a comparison of the sensitivities of WISE and Spitzer disc surveys. We estimate that WISE can be incomplete for discs at spectral type later than M8 in distant clusters such as SOri. WISE should be able to recover the M8-M9 discs in the nearby young clusters.

\end{abstract}


\section{Introduction}

Recent surveys of nearby young clusters spanning a range in ages and masses have revealed extensive information on the disc frequencies and the disc dissipation timescales. The general picture of inner disc evolution that emerges from {\it Spitzer}/IRAC observations of several clusters shows a sharp decline in the disc frequency of young stars with ages between $\sim$1 and 3 Myr. By an age of $\sim$5 Myr, the disc fraction has dropped by more than 80\%, while the timescale for nearly all stars to lose their inner discs appears to be $\sim$10 Myr (e.g., Haisch et al. 2001; Hillenbrand 2005; Hernandez et al. 2007). Other than the overall disc fractions, the strength in the disc emission at mid-infrared wavelengths also shows a decrease with age. A majority of young disc sources in Taurus ($\sim$1 Myr) possess optically thick inner discs, while most stars by an age of $\sim$5 Myr possess ``anemic'' discs which display weaker IRAC excesses than optically thick disc systems (e.g., Lada et al. 2006; Carpenter et al. 2009). The age of $\sim$5 Myr is thus critical, in the sense that nearly all optically thick primordial disc material in the inner $\sim$1 AU region has dissipated by this age, and most discs begin to show a transition from a primordial to a debris disc system. 

The disc dissipation timescales, however, show a dependence on the stellar mass and the stellar density in the region, such that longer disc lifetimes are observed for the sub-stellar sources, and disc fractions are found to be higher in less dense environments (e.g., Riaz \& Gizis 2008; Riaz et al. 2009). In Riaz et al. (2009), we had determined a BD disc fraction of $\sim$10\% in the $\sim$5 Myr old Upper Scorpius OB association (hereafter, USco). In comparison with younger $\sim$1-3 Myr old clusters that show a $\sim$40-50\% disc fraction, this indicated significant disc dispersal for sub-stellar sources over a short $\sim$2 Myr time span. In contrast, the BD disc fraction in the $\sim$10 Myr old TW Hydrae Association (hereafter, TWA; Webb et al. 1999) is much higher (50$\pm$20\%; Riaz \& Gizis 2008). We had argued that the BD density in the TWA is at least $\sim$100 times less than in USco, which could prolong the disc lifetimes. Thus age may not be a dominant factor but a less dense environment could explain the higher disc fractions observed. 


In the present work, we have further investigated the dependence of the BD disc fractions on the age, density and the BD formation mechanisms in a cluster. We have focused again on the USco association, since it is at a critical intermediate age of $\sim$5 Myr which is important to probe inner disc dissipation timescales. The USco association is part of the Scorpius Centaurus complex (de Geus et al.\ 1989), located at a distance of 145$\pm$2 pc (de Bruijne et al.\ 1997). The age of USco is $\sim$5 Myr, with little scatter of $\pm$2 Myr (e.g., Preibisch \& Zinnecker 1999). The region is relatively free of extinction (Av $\leq$ 2 mag) and star formation has already ended (Walter et al.\ 1994). The association has been targeted at multiple wavelengths over the past decade starting with X-rays (e.g.\ Preibisch et al.\ 1998), with Hipparcos (de Zeeuw et al.\ 1999), in the optical (Preibisch et al.\ 2001; 2002; Ardila et al. 2000; Mart\'{\i}n et al.\ 2004; Slesnick et al.\ 2006), and in the near-infrared (e.g., Lodieu et al.\ 2007), resulting in several low-mass stars and BDs being confirmed as genuine spectroscopic members (Ardila et al. 2000; Mart\'{\i}n et al.\ 2004; Slesnick et al.\ 2006; Lodieu et al.\ 2008; Lodieu et al.\ 2011). The disc population in USco has been studied by Carpenter et al. (2006; 2009), Scholz et al. (2007) and Riaz et al. (2009). The B/A and F/G type stars are all found to be debris disc systems, while primordial discs with inner disc clearing (no excess shortward of $\sim$8$\micron$) are more abundant among the K/M types and the sub-stellar sources. We have compiled a sample of all spectroscopically confirmed BD members in USco, and have made use of the recently released Wide Field Survey Explorer (WISE; Wright et al.\ 2010) catalog to identify the disc candidates. Our sample also includes some M dwarfs in USco that have not been previously studied in the mid-infrared. In Section \S\ref{WISE}, we describe our WISE spectral energy distribution (SED) classification scheme and the WISE colors that can best distinguish between the discless and the disc population. Section \S\ref{results} presents the results on new disc sources found in USco, and Section \S\ref{halpha-disc} compares the H$\alpha$ emission strength with the disc emission. Section \S\ref{discussion} discusses the various factors that may affect the disc fractions in a cluster.

\section{Sample Selection}
\label{sample}

We have included in our sample the optical ($RI$) survey complemented by 2MASS infrared photometry conducted by Slesnick et al.\ (2006; 2008) over the full association down to $R \sim$ 20 mag (corresponding to masses of $\sim$0.02 M$_{\odot}$). This is the most complete survey of the entire USco region. A large number of photometric candidates were selected and a total of 107 were confirmed as spectroscopic members based on strong H$\alpha$ emission and weak gravity-sensitive features. The spectral types (SpT) for these sources are between M4 and M8.5. We note that the disc properties of 28 USco members with SpT later than M6 identified by Slesnick et al.\ (2006) are already discussed in Riaz et al. (2009) using {\it Spitzer} archival data. The USco association has been independently targeted by the UKIRT Infrared Deep Sky Survey (UKIDSS) Galactic Clusters Survey (GCS) in two optical filters ($ZY$) and three infrared passbands ($JHK$) (Dye et al. 2006). The depth of GCS in terms of mass is roughly 10 Jupiter masses. Lodieu et al.\ (2007) identified over 140 photometric candidates in 6.5 square degrees imaged by the GCS based on their photometry, colours, and proper motions. Over 90\% of these candidates were confirmed spectroscopically as USco members using cross-dispersed near-infrared spectroscopy (Gemini/GNIRS; Lodieu et al.\ 2008) and multi-fibre optical spectroscopy (AAT/AAOmega; Lodieu et al. 2011). The SpT for these members range between M3.5 and M9. The GCS USco surveys are complete down to the SpT of M9. 


We have cross-matched this list of USco spectroscopic members with the WISE catalog. We adopted a matching radius of 6.5 arcsec to take into account the spatial resolution of WISE, which is $\sim$6 times poorer than the UKIDSS seeing-limited images. We note that all WISE counterparts to the UKIDSS sources are within one arcsec (or less) as the result of the good astrometric precision of WISE (0.15 arcsec; Wright et al.\ 2010). We will refer to the WISE 3.4, 4.6, 12 and 22$\micron$ bands as WISE band 1, 2, 3 and 4, respectively. Our final USco-WISE matched catalog consists of 215 sources and covers spectral types (SpT) between M3.5 and M8.5. We have considered the sources in our sample with SpT $\geq$M6 to be BDs or sub-stellar sources, as described in Section \S\ref{appendix}. Out of the 215 sources in our sample, 43 are BDs and 146 are M dwarfs. The USco samples previously studied by Carpenter et al. (2006; 2009) covered SpT of B0-M5. Our sample is complete for M4 and later types in USco.

As discussed in Section \S\ref{WISE}, we have considered the WISE 3.4, 4.6, and 12$\micron$ bands for SED classification, and have therefore required that all sources are detected at a $\geq$3-$\sigma$ level in these bands. A majority of the USco sources lack detections in the WISE band 4 at a $>$3-$\sigma$ level. Our $\geq$3-$\sigma$ quality criteria is to ensure good S/N detections, and to avoid spurious sources where the large excesses could be due to the background noise. This quality criteria imposed could introduce a potential bias to the disc fraction obtained, since the full sample detected in WISE at a $>$3-$\sigma$ level would be smaller in size than the number detected in e.g., a deep Spitzer survey. A disc fraction obtained with WISE could thus be higher than the ``true'' fraction inferred from the higher signal-to-noise Spitzer observations. This difference, however, is found to be small ($\sim$3\%), and the disc fractions obtained with WISE in USco and the other comparison regions are found to be consistent within the uncertainties, as discussed in detail in Sections \S\ref{USco} and \S\ref{appendix}. 


What is the smallest excess emission that we could detect for the brown dwarfs with WISE? This can be determined by calculating the ratio of the observed flux to the predicted photospheric flux, (Fobs/Fphot). As we discuss in Section \S\ref{USco}, the 12$\micron$ point is vital in the confirmation of excess sources. The predicted photospheric flux at 12$\micron$ and 22$\micron$ for an M6 dwarf is $\sim$0.25 mJy and $\sim$0.12 mJy, respectively. For the latest type objects in our sample at spectral type of M8-M8.5, the predicted photospheric flux at 12$\micron$ and 22$\micron$ is $\sim$0.15 mJy and $\sim$0.04 mJy, respectively. The WISE 5-$\sigma$ point source sensitivity at 12$\micron$ and 22$\micron$ is $\sim$0.7 mJy and $\sim$5 mJy, respectively (WISE Explanatory Supplement). It is therefore impossible to detect the photospheres for the BDs at these wavelengths, and our survey in these bands is only sensitive to disc detections. If we consider the WISE 5-$\sigma$ detection limit of $\sim$0.7 mJy at 12$\micron$ to be the lowest detectable observed flux, then this implies Fobs/Fphot $\sim$ 2.8 for the M6 sources and Fobs/Fphot $\sim$ 4.6 for the later type M8-M8.5 targets. Therefore, the minimum 12$\micron$ excesses that we can detect will be at the level of Fobs/Fphot $>$ 3--5 for our M6--M8.5 targets. At 22$\micron$, the minimum detectable excess would be Fobs/Fphot $>$ 40 for the M6 sources and Fobs/Fphot $>$ 125 for the later type targets. These large limits would explain why most of our disc sources were not detected in the 22$\micron$ band (Section \S\ref{USco}). At 3.4$\micron$ and 4.6$\micron$, the predicted photospheric fluxes are between $\sim$3 and 0.7 mJy for the M6-M8.5 spectral type range. These fluxes are higher than the WISE 5-$\sigma$ sensitivity of 0.04 mJy and 0.08 mJy in the 3.4$\micron$ and 4.6$\micron$ bands, respectively. Photospheric detection is thus possible in these shorter wavelength bands, and we should be able to detect any small excess above the photosphere. We note that the minimum excess that can be detected at any wavelength would also be dependent on the calibration uncertainty, the finite signal to noise, and the intrinsic spread in the stellar colors. To summarize, the two potential biases in our WISE disc survey are the $\geq$3-$\sigma$ quality criteria, and the minimum excess that can be detected at 12 and 22$\micron$. We have discussed the completeness level of our WISE disc survey in USco in Section \S\ref{USco}, and a comparison with the Spitzer surveys in other young regions is provided in Section \S\ref{appendix}.

\section{WISE SED Classification}
\label{WISE}

\begin{figure*}
\includegraphics[width=120mm]{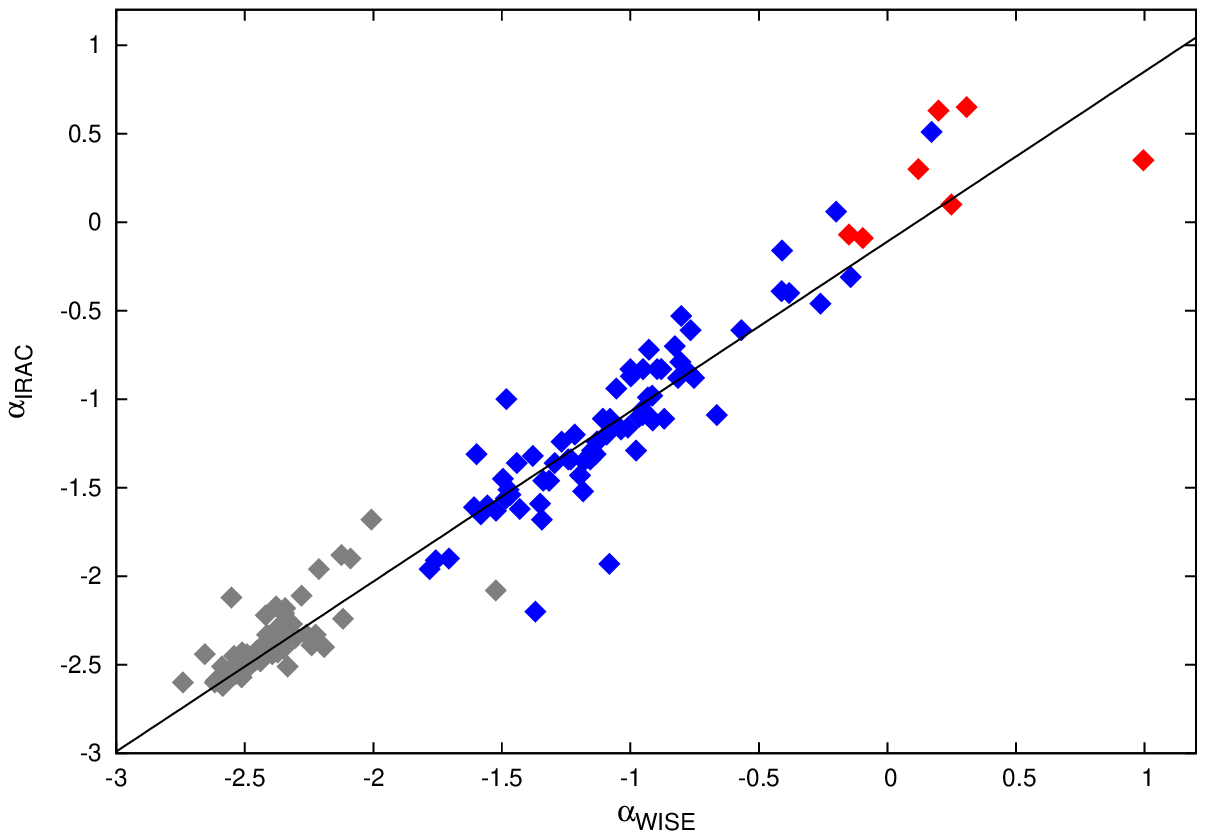}
\caption{The IRAC 2-8$\micron$ slope plotted against WISE 2-12$\micron$ slope for Taurus sources. Symbols are: grey--Class III; blue--Class II; red--Class I  }
   \label{slopes}
 \end{figure*}

\begin{figure*}
\includegraphics[width=120mm]{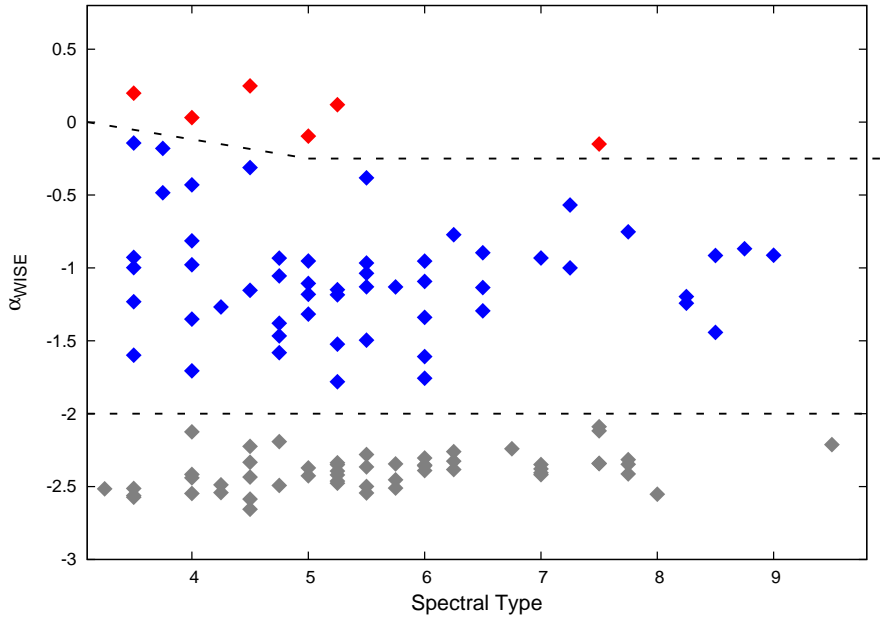}
    \caption{The WISE 2-12$\micron$ slope plotted against spectral type for Taurus sources. Symbols are the same as in Fig.~\ref{slopes}. The SpT of 0-9 imply M0-M9 types. Dashed lines mark the thresholds used for SED classification. }
   \label{slope-spt2}
 \end{figure*}

In order to devise a SED classification scheme based on the observed fluxes in the WISE bands, we started with the {\it Spitzer}/IRAC catalog of disc population in the Taurus star-forming region provided in Luhman et al. (2010). We cross-matched the Taurus sources in the WISE catalog and selected the ones that were detected at a $\geq$5-$\sigma$ level in WISE 1, 2, and 3 bands. This Taurus-WISE catalog consists of 194 sources. Luhman et al. (2010) have computed spectral slopes [$\alpha$ = {\it d} log($\lambda F_{\lambda}$)/{\it d} log($\lambda$)] using four different wavelength ranges: ($K_{s}$/[8.0], $K_{s}$/[24], [3.6]/[8.0], and [3.6]/[24]). They have then marked certain thresholds in the spectral slope versus SpT diagrams to distinguish between the red population of Class I and II sources and the sequence of Class III stars. For most of the sources, the 2-8$\micron$ ($K_{s}$/[8.0]) slope has been used for classification purposes. This particular spectral slope shows a rise in values towards later types of $>$M4. As mentioned, a majority of the USco sources lack detections in the WISE 4 band at 22$\micron$. We have therefore computed the 2-12$\micron$ ($K_{s}$/[12]) spectral slopes to classify these sources. This wavelength range covers four data points at $\sim$2.16, 3.4, 4.6 and 12$\micron$. As we discuss later in this section, including the 12$\micron$ point is essential as it is the only point that can identify the disc candidates. Nearly all of the USco sources show photospheric emission at shorter wavelengths and most are undetected at 22$\micron$. We could use the 3.4-12$\micron$ wavelength range, although it would provide lesser number of points to determine the spectral slopes. We first computed WISE 2-12$\micron$ spectral slopes ($\alpha_{WISE}$) for the Taurus sources, and compared these to the IRAC spectral slopes of 2-8$\micron$ from Luhman et al. (2010). Fig.~\ref{slopes} shows that the two slopes are well correlated, as indicated by the straight line fit to the plotted points. We have plotted with different colors the Taurus Class I/II/III sources. These are the SED classes assigned by Luhman et al. (2010) using the IRAC 2-8$\micron$ slopes. The Class III sources in Fig.~\ref{slopes} have $\alpha_{WISE}$ of $\leq$ -2, the Class II sources are mainly concentrated between $\alpha_{WISE}$ of $\sim$ -1.8 and $\sim$ -0.2, while the Class I sources have $\alpha_{WISE}$ $\geq$ -0.2. The Class III sources show a more clear separation from the Class II systems. The boundary between Class I and II is less distinguished. Fig.~\ref{slope-spt2} shows $\alpha_{WISE}$ plotted against the SpT for the Taurus sources. There is a lack of sources with SpT later than M8 that have good S/N detections in the 12$\micron$ band, and thus do not have $\alpha_{WISE}$ estimates. The threshold of $\alpha_{WISE}$ = -2 provides a good separation between the Class II and III sources. There may be a rise in this threshold at later types, such as the threshold may be at -1.5 at SpT of M8.5 and later. However, we do not have enough points at later types to confirm a rise in the Class II/III threshold. The threshold between Class I and II is around $\sim$ -0.2, although a higher value of $\alpha_{WISE}$$\sim$ 0.0 is required to distinguish between these two classes at earlier types. 



\begin{figure*}
\includegraphics[width=75mm]{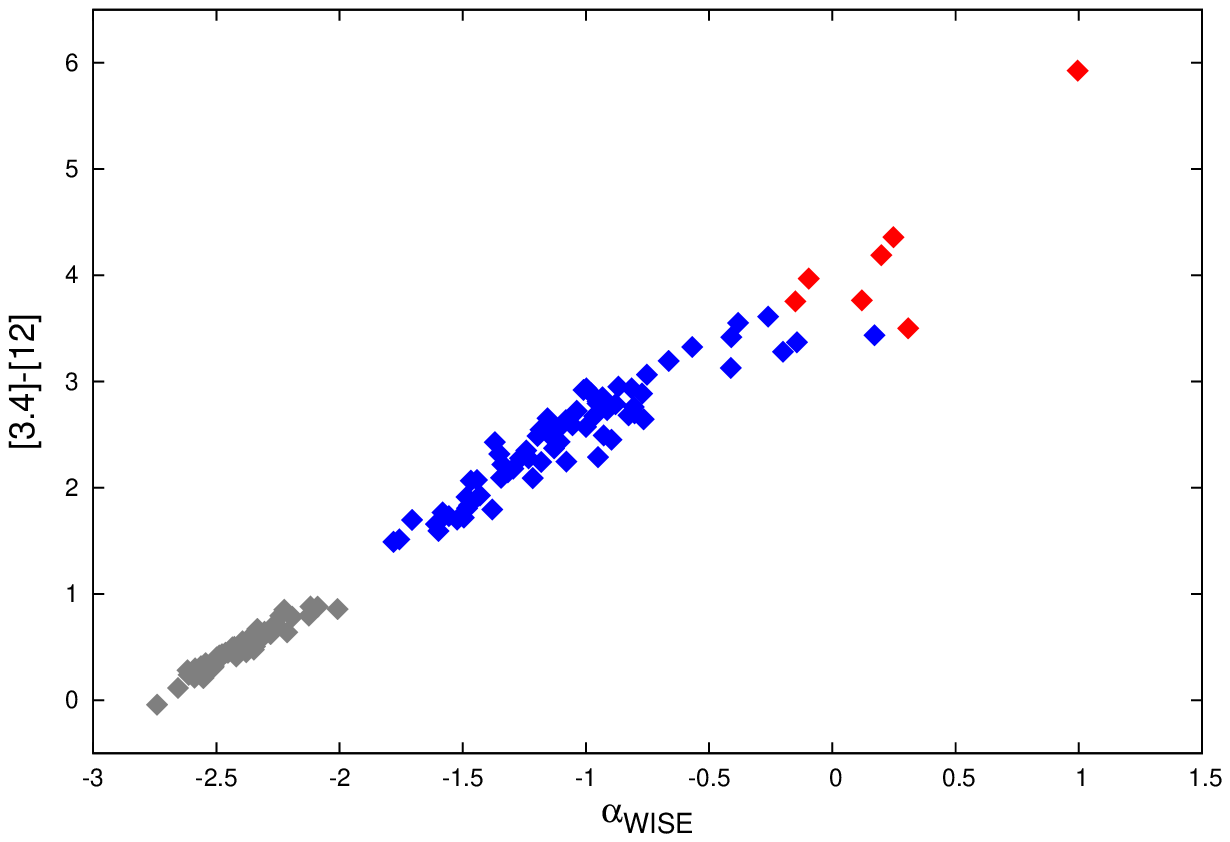} 
\includegraphics[width=75mm]{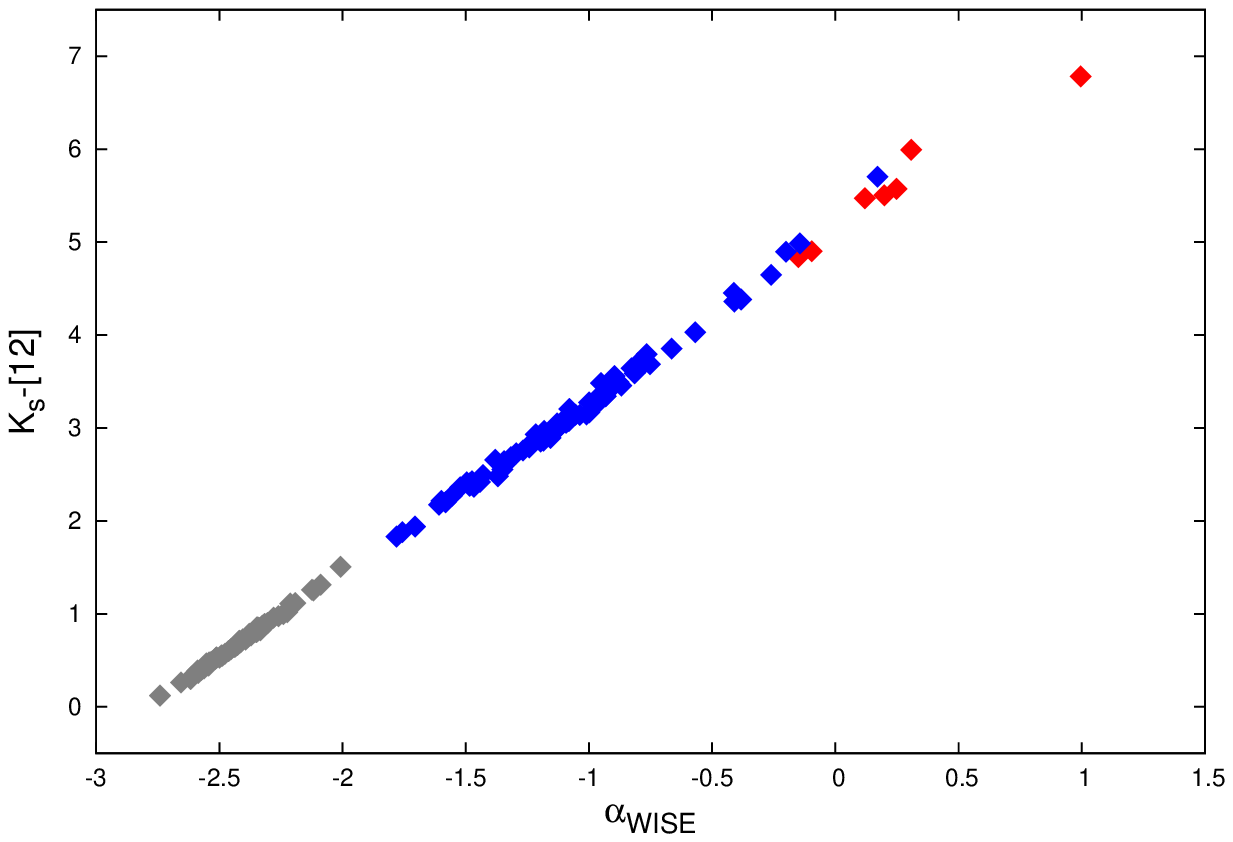} \\
\includegraphics[width=75mm]{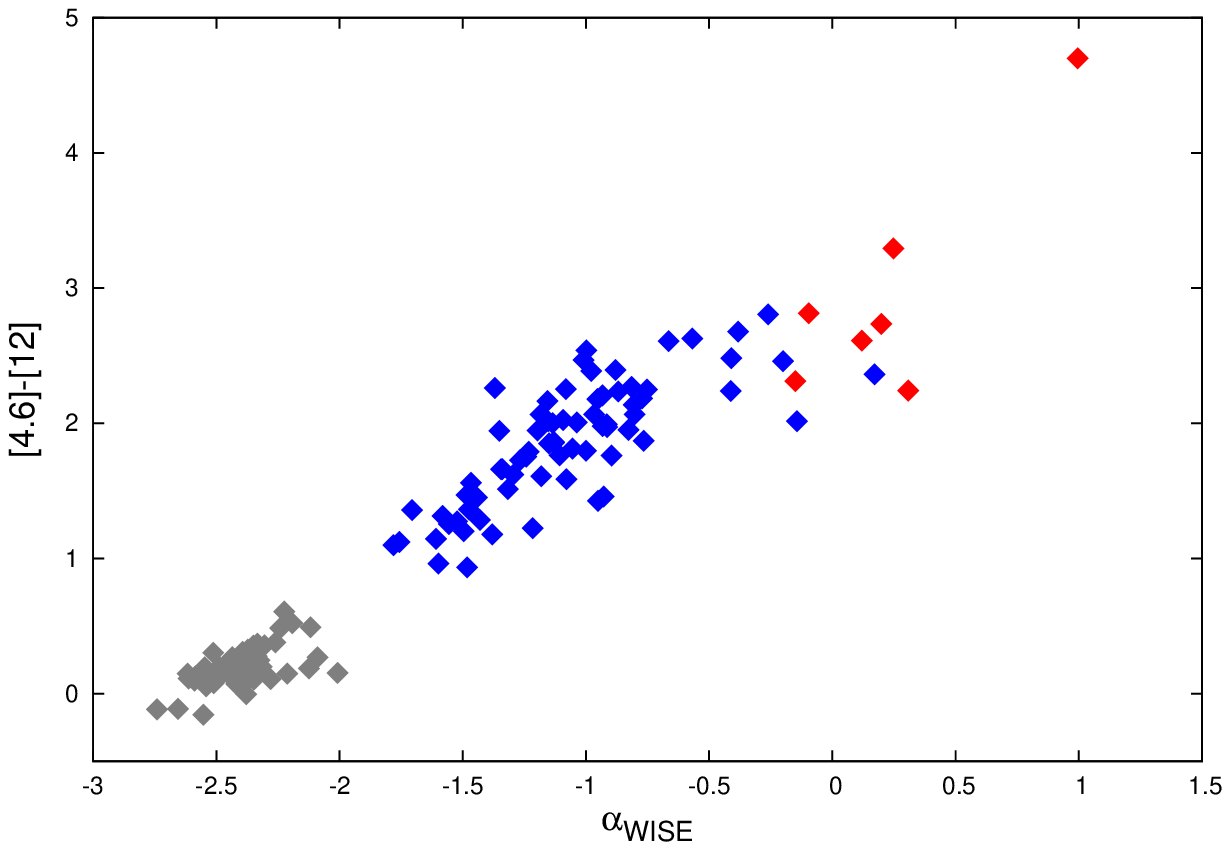} 
 \includegraphics[width=75mm]{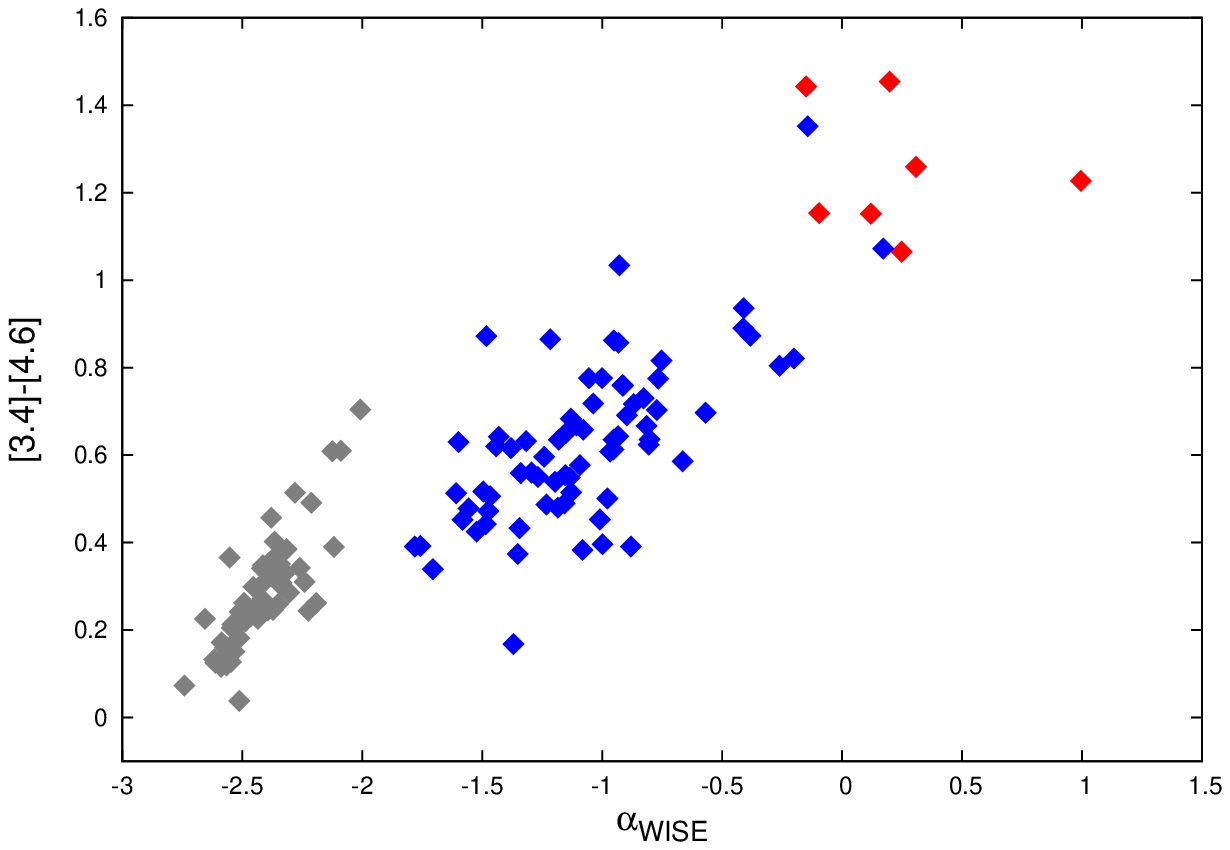} \\
  \includegraphics[width=75mm]{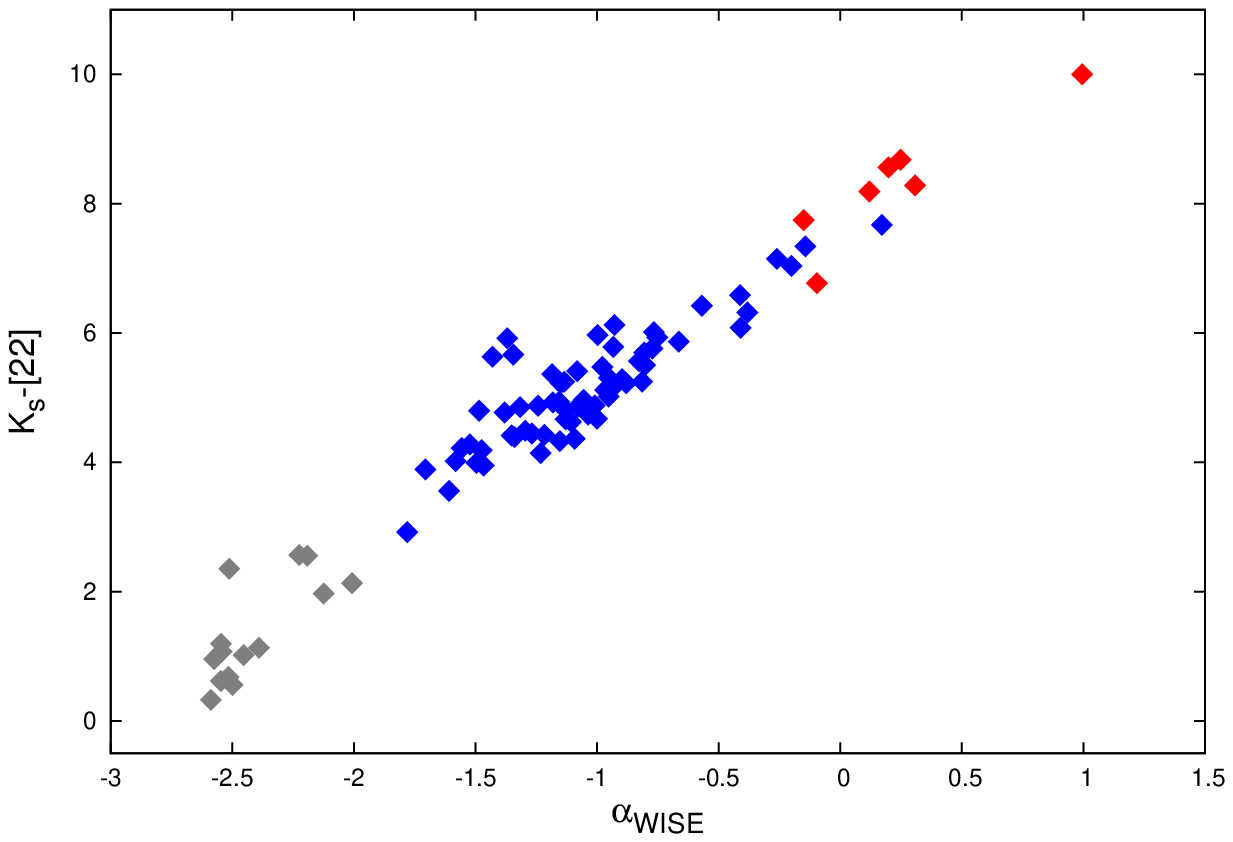}
   \includegraphics[width=75mm]{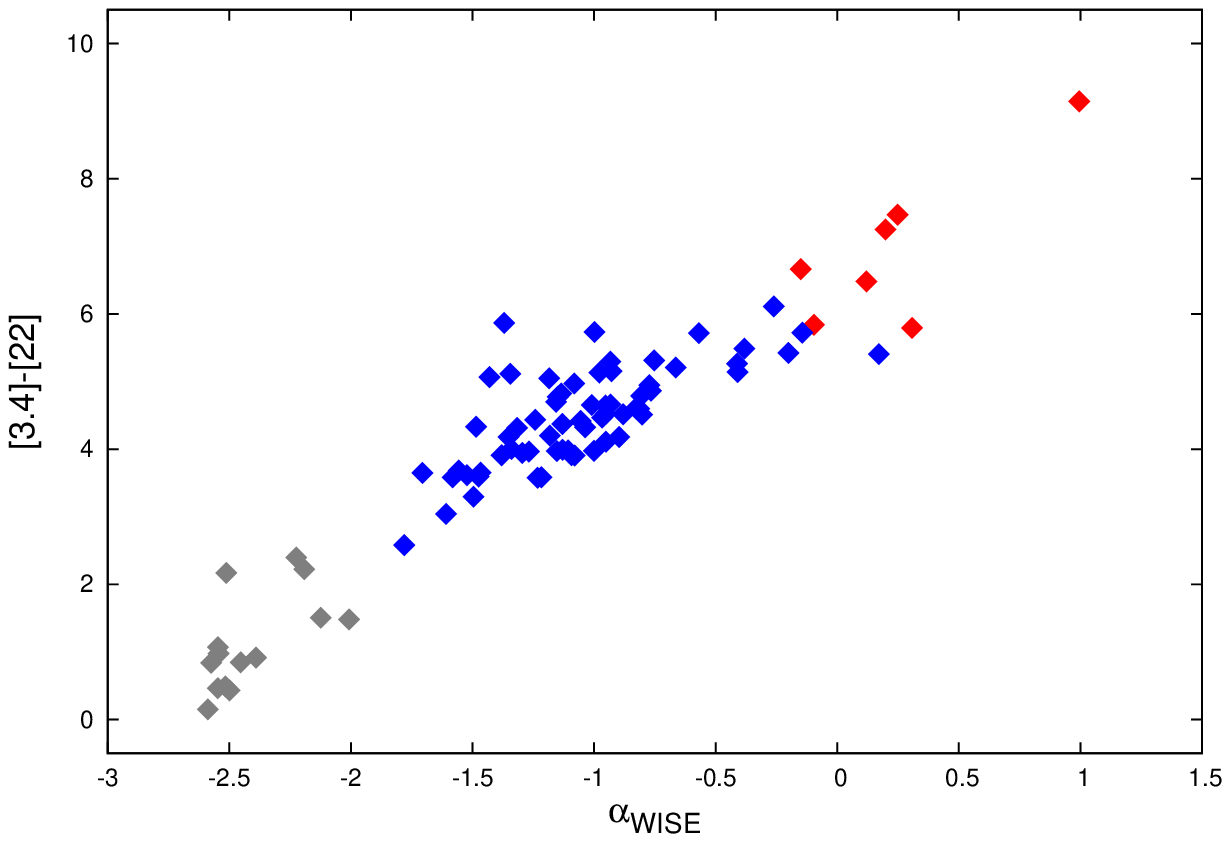} \\
    \includegraphics[width=75mm]{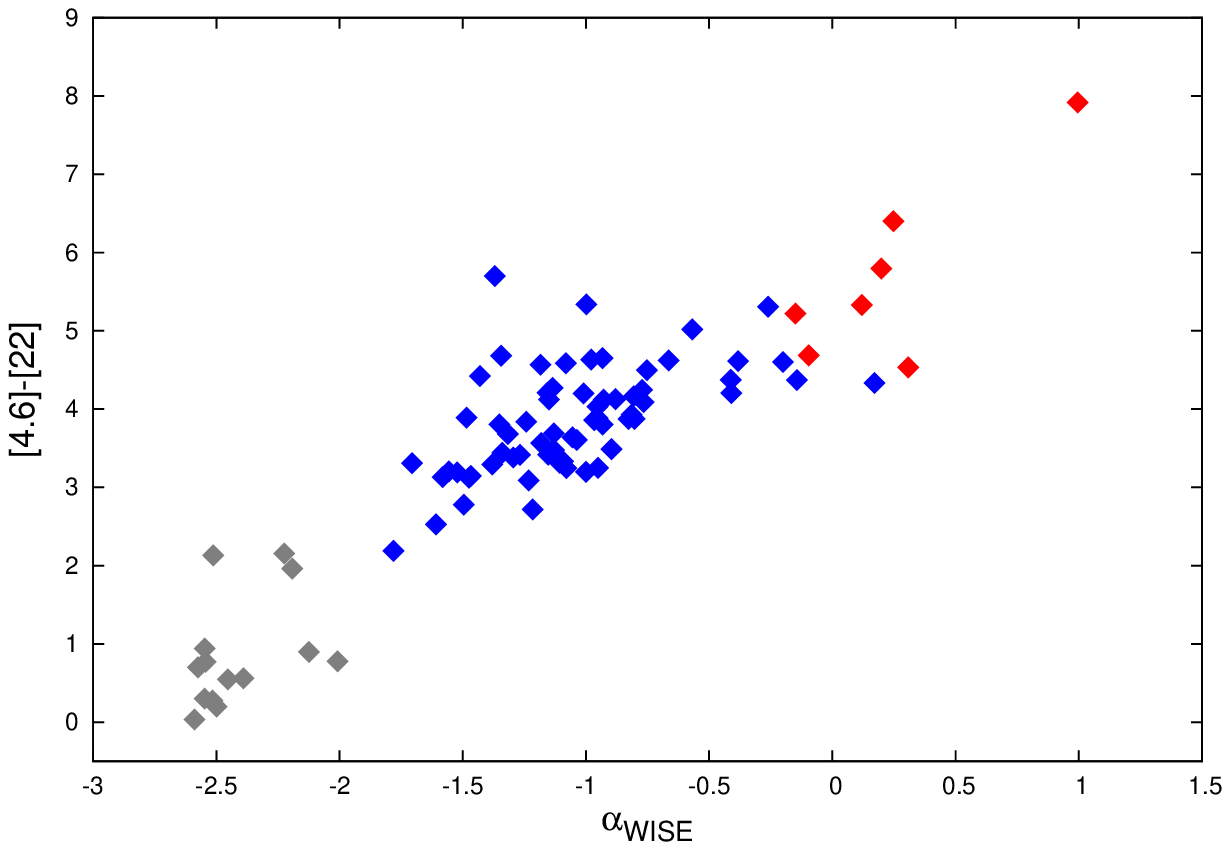}
     \includegraphics[width=75mm]{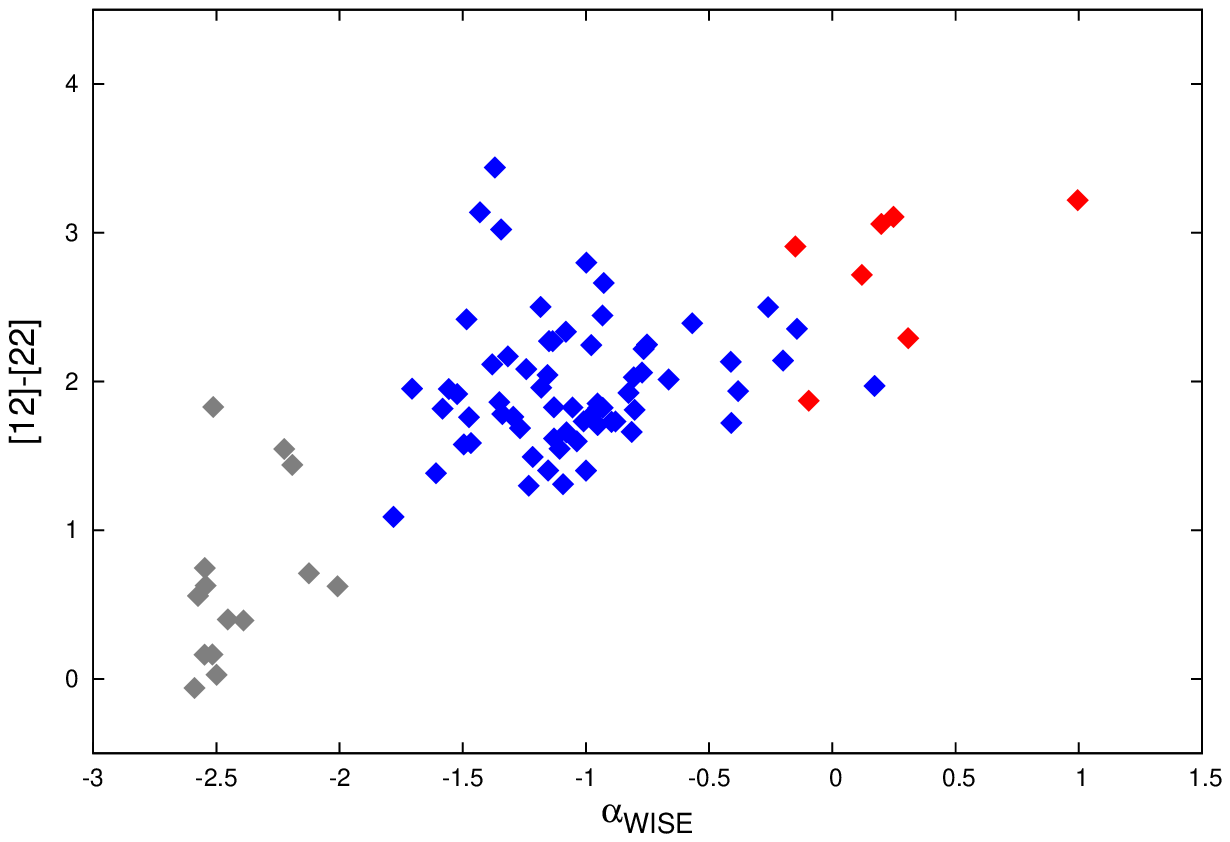} \\  
    \caption{$\alpha_{WISE}$ versus various WISE colors for Taurus sources. The [3.4]-[12] and $K_{s}$-[12] colors show the strongest correlation with the slope ({\it top panel}). Typical uncertainties in the colors are between 0.01 and 0.08 mag.  }
   \label{W-slopes}
 \end{figure*}

In Fig.~\ref{W-slopes}, we have compared the WISE spectral slopes with various WISE and WISE-NIR colors for the Taurus sources. For colors including the 22$\micron$ photometry, we have only plotted the objects with a S/N$\geq$3 in the 22$\micron$ band. The [3.4]-[12] and $K_{s}$-[12] colors ({\it top panel}) are well correlated with the WISE slopes and can best distinguish between the different SED classes. We can consider a boundary of [3.4]-[12] $\sim$1.2 and $K_{s}$-[12] $\sim$1.8 to separate the Class III sources from the disc population of Class I and II systems. The color [4.6]-[12] ({\it 2nd panel}) also provides a good separation between Class II and Class III sources (boundary of [4.6]-[12]$\sim$1). The color [3.4]-[4.6] is vague in separating the Class II and III sources, but can be used to identify the protostellar systems ([3.4]-[4.6]$>$1). The bottom two panels in Fig.~\ref{W-slopes} plot the WISE colors which include the 22$\micron$ photometry. The typical range in these colors for the photospheric sources is between 0 and 1. However, there are a few evolved/transition discs among the Class III objects which show redder colors between $\sim$1 and 3. A transition disc exhibits photospheric emission at wavelengths shortward of $\sim$10$\micron$ due to inner disc clearing, but retains excess emission at longer wavelengths due to the presence of circumstellar material in the outer disc regions. Such transition discs can be more clearly separated from the main Class III locus in the [4.6]-[22] and the [12]-[22] colors (boundary of $>$1). To summarize, colors which include the 12$\micron$ magnitudes can more clearly make a distinction between the photospheric and the disc population, whereas the colors which include the 22$\micron$ magnitudes can be important in identifying the transition disc candidates. 



\begin{figure*}
\includegraphics[width=190mm]{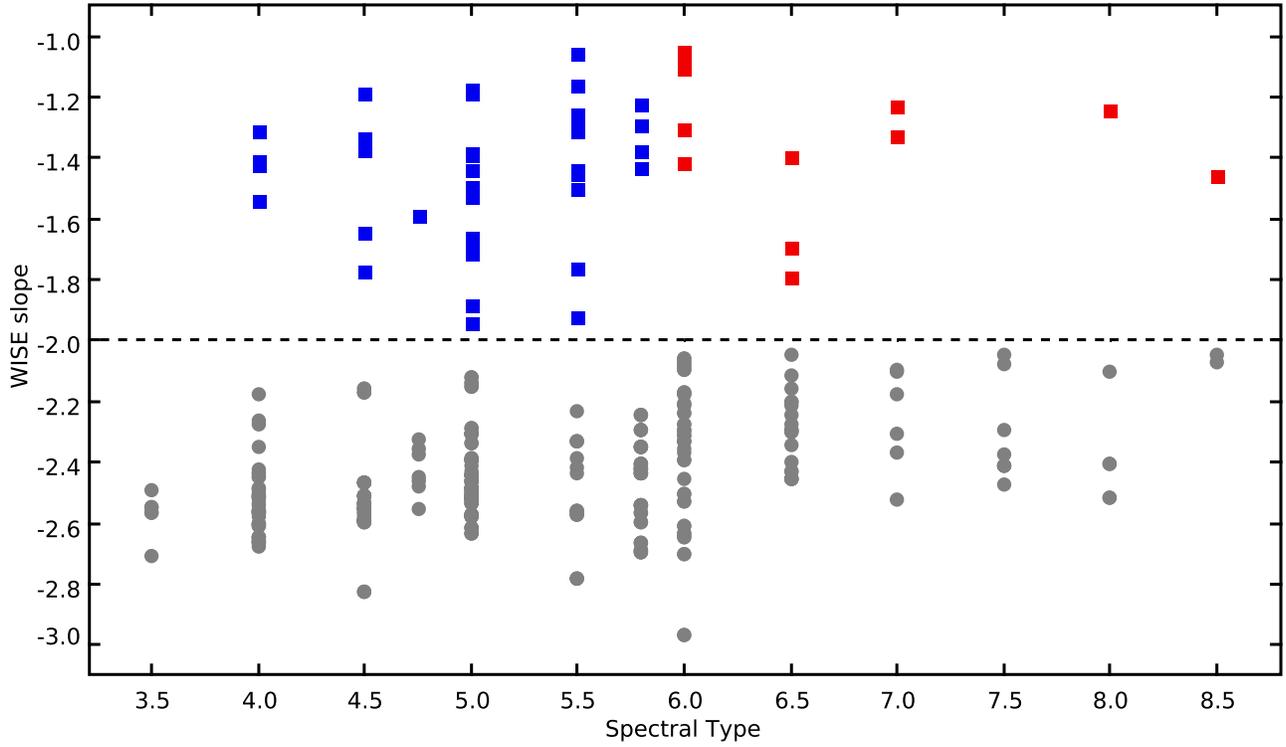}
    \caption{The WISE 2-12$\micron$ slope plotted against spectral type for our USco sample. The BD and M dwarf discs found in this work are indicated by red and blue symbols, respectively.  The SpT of 3.5-8.5 imply M3.5-M8.5 types. }
   \label{slope-spt-USco}
 \end{figure*}

Based on the analysis above, we have classified the USco sources as Class III if $\alpha_{WISE}$ $\leq$ -2.0, and as Class II if -2.0 $<$$\alpha_{WISE}$$<$ -0.2. Figure~\ref{slope-spt-USco} shows the SpT vs. $\alpha_{WISE}$ diagram for the full USco sample. The disc sources identified are above the threshold of $\alpha_{WISE}$ = -2.0, and have been classified as Class II systems. The extinction estimate in USco is A$_{V}$ $\leq$ 2 mag (Walter et al. 1994). We have therefore not applied any reddening corrections to the spectral slopes. The values for $\alpha_{WISE}$ and the SED classes assigned to the USco sources are listed in Table~\ref{USco-phot}.







\section{Results}
\label{results}

\subsection{New Discs in USco}
\label{USco}

\begin{figure*}
\includegraphics[width=50mm]{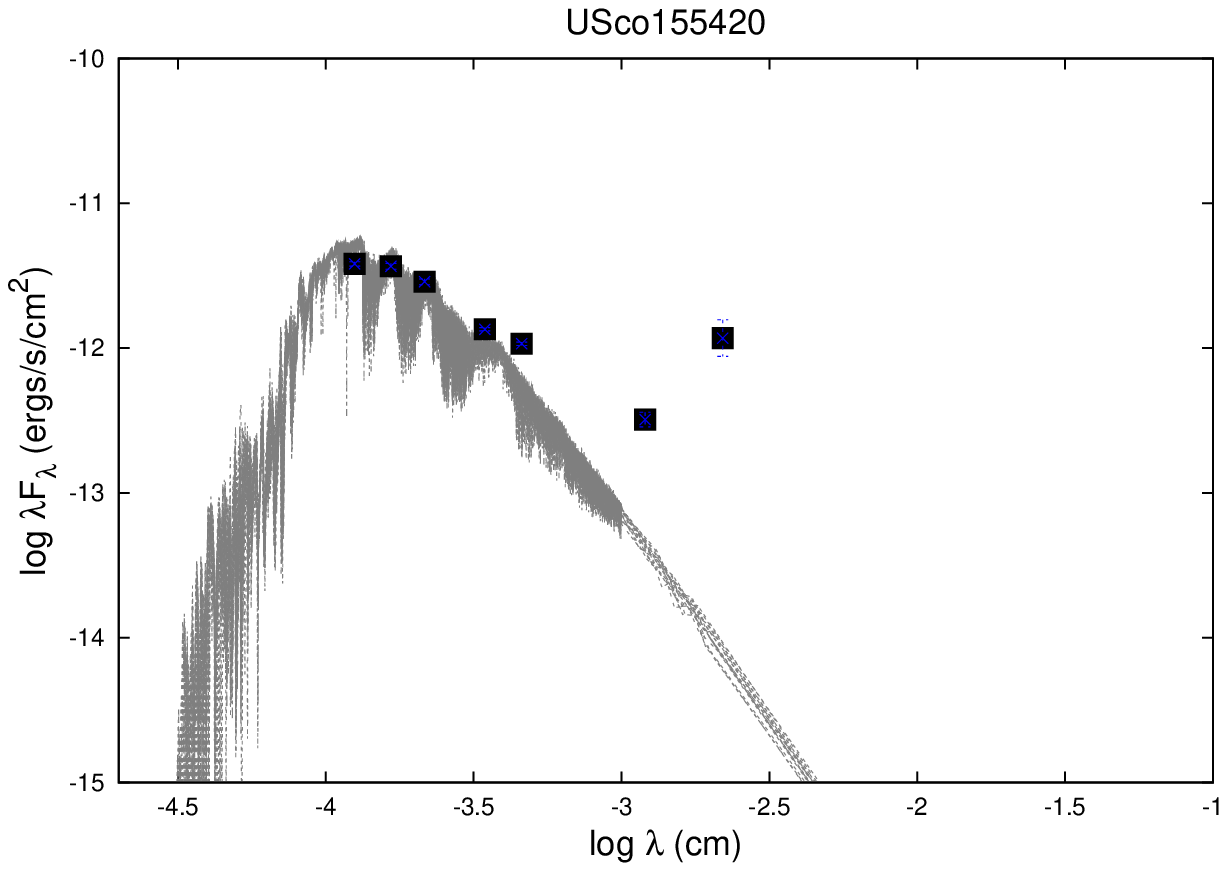} 
\includegraphics[width=50mm]{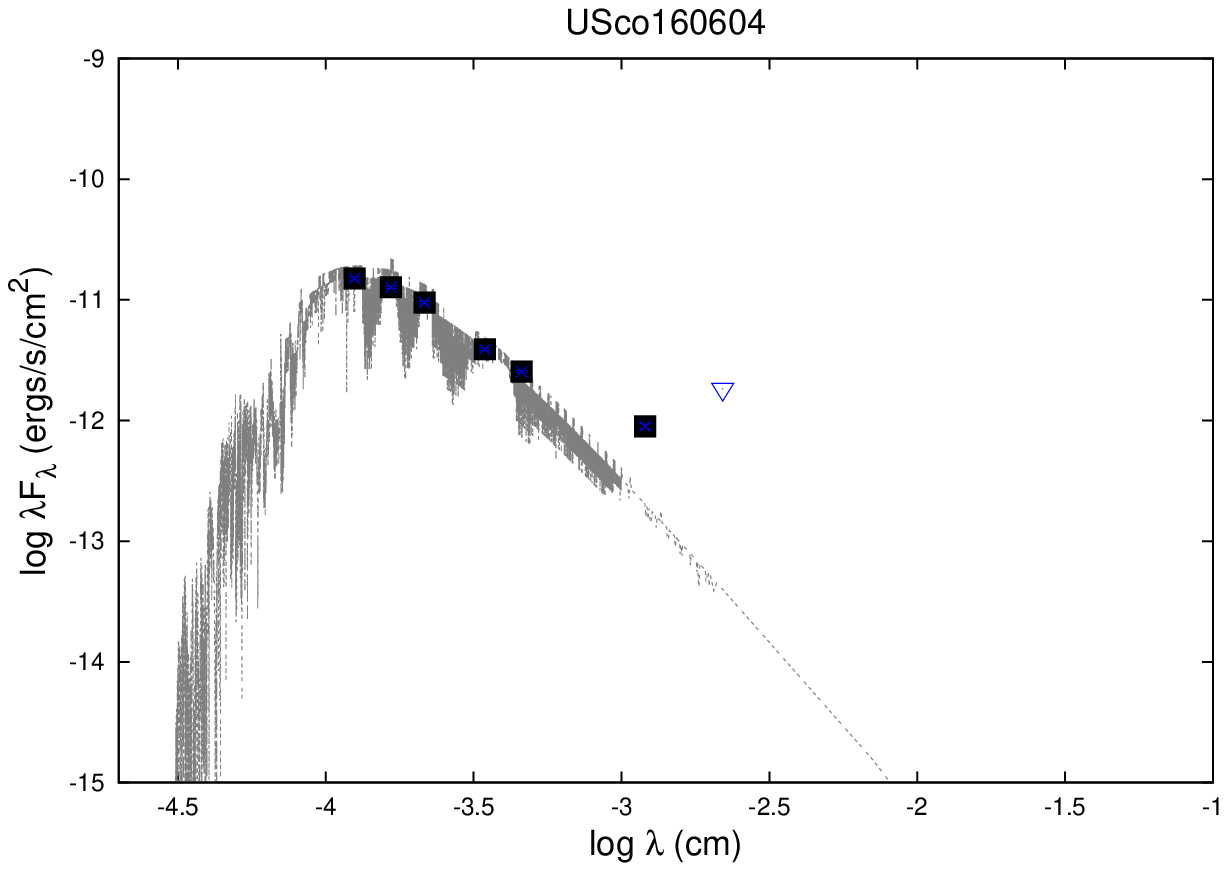} 
\includegraphics[width=50mm]{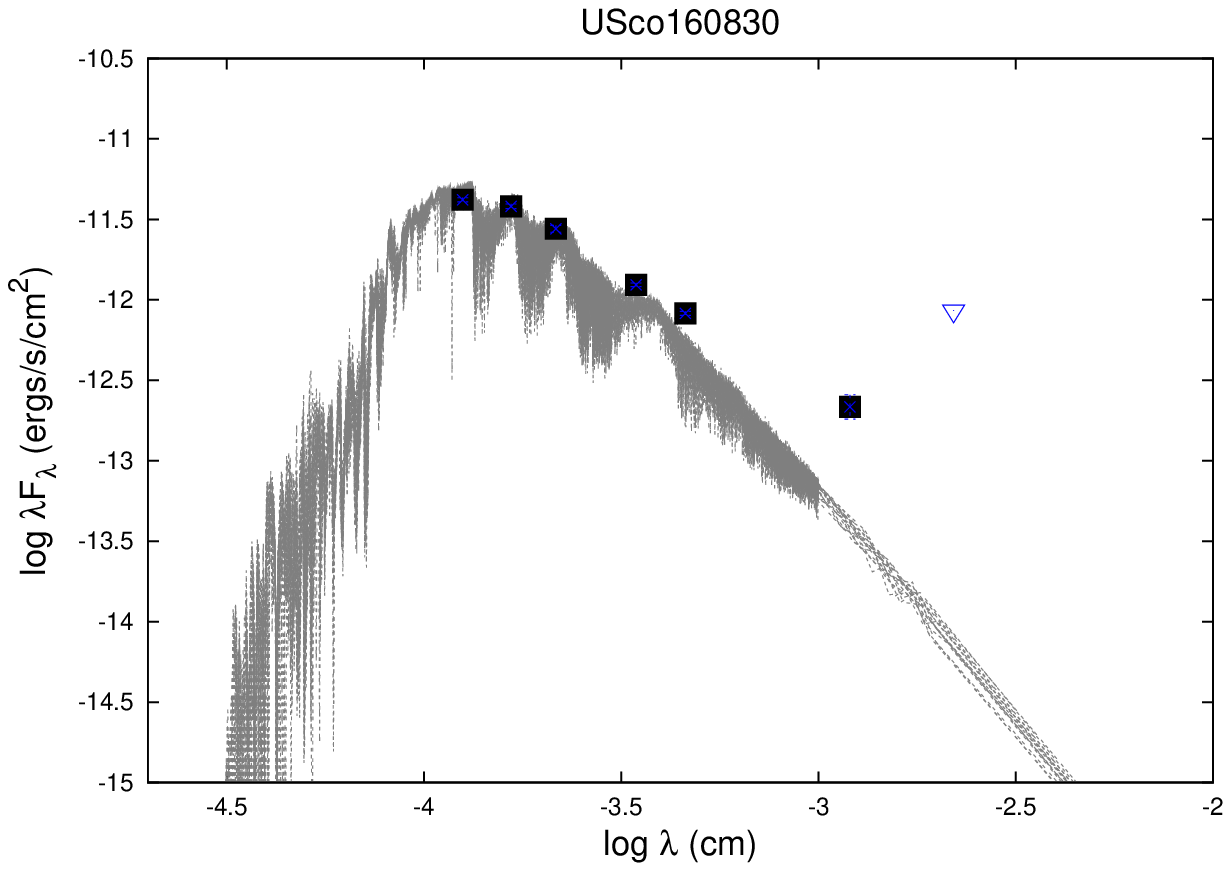} \\
\includegraphics[width=50mm]{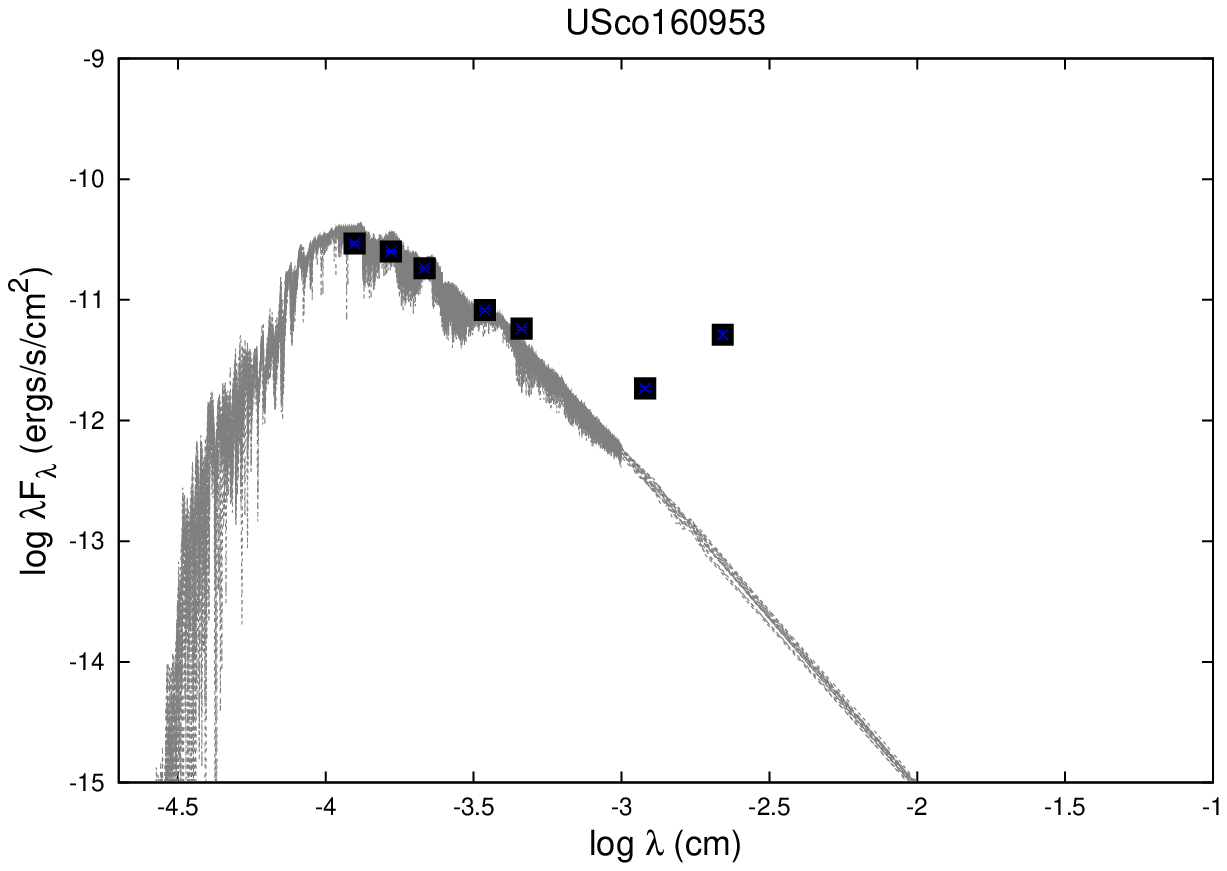} 
\includegraphics[width=50mm]{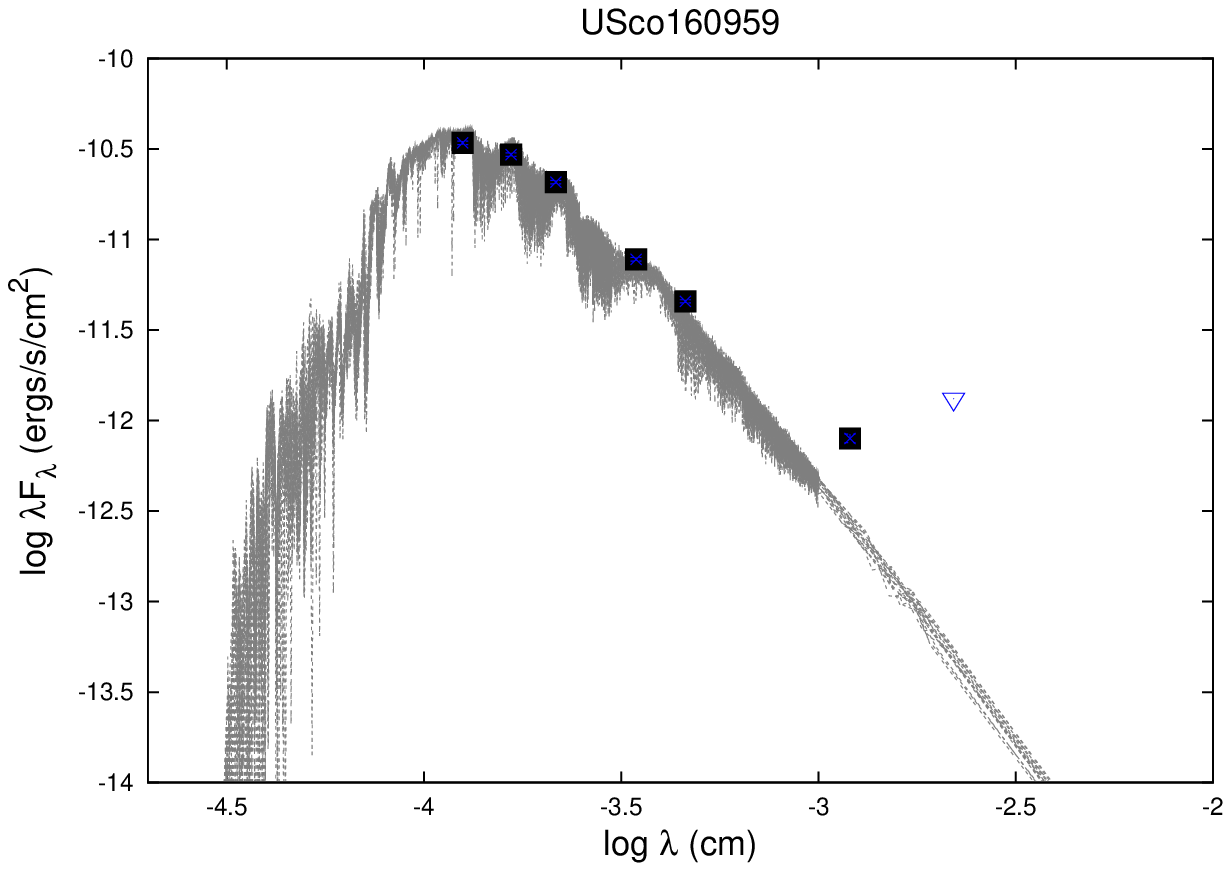} 
 \includegraphics[width=50mm]{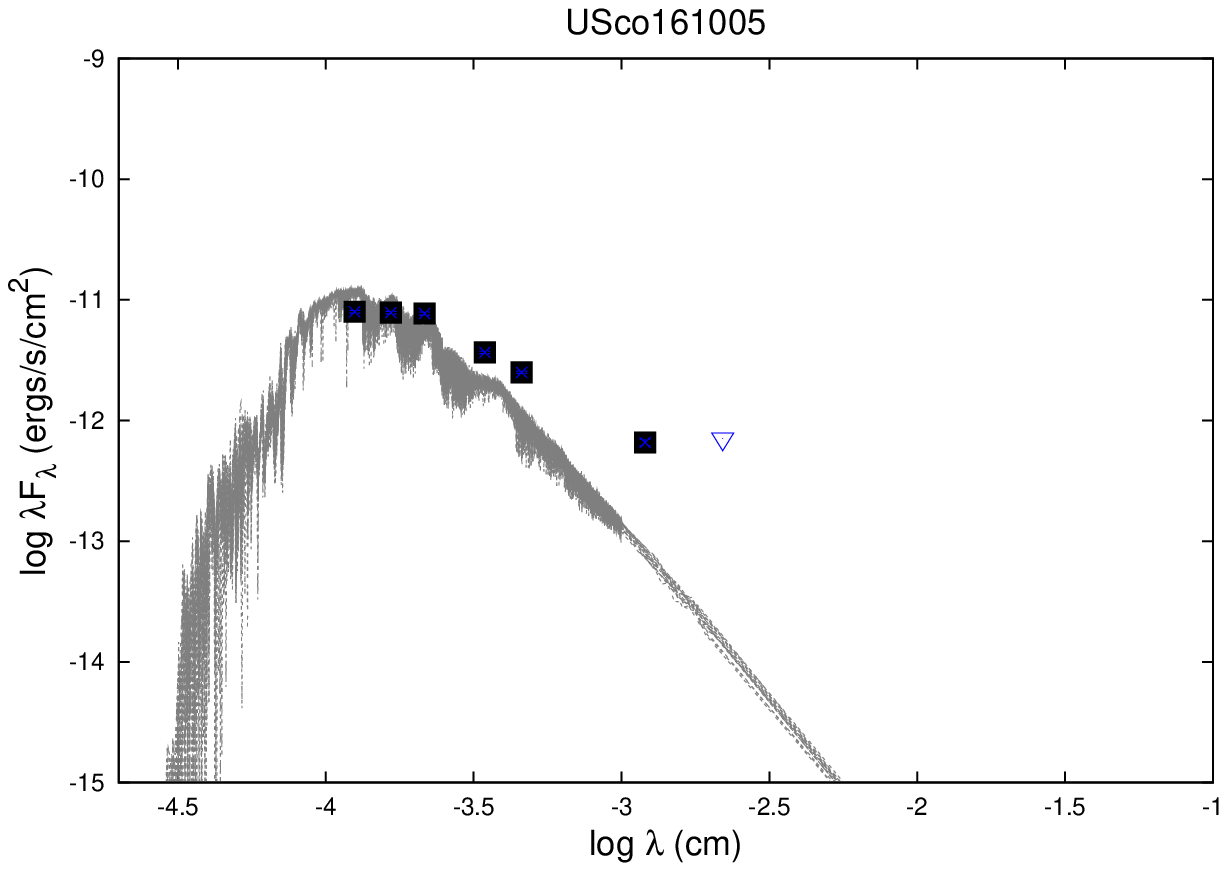} \\
\includegraphics[width=50mm]{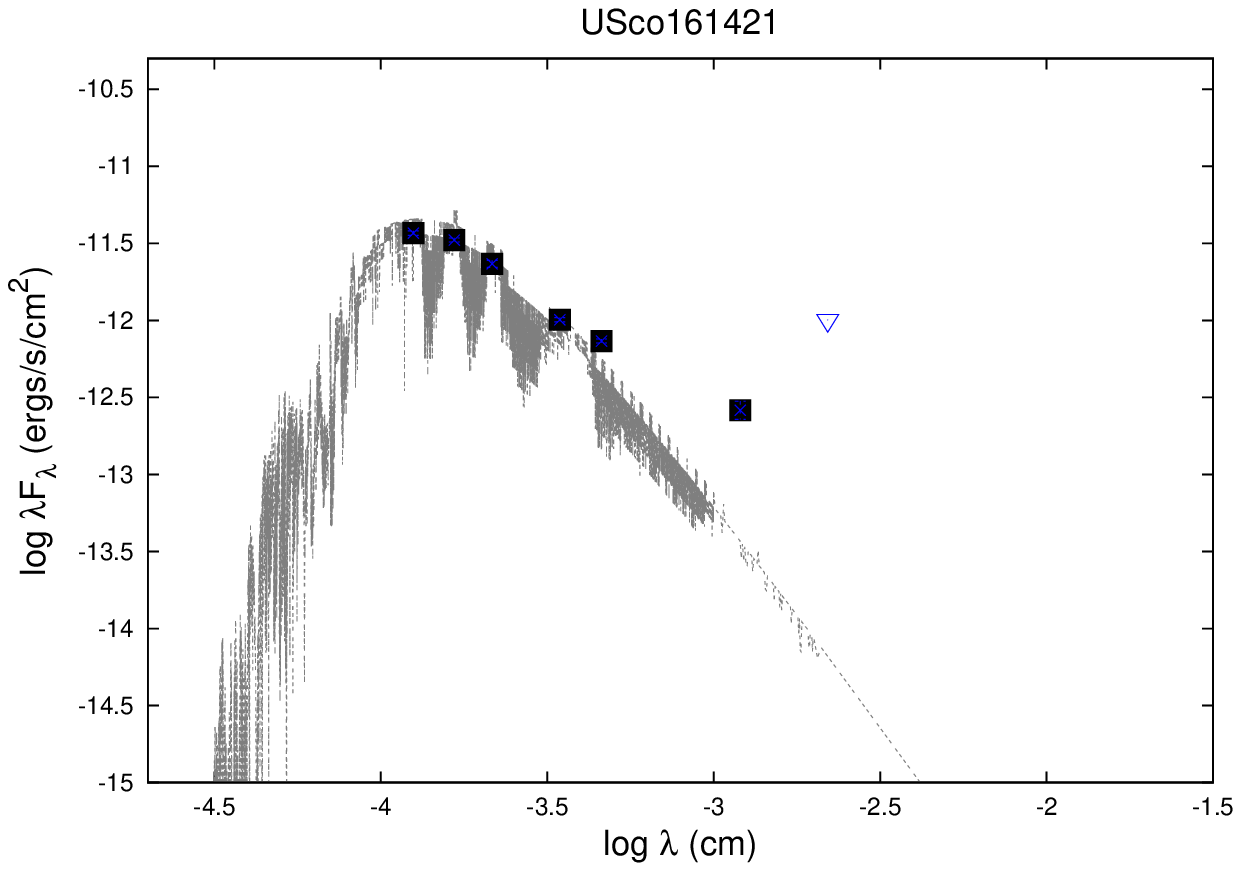} 
\includegraphics[width=50mm]{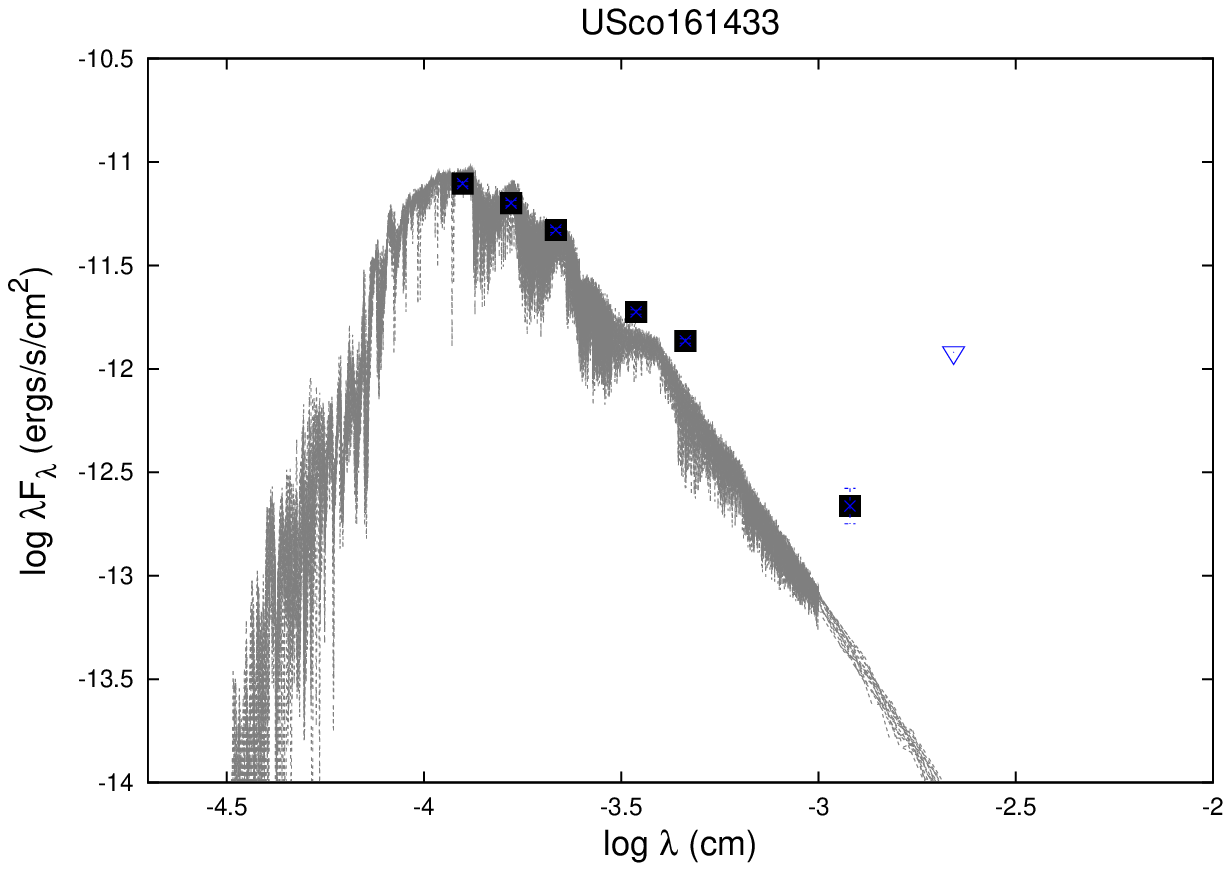} 
\includegraphics[width=50mm]{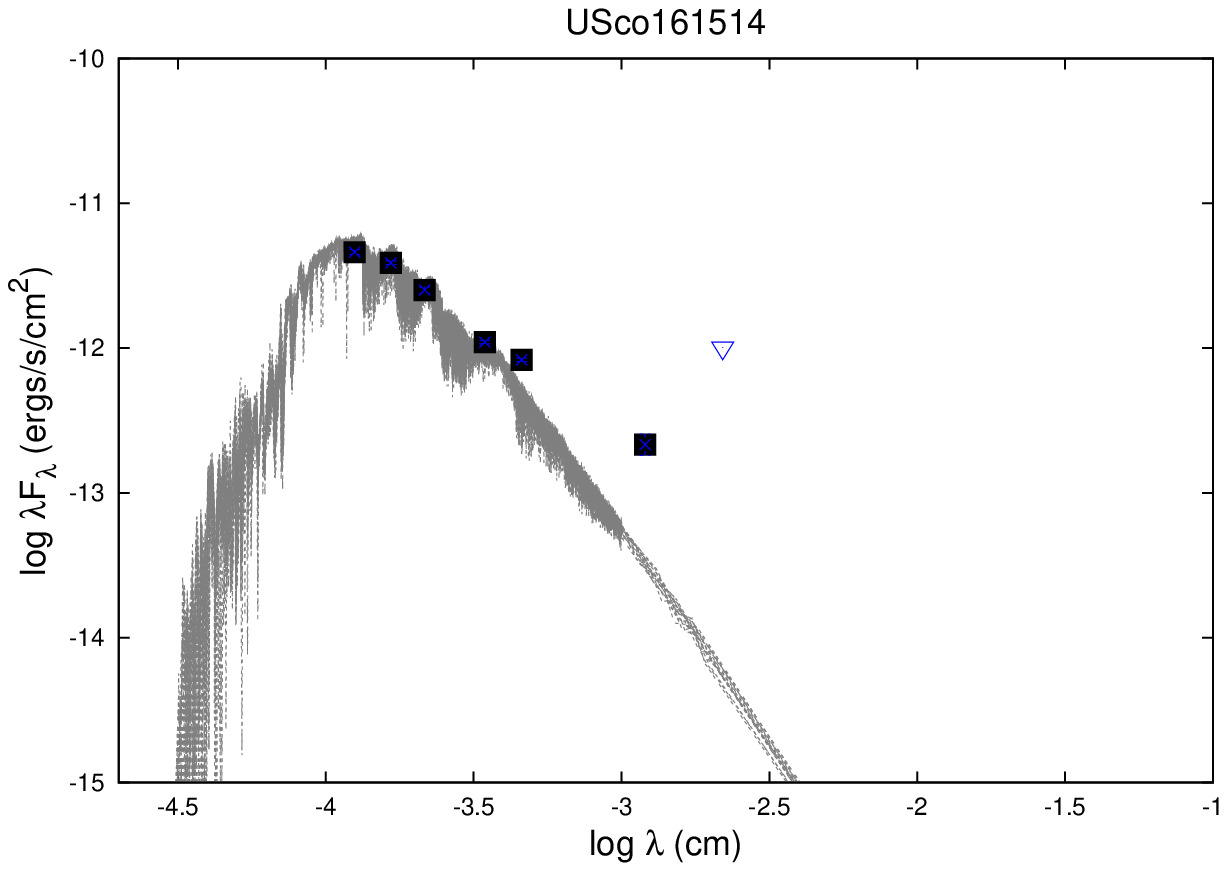} \\
\includegraphics[width=50mm]{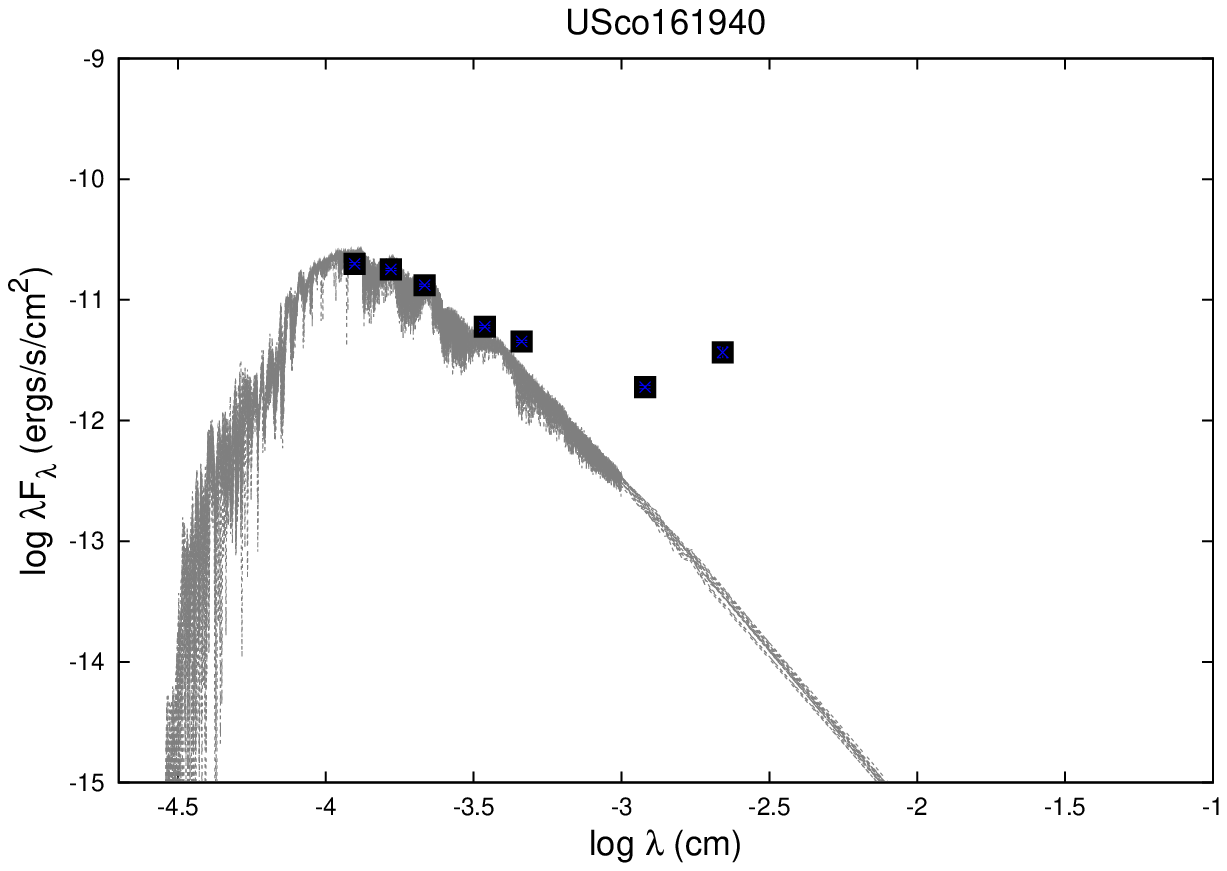} 
 \includegraphics[width=50mm]{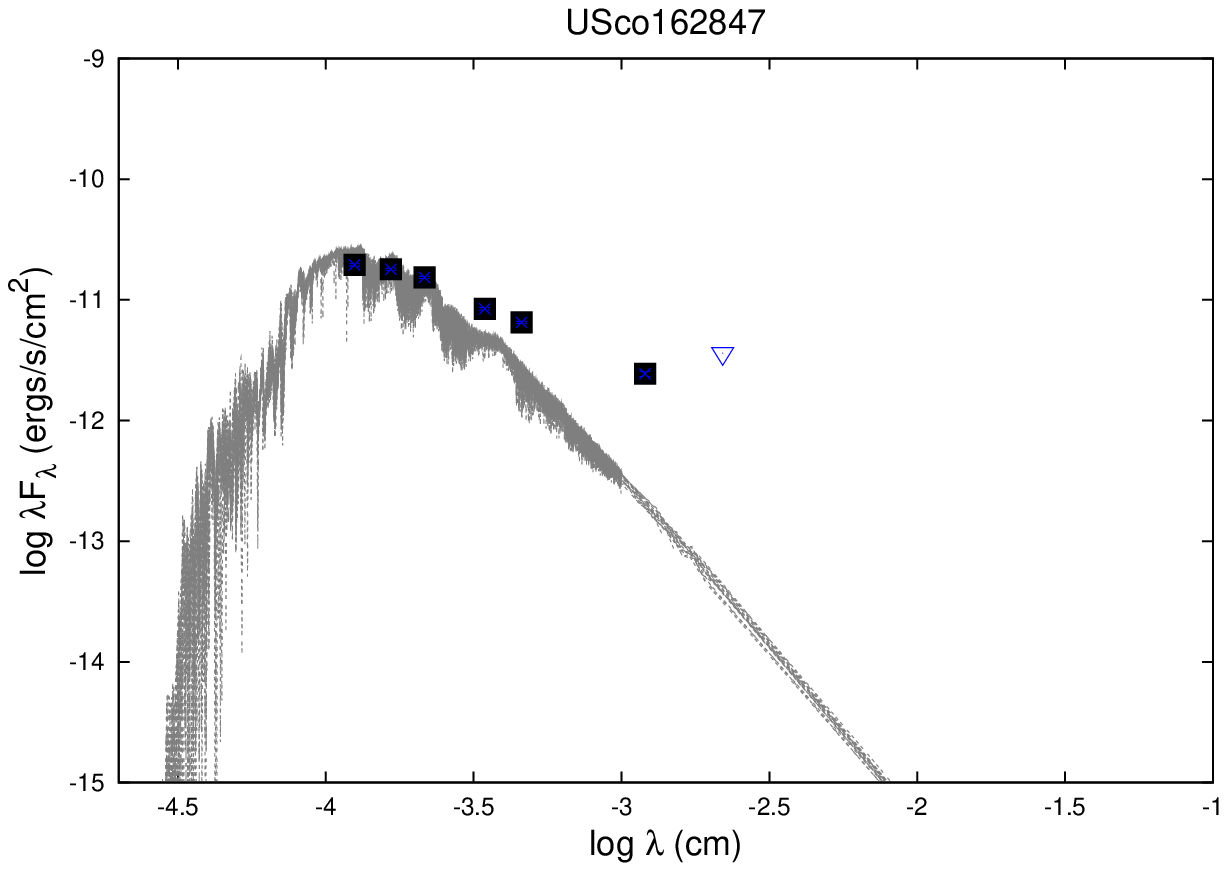} 
\includegraphics[width=50mm]{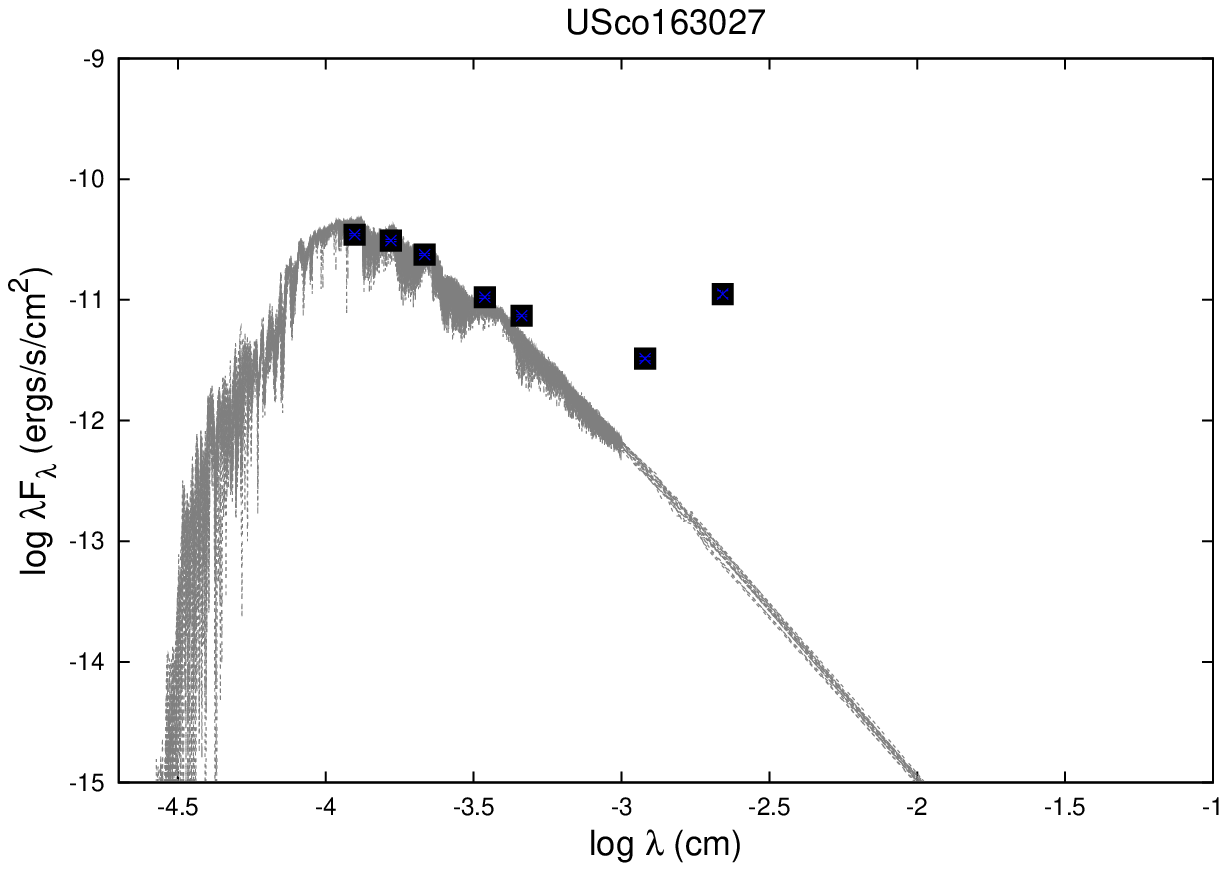} \\
    \caption{The BD discs in USco.}
   \label{USco-BD}
 \end{figure*}

Out of a sample of 215 sources, we have found 48 new discs in USco, 12 of which are BD discs. The SEDs for the disc sources are shown in Figs.~\ref{USco-BD} and \ref{USco-Mdwarf}. To fit the stellar photosphere, we have considered the NextGen models (Hauschildt et al. 1999) for solar metallicity and log {\it g} = 3.5. The effective temperature was obtained from the SpT using the relation for young sources in Luhman et al. (2003a). We have checked the various flags associated with the WISE photometry. Out of the 48 disc sources, 40 have `cc\_flag' of '0000', indicating that these are not affected by any artifacts. The other disc sources either have cc\_flag=h000 or d000, implying that the photometry in band 1 may be affected by the halo of a nearby bright source or a diffraction spike. However, we have checked the WISE and UKIDSS images and do not find the sources to be extended or blended with any bright nearby star, or a spike located close to or across the object. The photometric quality flag, ph\_qual, for these sources is either 'AAAA' or 'AAAB'. Thus even if band 1 may be affected by some artifact, the photometric quality is the 'highest', so these should be reliable detections. One source (USco162847) is blended with a bright nearby star in bands 1, 2 and 3 (cc\_flag is 'h' in these 3 bands). This object has ph\_qual = 'AAAB', indicating 'highest' quality photometry in bands 1, 2, and 3. This source is resolved into 2 bright stars in the UKIDSS GCS images but lies close to a nearby bright star in the WISE images. The WISE resolution is 6.5$\arcsec$ which is large compared to seeing-limited images taken by UKIDSS. Since the composite source shows an excess emission at 12 and 22$\micron$, we have considered this source as a disc candidate. The WISE photometry for the disc sources is listed in Table~\ref{USco-phot}. The value of `99.999' in Table~\ref{USco-phot} indicates a non-detection in that band.

\begin{figure*}
\includegraphics[width=160mm]{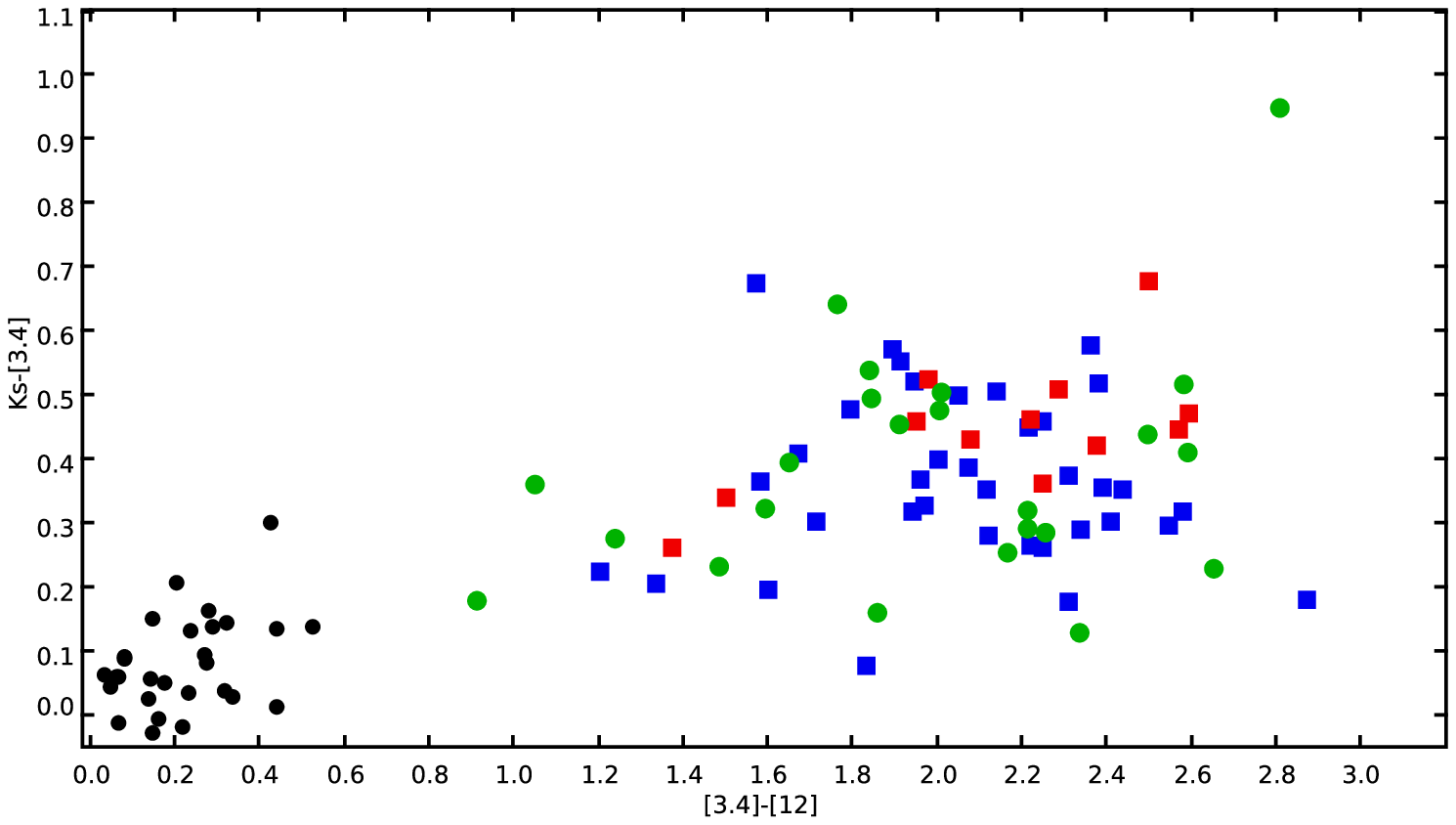} \\
\includegraphics[width=160mm]{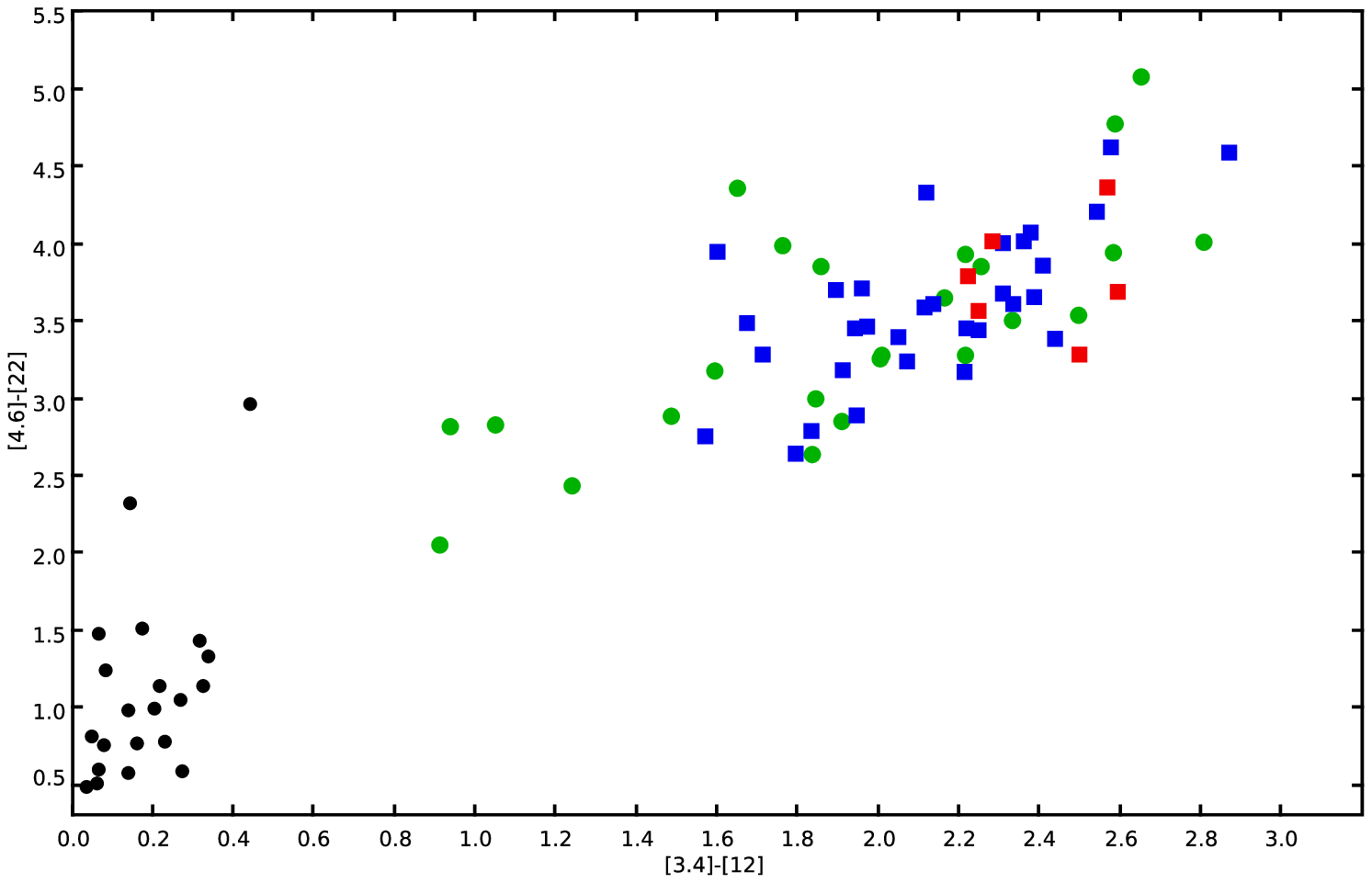} \\
   \caption{WISE and WISE-NIR ccds for the USco sources. The BD and M dwarf discs found in this work are indicated by red and blue symbols, respectively. The primordial discs from Carpenter et al. (2009) are indicated by green circles. Black symbols denote the debris discs from Carpenter et al. survey. The range in uncertainties  are 0.03-0.06 mag for the $K_{s}$-[3.4] color, 0.03-0.15 mag for the [3.4]-[12] color, and 0.04-0.2 mag for the [4.6]-[22] color. In the bottom panel, only the sources with a $\geq$3-$\sigma$ detection in the 22$\micron$ band have been plotted. } 
   \label{USco-color}
 \end{figure*}

In Fig.~\ref{USco-color}, we have compared the WISE colors for the disc sources from our work with the primordial and debris discs in USco found by Carpenter et al. (2009). The range in uncertainties  are 0.03-0.06 mag for the $K_{s}$-[3.4] color, 0.03-0.15 mag for the [3.4]-[12] color, and 0.04-0.2 mag for the [4.6]-[22] color. The colors for all primordial discs are much redder than the debris disc sources. Carpenter et al. have noted a high fraction of primordial discs among the M-type stars, the K/G types are mainly all debris discs with a few primordial discs, while the higher mass B/A/F stars are all debris disc systems. Their categorization of a primordial disc is based on excess emission observed at 8 and 16$\micron$, while a debris disc is classified based on excess emission at $\geq$24$\micron$ (Section \S\ref{appendix}). Thus these primordial discs can be considered as primordial transition discs. The age of $\sim$5 Myr for USco is considered to be a transitional age, when circumstellar discs are in a transition from a primordial to a debris disc system. A majority of disc sources at this age are found to exhibit inner disc clearing. In the literature, different definitions have been used to distinguish between a primordial transition disc and a debris disc. A primordial transition disc is basically defined to be a disc that shows photospheric emission shortward of $\sim$10$\micron$, indicating that the inner $\sim$0.1-1 AU disc region (depending on the mass of the central source) is devoid of optically thick material, but it retains strong excess emission at longer wavelengths of $>$10$\micron$, which indicates the presence of primordial material in the outer disc regions. The color excesses at longer wavelengths for the primordial transition discs would be comparable in strength to the main Class II population in the cluster. 

\begin{figure*}
\includegraphics[width=140mm]{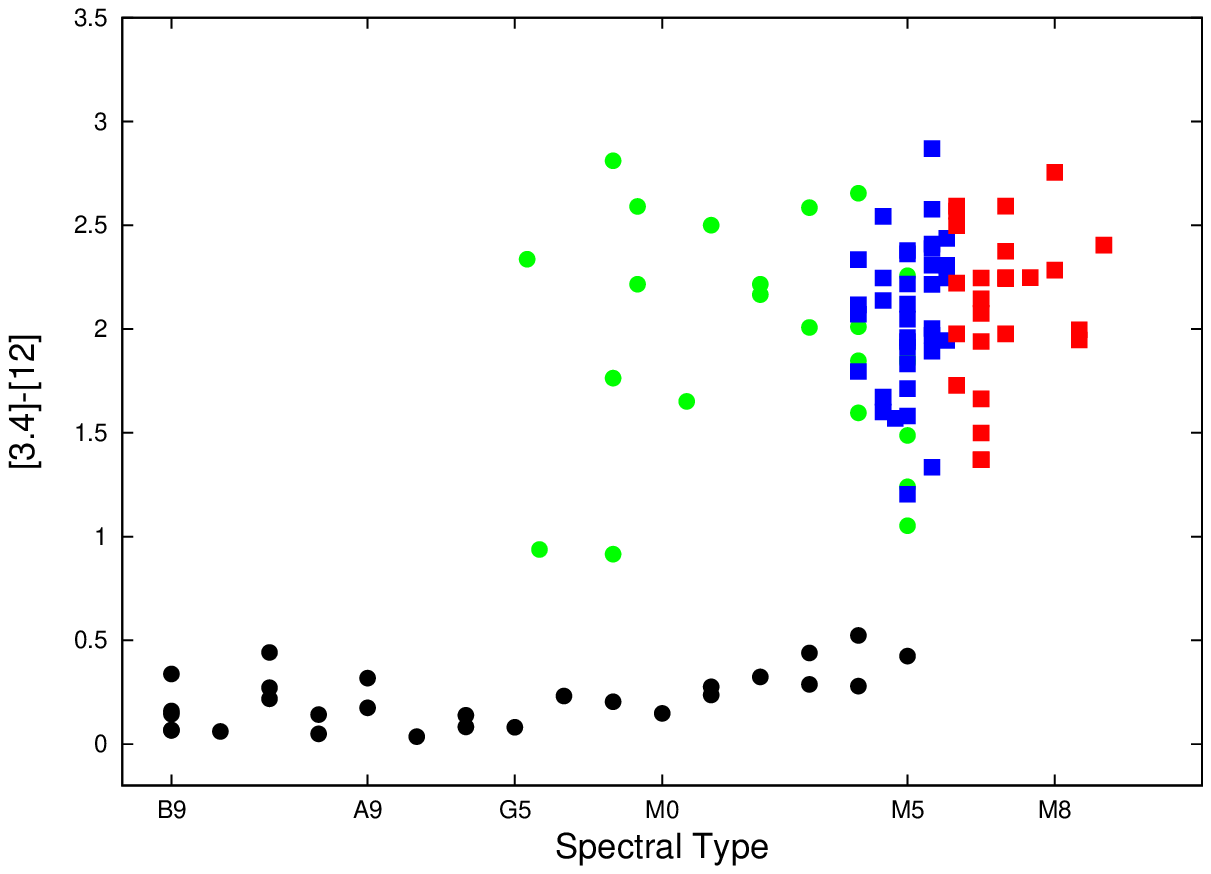} \\ 
 \includegraphics[width=140mm]{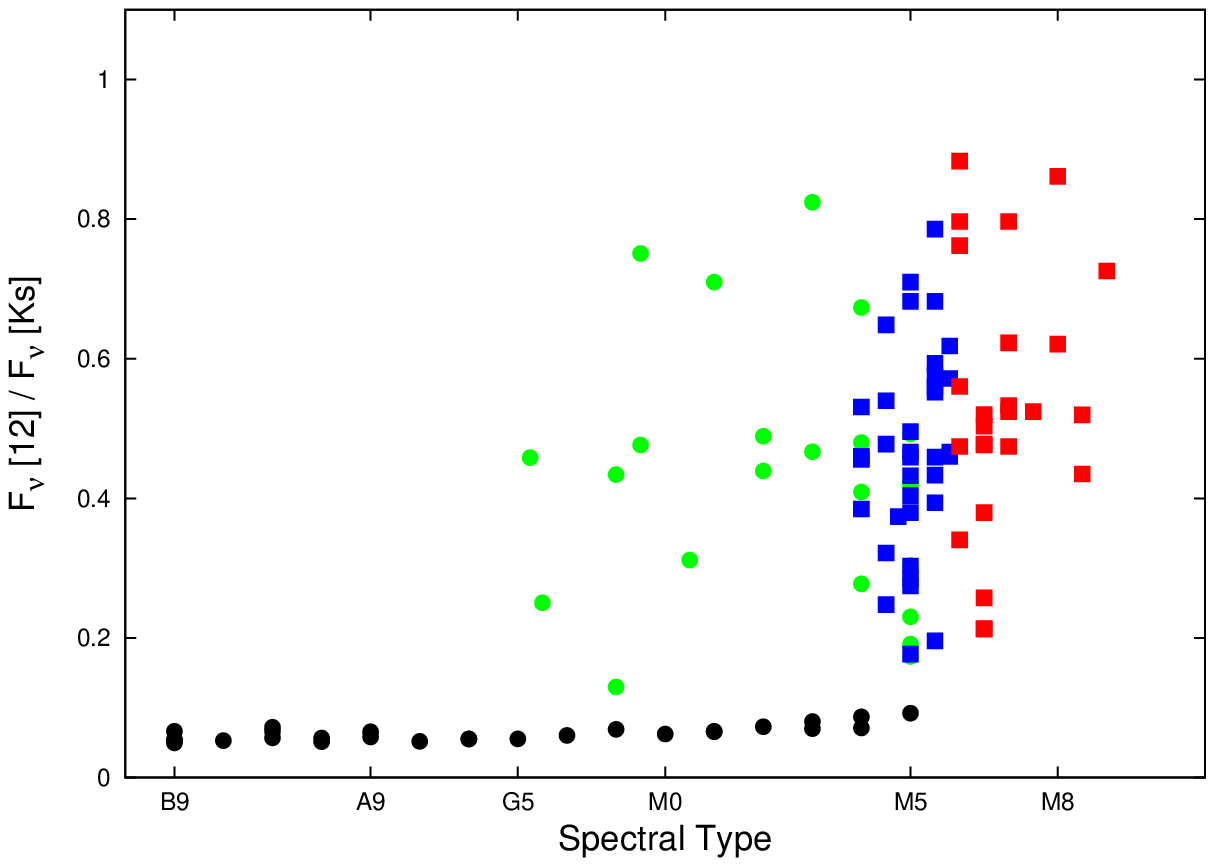} \\ 
    \caption{{\it Top}: The WISE [3.4]-[12] color versus spectral type. Symbols are the same as in Fig.~\ref{USco-color}. Data for the earlier type discs is from Carpenter et al. (2009). {\it Bottom} The 12$\micron$ to $K_{s}$ flux ratio versus the spectral type.  We have also included the BD discs from Scholz et al. (2007) and Riaz et al. (2009). These are also denoted by red symbols.   }
   \label{USco-color2}
 \end{figure*}

Fig.~\ref{USco-color} shows that the WISE colors for our M dwarf and BD discs are similar to the primordial transition discs in USco. If we consider the Luhman et al. (2010) boundary of K-[3.6] $\sim$0.4 for photospheric colors of M4-M5 stars, and K-[3.6] $\sim$0.5 for M6-M8 objects, then we have about 11 M dwarf discs and 3 BD discs that lie above this boundary. Among these sources, we have two BD discs, USco161005 and USco162847, and 6 M dwarf discs, USco155140, USco161117, USco161349, USco162222, USco161816, and USco161819, which have Fobs/Fphot $\sim$2-2.5 at 3.4 and 4.6$\micron$, while the rest have Fobs/Fphot $\sim$ 1.1-1.4 (Figs.~\ref{USco-BD}, \ref{USco-Mdwarf}). A large fraction of USco discs thus show signs of inner disc clearing, with weak or no excess emission shortward of 12$\micron$. We have therefore classified the new discs found in our survey as primordial transition discs. Among the BD discs, four sources, USco155420 (M8), USco160953 (M6), USco161940 (M6) and USco163027 (M6), show flaring between 12 and 22$\micron$ (Fig.~\ref{USco-BD}). For most of the previously known USco brown dwarf discs, the 24$\micron$ flux density is between $\sim$2 and 5 mJy, and only a few discs have been found with higher fluxes of 15-20 mJy at 24$\micron$ (Scholz et al. 2007, Riaz et al. 2009). A $\sim$2-5 mJy flux density will be below the WISE 5-$\sigma$ point-source sensitivity of 6 mJy at 22$\micron$ (WISE Explanatory Supplement), which could explain why most of these discs were not detected in the 22$\micron$ band. In Fig.~\ref{USco-color2}, the USco BD discs show larger color excesses compared to the earlier type discs. The 12$\micron$ to $K_{s}$ flux ratio for most BD discs is also larger in comparison to earlier type discs, indicating larger 12$\micron$ flux excess for the BD discs. Since these disc sources are presumably at the same age, this suggests that discs persist for a longer time around later type stars. We discuss this further in Section \S\ref{discussion}.

All of the disc sources show excess emission at 12$\micron$, including the BD discs, and there are no discs that have excesses at 22$\micron$ but none at 12$\micron$. Based on WISE data, a debris disc can be defined as a source with excess at 22$\micron$ only. The minimum excess that we can detect with WISE for a brown dwarf disc at 22$\micron$ would have to be at least $\sim$40 times larger than the photospheric flux at this wavelength. Therefore WISE may have missed brown dwarf debris discs due to low sensitivities. However, we note that none of the deeper Spitzer surveys have found BD debris discs, i.e. discs that showed excess emission at 24$\micron$ but none at shorter wavelengths. In comparison, for an M4 dwarf, an excess $\sim$5 times larger than the photospheric flux could have been detected with WISE. As mentioned, a high fraction of M dwarf discs in USco are found to be in the primordial stage, compared to earlier-type stars (Carpenter et al. 2009). The absence of debris systems among our M dwarf and BD targets is thus consistent with previous results.





From our present survey in USco, we find a BD disc fraction of 28$\pm$9\% (12/43). The error bars are the 1-$\sigma$ Gaussian distribution uncertainties. This fraction is comparable to the 37$\pm$9\% fraction reported by Scholz et al. (2007) from their $\sim$8-14$\micron$ {\it Spitzer}/IRS spectral observations. In Riaz et al. (2009), we had reported a much lower USco BD disc fraction of 11$\pm$4\% (3/28). This fraction was based on {\it Spitzer}/IRAC and MIPS 24$\micron$ observations, and was mainly determined from the excess emission observed at 8$\micron$. Nearly 90\% of the sources in that survey were undetected at 24$\micron$. As we have found in the present work, a majority of the discs show photospheric emission shortward of 12$\micron$. A small disc fraction based on 8$\micron$ excess emission is thus expected. Combining our present work with Scholz et al. (2007) and Riaz et al. (2009), we have a USco BD disc fraction of 26$\pm$9\%. For the M dwarfs in USco, we have a disc fraction of 25$\pm$8\% (36/146). Our M dwarf disc fraction is comparable (within the uncertainties) to the 17$\pm$6\% fraction reported by Carpenter et al. (2009) for the M type primordial discs in their sample. 

We note that due to lower sensitivities, a disc survey conducted with WISE could be incomplete as compared to earlier deep Spitzer surveys. There are 16 known brown dwarf discs in USco detected by previous Spitzer surveys (Scholz et al. 2007; Riaz et al. 2009). The 1-$\sigma$ sensitivity at 8-11$\micron$ from these Spitzer surveys is $\sim$0.09 mJy. These surveys were thus deep enough to detect the photospheres for the brown dwarf targets. Out of these 16 Spitzer BD discs, 15 were detected in the WISE 1, 2, and 3 bands at S/N $>$ 3. One of the discs (usd161916) was detected in the WISE 1 and 2 bands, but was undetected at 12$\micron$. The observed flux density for this object at 11$\micron$ is 0.65 mJy (Scholz et al. 2007), which is below the WISE 5-$\sigma$ sensitivity at 12$\micron$ ($\sim$0.7 mJy). This is an M8 disc. However, the M9 discs from Scholz et al. (2007) Spitzer survey were also recovered by WISE at a $>$3-$\sigma$ level in the 1, 2, and 3 bands. Half of these Spitzer discs are detected in the WISE 22$\micron$ band at a S/N$>$3, while the rest are undetected at this wavelength (S/N $<$ 2). Other than the WISE lower sensitivities, another potential bias could be due to the quality criteria that we have applied, in terms of the S/N to be at least 3 in bands 1, 2 and 3, which would reduce the size of the original sample and bias the results towards a higher disc fraction. We have checked for WISE matches for the full Spitzer sample of 63 sources, including the discless objects, from the work of Riaz et al. (2009) and Scholz et al. (2007). There are 11 objects which have S/N $<$ 3 in the WISE 12$\micron$ band, while the rest were detected at higher S/N in all three WISE bands. The BD disc fraction in USco from Spitzer surveys was 16/63 = 25.4\%$\pm$ 9\%. Based on the WISE-recovered sample, the BD disc fraction will be 15/52 = 28.8\% $\pm$ 9\%. These fractions are still consistent within the uncertainties and the differences are not significant. We can therefore estimate that the lower WISE sensitivities and our quality criteria would result in a BD disc fraction in USco which is slightly higher (by $\sim$3\%) than the fraction determined from the higher S/N Spitzer data. 

\subsection{H$\alpha$ versus Disc Emission in USco}
\label{halpha-disc}

\begin{figure*}
\includegraphics[width=170mm]{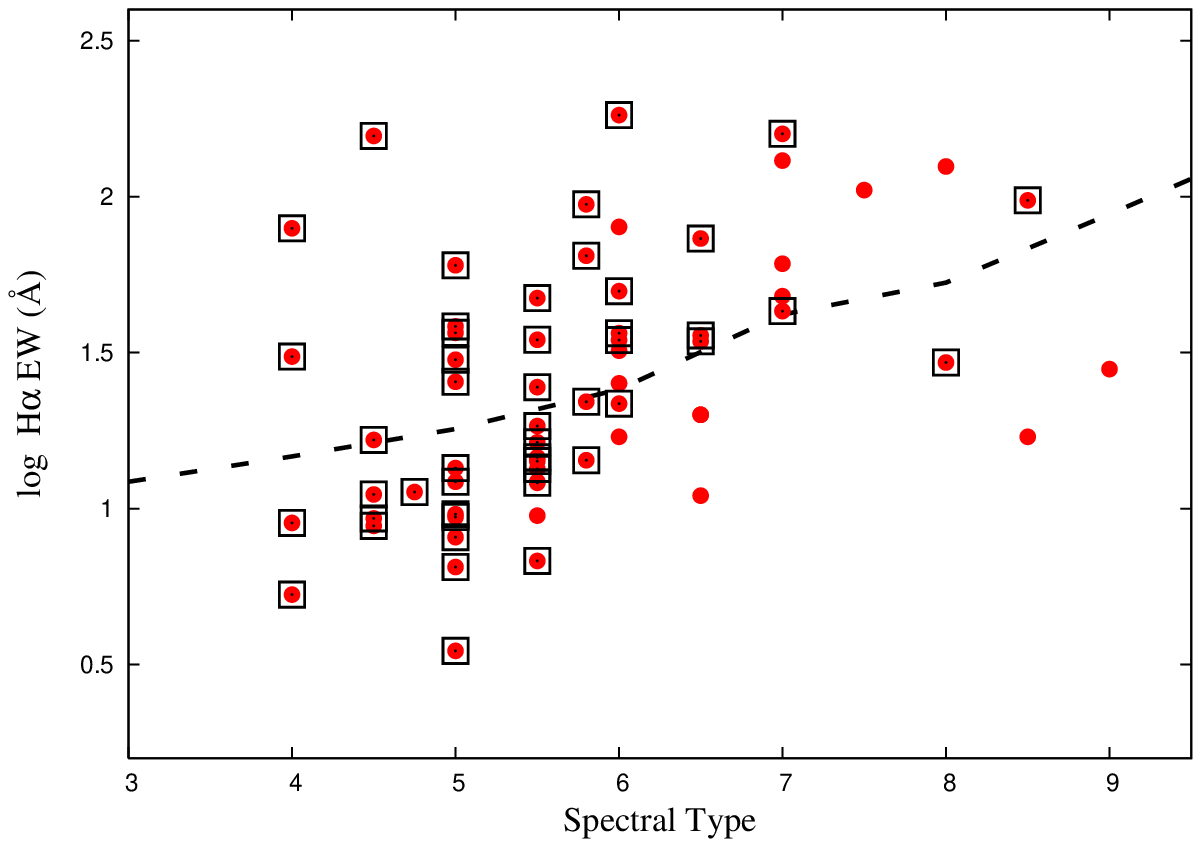} 
    \caption{H$\alpha$ equivalent width versus the spectral type for the USco disc sources from the present work (black open squares). Also included are the BD discs reported in Scholz et al. (2007) and Riaz et al. (2009). Dashed line is the empirical accretor/non-accretor boundary from Barrado y Navascu\'{e}s \& Mart\'{i}n (2003). The SpT of 3-9 imply M3-M9 types.  }
   \label{Halpha}
 \end{figure*}

Figure~\ref{Halpha} compares the H$\alpha$ emission strength for the disc sources in USco. The new discs we have found in this work are marked by black squares. The H$\alpha$ equivalent width and SpT are from Slesnick et al. (2006; 2008) and Lodieu et al. (2008; 2011). We have included in this figure previously known USco discs from Scholz et al. (2007) and Riaz et al. (2009). The dashed line in this figure is the empirical accretor/non-accretor boundary from Barrado y Navascu\'{e}s \& Mart\'{i}n (2003). Sources that lie above this boundary can be considered as CTTS, and the ones below it as the non-accreting weak-line T Tauri stars (WTTS). We find a nearly equal number of disc sources above and below the accretor boundary, which indicates that the presence of passive or non-accreting discs is as common in USco as accreting disc sources. The disc bearing sources among the BDs (SpT$\geq$M6) are mostly all accreting systems. Emission in H$\alpha$ is known to rise towards the later types, with a peak near SpT of M6-M7 (e.g., Riaz et al. 2006). It is thus more likely to find disc sources among the later types that also show strong emission in H$\alpha$. 

A lack of correlation between Class II and CTTS systems, or Class III and WTTS systems, is expected at this transition age of $\sim$5 Myr. A large fraction of USco discs show photospheric emission shortward of 12$\micron$, indicating the inner $\sim$0.1-1 AU disc region to be cleared of circumstellar material. The H$\alpha$ emission line traces the gas in the inner disc regions and is not a diagnostic of the total gas content in the disc. It thus seems reasonable to find an anti-correlation between the presence of disc emission and lack of H$\alpha$ emission, since the disc emission is being detected at longer wavelengths and not from the inner part of the disc close to the central source. Fig.~\ref{Halpha} thus indicates that selecting young stellar objects based on the strength in H$\alpha$ emission could result in a rejection of young disc sources with inner disc clearings.

\section{Discussion}
\label{discussion}
\subsection{Disc Fractions versus Stellar Mass, Age and Density}


In Fig.~\ref{fractions}, we have plotted the disc fractions for clusters at ages of $\sim$1-10 Myr in 3 mass bins: the sub-stellar sources (SpT=M6-M9; $<$0.1 $M_{\sun}$), the low-mass stars (SpT=K5-M5; $\sim$0.1-1 $M_{\sun}$), and the high-mass stars (SpT$<$K5; $\sim$1-4 $M_{\sun}$). Section\S\ref{appendix} provides details on the compilation of the disc fractions as well as the fraction of discs that WISE might have missed around M6-M8.5 type objects due to limited sensitivity. The disc fractions plotted here imply the `inner' disc fractions, since these have been obtained from observations at $\sim$3-24$\micron$, which probe the inner disc regions of radii within $\sim$1 AU in BD discs and within $\sim$5 AU in higher mass T Tauri discs (e.g., Hillenbrand 2005). It is important to determine the evolutionary stage of these discs, i.e. whether these are in the primordial, primordial transition, or debris phase. As discussed in detail in Section \S\ref{appendix}, nearly all of the discs in Taurus, Cha I, IC 348 and SOri are in the primordial/primordial transition stage, and a small $\sim$1-2\% fraction is found to be of debris discs. On the other hand, a majority of the high- and low-mass discs in USco and TWA are in the debris phase, while the BD discs in these older clusters can all be classified as primordial/primordial transition discs. Thus the later type discs tend to remain in the primordial stage for a longer time. We note that the uncertainties on the disc fractions are quite large for some of these young clusters, which would make the fractions comparable to each other. Nevertheless, we find some obvious differences in the disc frequency trends for the different mass bins, as discussed further below. In Fig.~\ref{density}, we have plotted the disc fractions versus the stellar density in a cluster. By stellar density, we imply the total number of objects (stars + BDs) detected in a given survey area. Details on the compilation of stellar densities are provided in Section\S\ref{appendix}. Our results for stellar densities indicate SOri and Cha I to be denser than Taurus and USco, which is in agreement with what has been previously noted by several surveys.

The disc fractions in Fig.~\ref{fractions} are the lowest and nearly constant in USco over a wide range in stellar masses, unlike the SOri or Taurus clusters. The disc fractions decrease with stellar mass in Taurus, while the opposite trend is observed in SOri. IC 348 also shows a slight increase in disc fractions with decreasing stellar mass. In comparison, Cha I which is at a similar age of $\sim$2 Myr as IC 348, shows a higher disc fraction for the high-mass stars. Luhman et al. (2010) had noted that IC 348 and SOri are denser than Cha I or Taurus, and that a denser environment could result in a lower disc fraction for the high-mass stars in a cluster. This can be seen in Fig.~\ref{density} (top panel), where the fractions for the high-mass stars in the denser regions of SOri and IC 348 are lower than in Taurus or Cha I. However, the disc fractions are lower in USco than SOri, even though the two clusters are at similar ages ($\sim$3-5 Myr) and USco is even more sparse, with a stellar density $\sim$20 times lower than in SOri. A denser environment also does not seem to affect the low-mass and the sub-stellar sources (Fig.~\ref{density}). The disc fractions for these mass groups are similar in IC 348 and Cha I, which have about an order of a magnitude difference in stellar densities. Pfalzner et al. (2006) have simulated the relative disc mass loss induced by encounters between stars in a cluster. At an age of 2 Myr, their simulations indicate a disc mass loss of $\sim$30\% for intermediate-mass (1-10$M_{\sun}$) and low-mass ($<$1$M_{\sun}$) stars, whereas for stars with $M>$10$M_{\sun}$, it is $\sim$60\% or higher. As they explain, the average number of stellar collisions for high mass stars ($>$10$M_{\sun}$) is higher compared to intermediate- or low-mass stars, and so the disc dispersion rate will also be higher for the high mass sources. Also, if the disc sizes for BDs are smaller than high-mass stars, then frequent flyby events would be less likely to disrupt the sub-stellar discs. While the mass ranges from Pfalzner et al. are different than the ones considered here, their simulations could explain why we see similar low-mass and BD disc fractions in Cha I and IC 348, compared to a factor of $\sim$3 difference observed for the high-mass stars (Fig.~\ref{density}). A denser environment thus may be more effective in increasing the disc disruption rate for the high-mass stars.


An alternate explanation for the prominent difference between the Taurus and SOri/USco disc fractions for the high-mass stars could be their early transition from a primordial to a debris phase. By an age of $\sim$5 Myr, $\sim$86\% of the high-mass discs are in the debris phase. Thus nearly all massive sources have experienced significant inner disc clearing by this age, and are found to have excesses only at wavelengths of $\geq$24$\micron$, with photospheric emission at shorter wavelengths. The disc fraction then shows a rise again at older ages, from $\sim$20\% at 5 Myr to $\sim$50\% at 10 Myr. All of these older discs are in the debris phase, as these are mainly detected due to strong 24$\micron$ or 70$\micron$ excess emission. Debris disc particles undergo continual shattering collisions with planetesimals and larger bodies in the system, due to which debris dust is continually generated. This results in an increase in the dust mass at the onset of the debris disc phase. The increase in the debris disc fraction is due to the detection of large mass of cold dust in the outer regions of the disc, while the inner warm regions have been dissipated. We note that the TWA disc fraction has a high uncertainty due to the sparse population of this region (Section \S\ref{appendix}). However, Siegler et al. (2007) and Reike et al. (2005) have reported disc fractions of $\sim$40-50\% for B-A type stars in 30-50 Myr old clusters, based on much larger sample sizes. To summarize, the dip in the USco fraction for the high-mass stars is due to the decline of primordial discs by this age, or the fewer number of discs detected at this age with excesses at short wavelengths ($\leq$24$\micron$).

In comparison to the high-mass stars, BD discs seem to evolve at a slower rate, as their inner disc fraction remains nearly constant until $\sim$3 Myr. Thereafter, the disc fractions show a steep decline by a factor of $\sim$2 between 3 and 5 Myr. The statistical significance of this sharp drop in fractions between 3 and 5 Myr is discussed in Section \S\ref{appendix}. This rapid inner disc dispersal over a short $\sim$2 Myr timescale thus seems to occur for both the high-mass and the sub-stellar sources, albeit at an earlier age of 2-3 Myr for the high-mass stars, but at an older age of 3-5 Myr for the BDs. The later drop for the BDs could be explained by their slower viscous evolution timescales. The viscous scaling time for BDs are expected to be of the order of $10^{6}$ yr, longer than the typical $10^{4}$-$10^{5}$ yr for T Tauri stars (e.g., Alexander \& Armitage 2009). In other words, if accretion of material onto the BD occurs at a slower rate than higher mass stars, then discs could persist for a longer timescale. The inner discs around higher mass stars may also be replenished faster by material coming in from the outer disc region. Such a process of outer disc accretion seems to occur either at a slower rate for BDs, or may not be that significant if the disc sizes are too small compared to the higher mass T Tauri discs. It is interesting to note that the high mass and sub-stellar sources provide two extreme cases of rapid inner disc dispersion, while the intermediate low-mass stars show a more linear decline in disc fractions with age (Fig.~\ref{fractions}). These sources have an almost 50/50 debris/primordial disc fraction at 5 Myr, which increases to 70/30 by 10 Myr (Table~\ref{frac-den}). The low-mass discs also show a rise at ages older than 10 Myr, due to an increase in the debris disc fraction (e.g., Siegler et al. 2007). Thus a large fraction of the low-mass discs are in the debris phase by $\sim$10 Myr, but the transition is more gradual than that seen for the high-mass stars, with no sudden drops in the disc fractions observed. 






The rapid inner disc dispersal could be due to processes such as dust coagulation or planetesimal formation. The processes of significant grain growth and dust settling in the inner disc regions could lead to rapid formation of kilometer-sized large bodies or planatesimals in the disc. From the simulations of Dullemond \& Dominik (2004), such processes occur on a faster timescale in the inner disc regions, and small grains can be removed from the inner regions on a timescale of less than 1 Myr. Photoevaporation could also result in rapid inner disc clearing. The process of photoevaporation sets in when the disc accretion rate falls below the photoevaporative wind rate, which is about 10$^{-10} M_{\sun}$/yr (e.g., Muzerolle et al. 2006). From the simulations of Alexander et al. (2009), giant planet formation and photoevaporation can both dominate inner disc clearing. At early times, planet formation/migration is the more dominating process as the accretion rates are too high for photoevaporation to set in. At ages of $\geq$2 Myr, the disc accretion rate falls below the photoevaporative wind rate and essentially all discs are being cleared out by photoevaporation. By an age of $\sim$5 Myr, the inner disc fraction in their simulations has dropped to $\sim$20\% from nearly 100\% at $\sim$2 Myr. Thus a coupling of the planet formation/migration and photoevaporation processes could result in rapid inner disc clearing over a $\sim$2-3 Myr timescale. Such processes would work for the high-mass stars where accretion rates measured from UV continuum emission are $\sim$10$^{-8} - 10^{-9} M_{\sun}$/yr at an age of $\sim$1 Myr (e.g., Herczeg et al. 2008). For BDs, the accretion rates are $<$ 10$^{-10} M_{\sun}$/yr at an age of $\sim$1 Myr (e.g., Muzerolle et al. 2005), which would imply that photevaporation can be effective even at these early ages. The inner disc fractions for BDs at $\sim$1-3 Myr should then be as small as the fraction found in the USco. Giant planet formation is also feasible for large disc masses of $\sim$0.1 $M_{\sun}$ as considered by Alexander et al. (2009), whereas typical disc masses for BDs in Taurus are found to be much smaller, of the order of $\sim$10$^{-4}$ $M_{\sun}$ (e.g., Scholz et al. 2007). From the simulations by Payne \& Lodato (2007), a close-in rocky planet with a mass of $\sim$0.1 $M_{\earth}$ can form around a BD at a semi-major axis as small as $\sim$0.2 AU, considering an average BD disc mass of 1.5 x10$^{-4} M_{\sun}$ and an outer disc radius of 100 AU. However, the growth of such planetary cores is at a much slower rate around BDs compared to the solar-type stars. Dust coagulation leading to planet formation thus would take longer for the sub-stellar case, and cannot explain the sharp drop observed between the SOri and USco inner disc fractions in Fig.~\ref{fractions}. 

\begin{figure*}
 \resizebox{170mm}{!}{\includegraphics[angle=0]{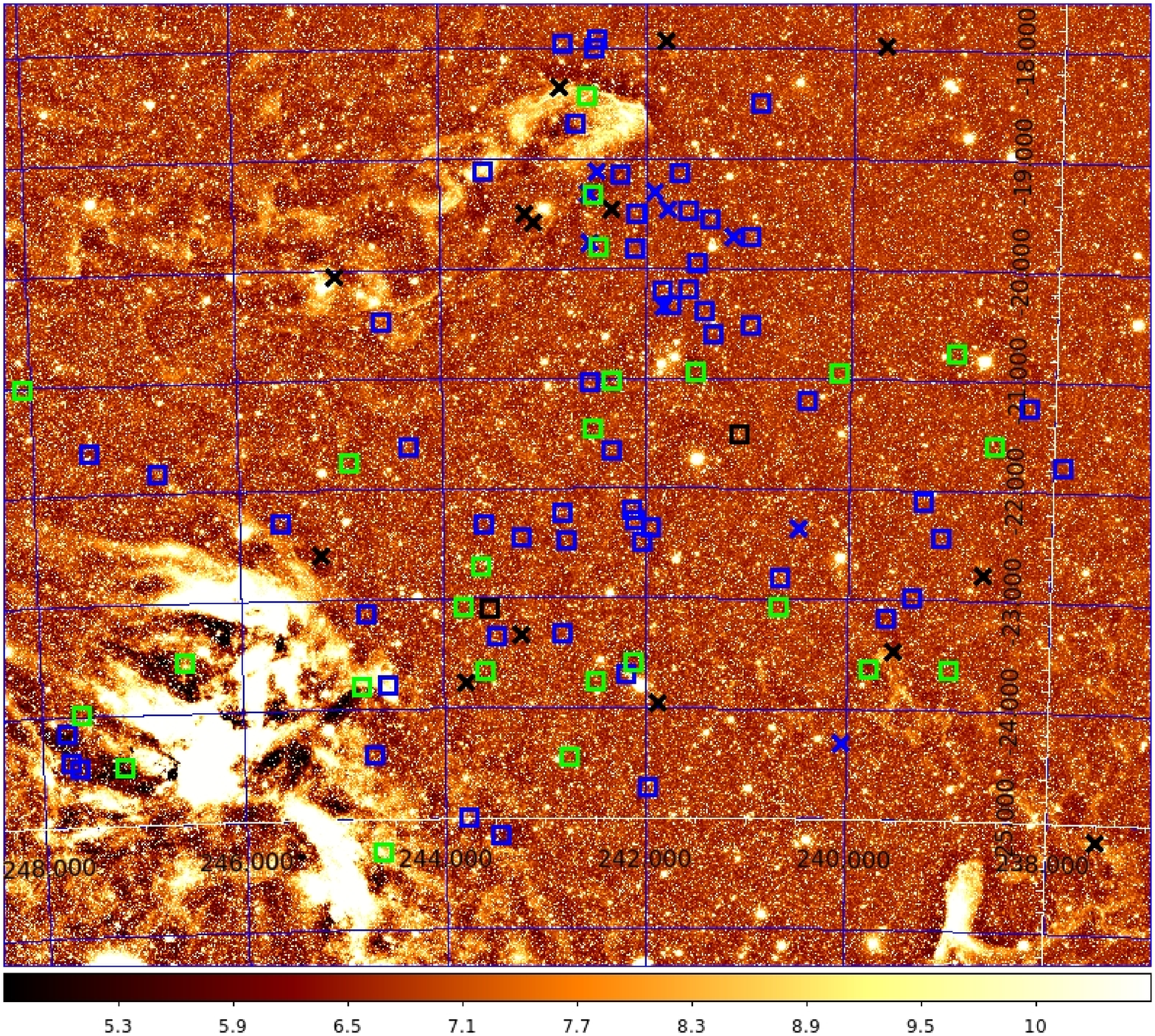}}  \\ 
    \caption{The spatial distribution of USco discs. This is a false-colored WISE 12$\micron$ image. North is up, east is to the left. Symbols are: black--high-mass stars; blue--low-mass stars; green--BDs. Crosses indicate debris discs, squares indicate primordial discs. The color bar in units of MJy sr$^{-1}$ is shown at the bottom.  }
   \label{spatial}
 \end{figure*}

Other processes such as photoionisation could also be in play. The ionizing radiation and winds from massive stars can affect their immediate surroundings, thereby dispersing the circumstellar material around stars. In Fig.~\ref{spatial}, we have plotted the known OB stars in USco, along with the disc sources from all mass groups. We do not find any particular concentration of the disc sources farther away from the massive stars, which might have suggested minimal interaction with the ionizing winds and so a higher probability of disc survival. Among the low-mass stars, there is a small concentration of disc sources near ($\alpha$, $\delta$)$\sim$(242$\degr$, -22$\degr$). Also, nearly all of the BD discs are located south of $\delta$ $\sim$ -21$\degr$, while most of the higher mass debris discs are located north of this declination. There may be a possible mass dependency in the spatial distribution of the discs. We note that these discs come from different surveys that may not cover the same region in USco. Overall, the discs are largely spread throughout the association, and we do not find any strong evidence of photoionisation in USco. Also, both SOri and USco are known to host OB stars. We would thus expect similar disc fractions in these regions if photoionization is the main cause of inner disc dissipation.

To summarize the discussion above, we can explain the rapid decline in the high-mass inner disc fraction between $\sim$2 and 3 Myr using planet formation and/or photoevaporation models. Accretion of material onto planets as well as photoevaporation would lead to a transition from a gas-rich optically thick disc to a secondary debris disc. This would explain the high fraction of debris disc systems found in USco for the high-mass stars (Table~\ref{frac-den}). Stellar density may also play a role and disc lifetimes may be prolonged in a less dense environment. But this is only notable for the high-mass stars. None of the processes above can explain the drop observed for the BD disc fractions between 3 and 5 Myr. In order to determine the significance of the observed decline between 3 and 5 Myr, we have built a probability distribution for the USco BD disc fraction using the method described in Burgasser et al. (2003). The distribution was constructed using a sample size of N=106, and number of discs n=28 (combining this work with Scholz et al. and Riaz et al. surveys). The peak is at 26.4\% with 1-sigma Gaussian uncertainties of 9\%. The 3 Myr SOri disc fraction is 57\%$\pm$ 19\%, i.e. the disc fraction could be between 38\% and 76\%. From the probability distribution, the probability that the USco sub-stellar disc fraction is between 38\% and 76\% is found to be $\sim$28\%, at a confidence level of 97.5\%. There is thus a $\sim$70\% probability that the 3 and 5 Myr disc fractions are different. We note that the USco point is not just 1-sigma below the SOri point, it is also 1-sigma lower than the Cha I, IC 348 and Taurus points. The question we have tried to address in this paper is not how likely it is that USco could actually be the same as any *one* of the others, but that it could be the same as *all* of the other clusters, which seems unlikely. Also notable for the BD discs is their persistent optically thick emission even at later ages. By $\sim$5 Myr, $\sim$86\% of the high-mass sources are in the debris phase, while for the BDs, even the $\sim$10 Myr old discs show excesses at wavelengths as short as $\sim$5$\micron$, and are in their primordial phase (Riaz \& Gizis 2008). In Fig.~\ref{fractions}, comparing the SOri and TWA disc fractions for BDs suggests that it may be nearly constant at $\sim$50\% over 1-10 Myr ages. The sharp drop between 3 and 5 Myr may not be a real trend with age, but perhaps the intrinsic or the original BD disc fraction in USco is much smaller than the younger or the older clusters.

\subsection{BD Formation Mechanisms}

Could it be possible that this particular low BD disc fraction in USco is due to the different formation mechanism of BDs dominant in these clusters? As noted earlier, Taurus, Cha I and USco are low density clusters and always have been, but IC 348 and SOri are more classic denser clusters. It may be that the denser clusters formed more BDs from collapsing low-mass cores rather than other mechanisms such as, by fragmenting massive discs or ejection of protostellar embryos, and so initially they had a higher disc fraction. In Fig.~\ref{ratio}, we have plotted the BD disc fraction versus the ratio of the number of BDs to the number of stars in these clusters. Details on the compilation of the BD/Star number ratios are provided in Section \S\ref{appendix}. USco has a factor of $\sim$2 higher ratio than the other clusters but the lowest disc fraction, which indicates that a higher number of BDs formed does not necessarily translate into a higher disc fraction. Also, the denser clusters such as SOri do not form a higher number of BDs, as the ratios are quite similar to the less dense Taurus and Cha I. 

Different formation mechanisms of BDs in these clusters could also lead to different disc properties, even if the total number of BDs formed is similar. There are five standard models considered for the formation of BDs:

\begin{enumerate}

\item From very low-mass cores, which should result in the presence of a large disc around the sub-stellar source formed (e.g., Padoan \& Nordlund 2004; Hennebelle \& Chabrier 2008; Machida et al. 2009)
\item Ejection from larger cores, which would result in a small, truncated disc at best (e.g., Reipurth \& Clarke 2001; Goodwin et al. 2004)
\item From massive discs; it is estimated that $\sim$20-70\% of the BDs formed by disc fragmentation can have discs (e.g., Stamatellos \& Whitworth 2009ab) 
\item Liberated from binaries, which should result in the presence of a disc (e.g., Goodwin \& Whitworth 2007)
\item In small 'cores' in filaments, the BD disc fraction should be the same as the low-mass disc fraction (e.g., Bonnell et al. 2008; Clarke et al. 2008; Bate 2009) 

\end{enumerate}

To clarify these models further, model (1) is based on a star-like formation of BDs, i.e. via the collapse of molecular cloud cores with sub-stellar masses. In this model, gravitationally unstable protostellar cores of BD mass are formed directly by the process of turbulent fragmentation (e.g., Padoan \& Nordlund 2004). Since this is a star-like formation, and since most stars are formed with circumstellar discs, BDs formed via this mechanism are expected to be surrounded by discs. Model (2) considers that BDs are not formed through the collapse of low-mass cores, but that rather they are ejected as a result of encounters in multiple systems made up of a small number of stellar embryos. The ejected BDs have suffered multiple encounters that will have pruned their circumstellar discs. As a result, this model predicts weaker accretion diagnostics and few BDs with discs (e.g., Reipurth \& Clarke 2001). In model (3), BDs are formed due to gravitational fragmentation of massive extended discs around Sun-like primary stars. At an early age of $\sim$20 kyr, this model predicts 23\% of the BDs liberated in the field to have discs. Many of the BDs that still remain bound to the central star are loosely bound, and so when they escape, they are expected to retain their discs. The disc fraction could thus increase up to a maximum of 70\% (Stamatellos \& Whitworth 2009ab). Model (4) is based on a binary disruption mechanism. In this model, BDs form as distant companions to low-mass stars, in particular M-dwarfs. Such systems are then readily disrupted by the mild perturbations of passing stars at relatively large distances (a few hundred AU or more). The result is a population of single BDs and low-mass hydrogen-burning stars. Since the disruption is gentle, both the BDs and the low-mass hydrogen-burning stars are able to retain their circumstellar discs (e.g., Goodwin \& Whitworth 2007). In model (5), BDs form in filamentary structures that develop as the gas falls in the gravitational potential of the forming cluster. In this model, there is no need for close interactions or ejections to ensure the formation of sub-stellar cores. Ejections can occur but are not fundamental to the process. This model predicts BDs to have the same disc properties as the low-mass stars formed, i.e., the circumstellar disc properties for BDs should form a continuum with low-mass stars. However, this mechanism would only be expected to be efficient in dense clusters. The abundance of BDs in stellar clusters is of the order of 25\% in highly clustered regions, whereas it decreases to the order of 10\% in isolated regions (Bonnell et al. 2008; Clarke et al. 2008).


Any, some, or all of these mechanisms could operate in a cluster, perhaps with the same relative efficiency, or perhaps different in different regions. Different mechanisms should give different disc properties relative to the other mechanisms, for e.g., very low-mass cores would probably produce fairly massive extended discs, while ejection would produce lower-mass truncated discs, or no discs at all. Once formed, we can probably assume that the same discs evolve in the same way independent of the formation mechanism. But the initial disc fraction may be different and discs could start significantly less massive/extended depending on the formation mechanism. The observations seem to suggest that BD discs do not have a nice decline in fraction with time. If different clusters form BDs via different mechanisms, then the initial BD disc populations will be different in these clusters. {\it The observed BD disc fraction thus may only partially be due to age evolution, and rather the disc fraction may be a signature of the BD formation mechanism dominant in the cluster}. For e.g., the mechanism (2) may be more active in USco than the rest. The ejection model (2) could result in an excess of BDs being formed, but very few of those BDs will have circumstellar discs. This mechanism may also be prevalent in Taurus that has a lower BD disc fraction than SOri despite being younger than this region. Model (3) may also be applicable in USco if the low disc fraction limit of $\sim$20\% is taken into account. In comparison, formation from very low-mass cores (1) would require higher turbulence and likely a denser region. Model (5) is also applicable in a dense region. A large fraction of the sub-stellar sources formed via these mechanisms could end up with discs, resulting in high disc fractions as found in the dense SOri or IC 348 regions. Thus while a dense environment could disrupt the discs around high-mass stars (e.g., Pfalzner et al. 2006), it could result in a high disc fraction for BDs. This can explain the large difference in the BD and high-mass disc fractions in SOri and IC 348 (Fig.~\ref{density}). On the other hand, if mechanisms such as (4) are in play, the disc disruptions are minimal and all BDs formed should be able to retain their discs, irrespective of the stellar density in the region. This could explain the difference in the BD disc fractions for Cha I and USco, which have similar stellar densities. 

The sharp decline in the BD disc fraction between 3 and 5 Myr thus may not be an age effect but likely indicates that USco has formed its BDs differently than SOri or the other young clusters. The one mechanism that seems the most applicable in USco to explain the low BD disc fraction is the ejection model. All of the other models imply similar BD disc properties as found for the stellar members in the cluster. We do find the disc fraction for BDs in USco to be similar to the stars. Then perhaps any of the other formation mechanisms could work as well, although density could be an issue. We could argue that the disc fractions for the stars may also be indicative of the respective formation mechanism. A difference in BD and star formation mechanisms has been argued for in e.g., Thies \& Kroupa (2007), and it has been suggested that stars and BDs are two correlated but disjoint populations with different dynamical histories. However, the high- and low-mass stars 'arrive' at this low USco fraction via a different track, i.e. via a transition to the debris phase. We have observational evidence of it in terms of the high debris disc fraction in USco for these higher mass bins, unlike the BDs. The similar disc fraction for stars and BDs in USco is probably not indicative of similar formation mechanisms for stars and BDs in USco,
rather a coincidence that at this time stellar discs have evolved to a low fraction whilst BD discs started with a low fraction. The absence of any clear age dependency would explain the primordial nature of older BD discs. The $\sim$10 Myr old TWA could be an exception where the extremely low-density environment might allow more BD discs to survive at early times of formation. A better argument could be that the disc fractions observed are a combination of evolution with age and the initial disc fraction due to the different formation mechanisms that are dominant in each cluster. 


\section{Summary}

We have studied the BD disc dissipation timescales in clusters at ages of $\sim$1-10 Myr, and have investigated the various factors that may affect disc dissipation for BDs. Combining with several previous surveys, we find no clear age dependence for the BD disc fractions. The two most notable points are a sharp drop in the BD inner disc fraction between 3 and 5 Myr, and the persistent primordial nature of these discs even at older ages. This indicates that while the overall BD inner disc fractions have decreased, the discs tend to remain in their primordial stages at ages as old as $\sim$10 Myr. We do not find any convincing evidence that any of the well-known inner disc dispersal mechanisms are active in BD discs. We have instead considered the different BD formation theories, and have argued that USco may have formed its BDs differently than SOri and other younger clusters. Thus the sharp decline in disc fractions may not be due to the difference in ages but may be indicative of the different BD formation mechanisms in these clusters, and there may not be a clear age dependence for the BD disc fractions. 

Our work also includes a SED classification scheme based on the $K_{s}$ and WISE bands of 3.4-12$\micron$. We find the WISE [3.4]-[12] color to provide a clear distinction between the photospheric and the disc population.

\section*{Acknowledgments}

This publication makes use of data products from the Wide-field Infrared Survey Explorer, which is a joint project of the University of California, Los Angeles, and the Jet Propulsion Laboratory/California Institute of Technology, funded by the National Aeronautics and Space Administration. NL was funded by the Ram\'on y Cajal fellowship number 08-303-01-02 and the program number AYA2010$\_$19136 from the Spanish ministry of Science and Innovation.

\begin{figure*}
\includegraphics[width=50mm]{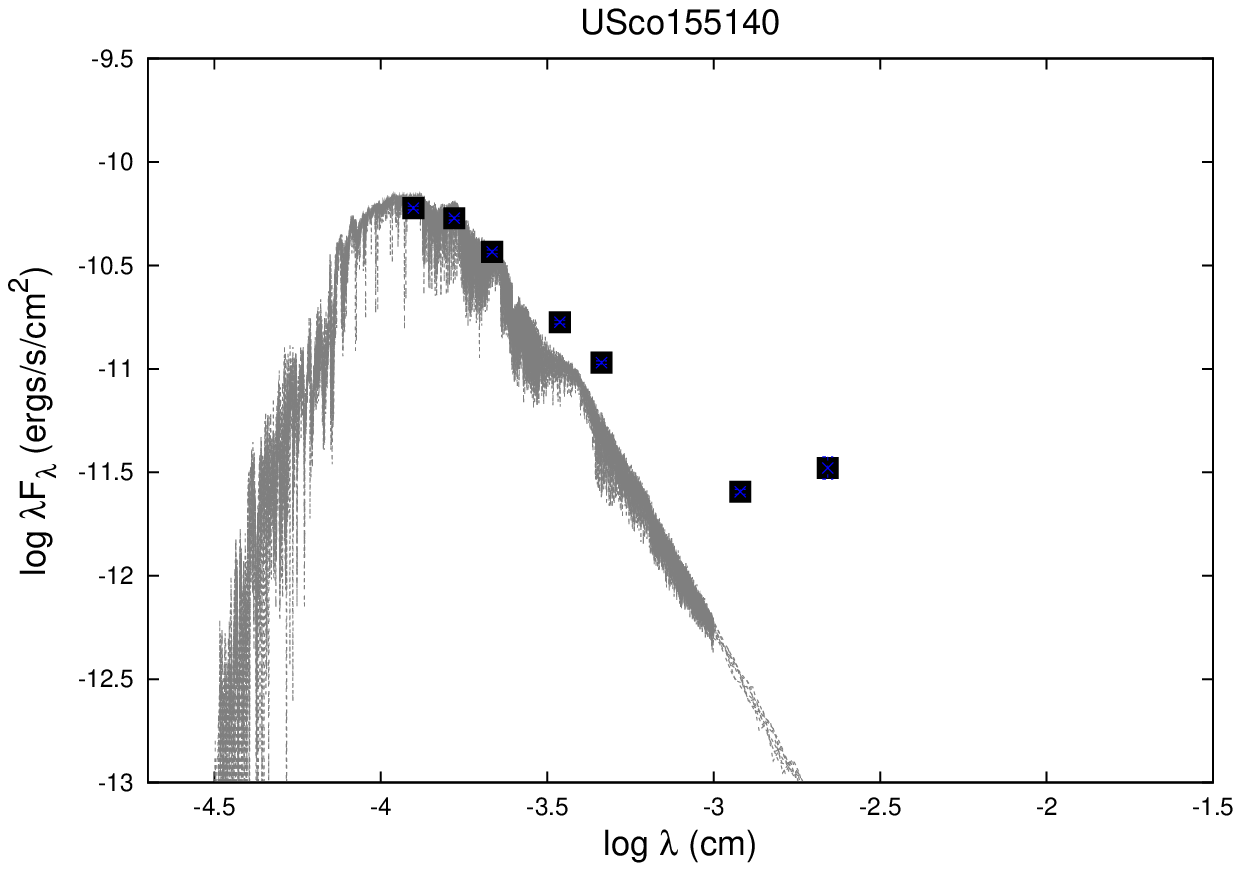} 
 \includegraphics[width=50mm]{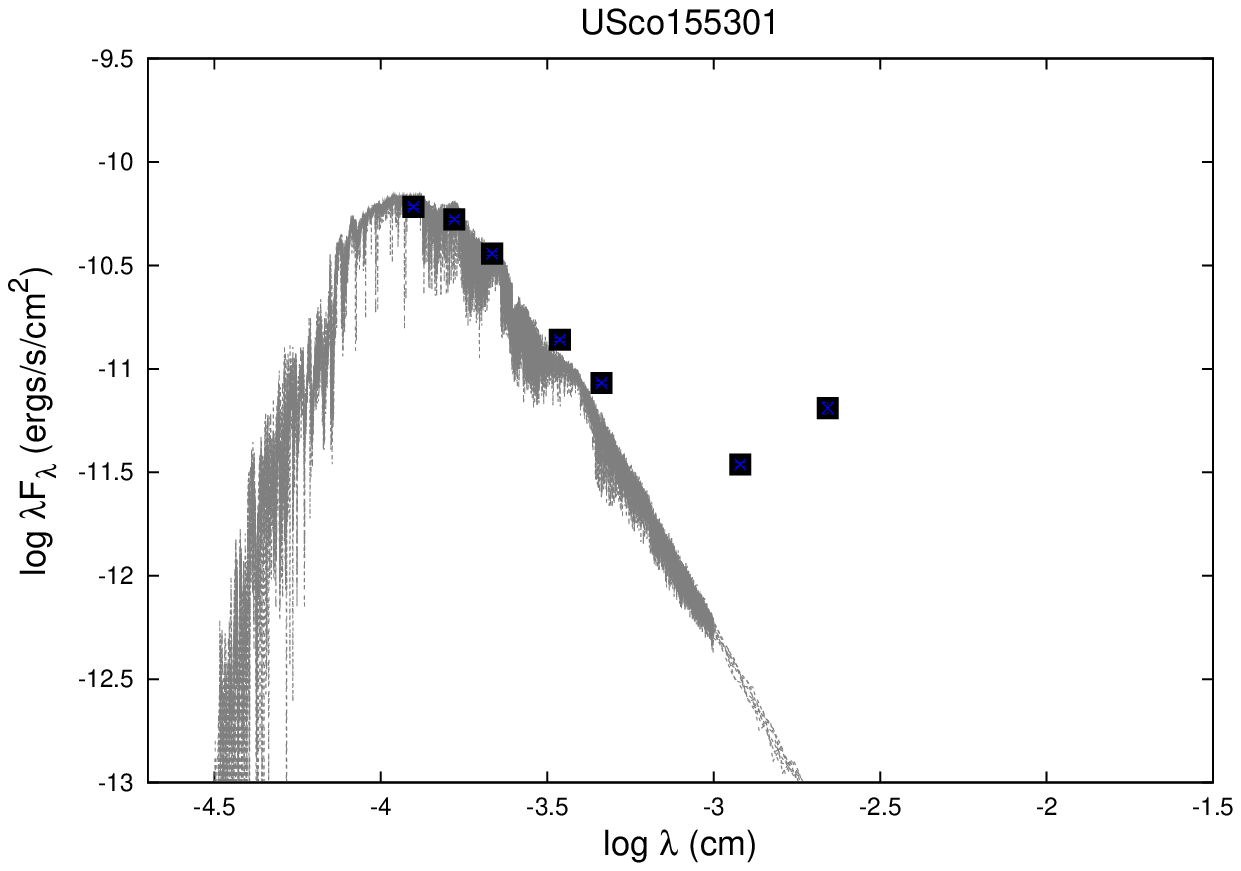} 
  \includegraphics[width=50mm]{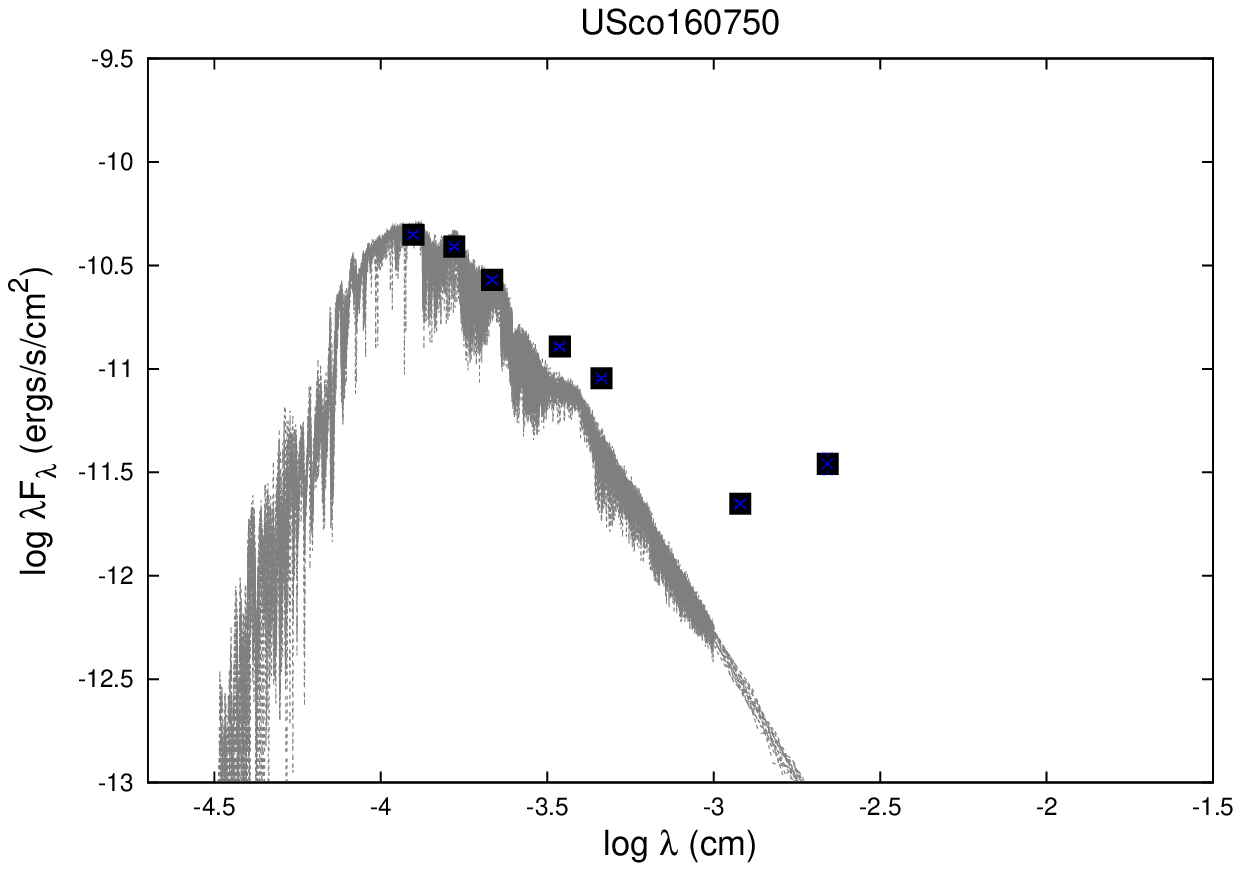} \\
   \includegraphics[width=50mm]{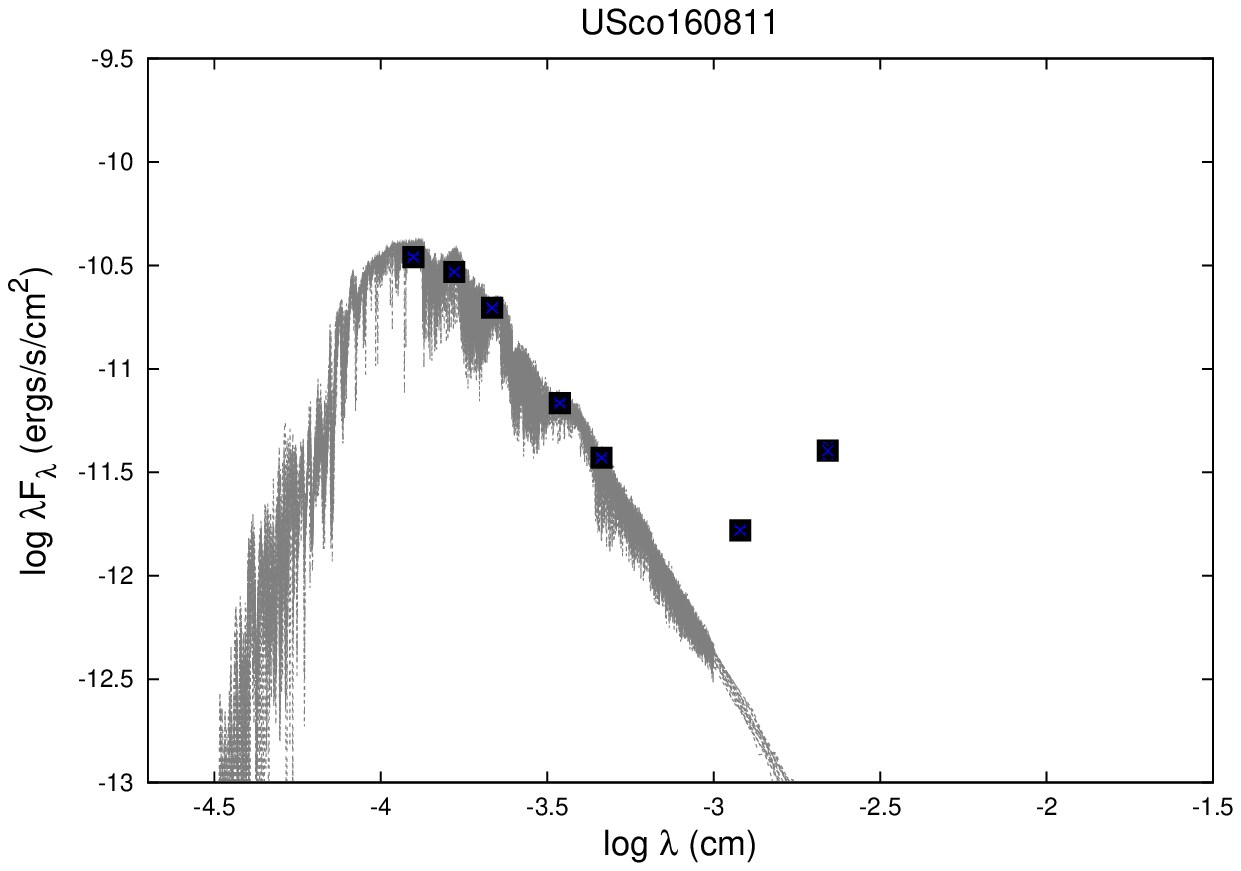} 
    \includegraphics[width=50mm]{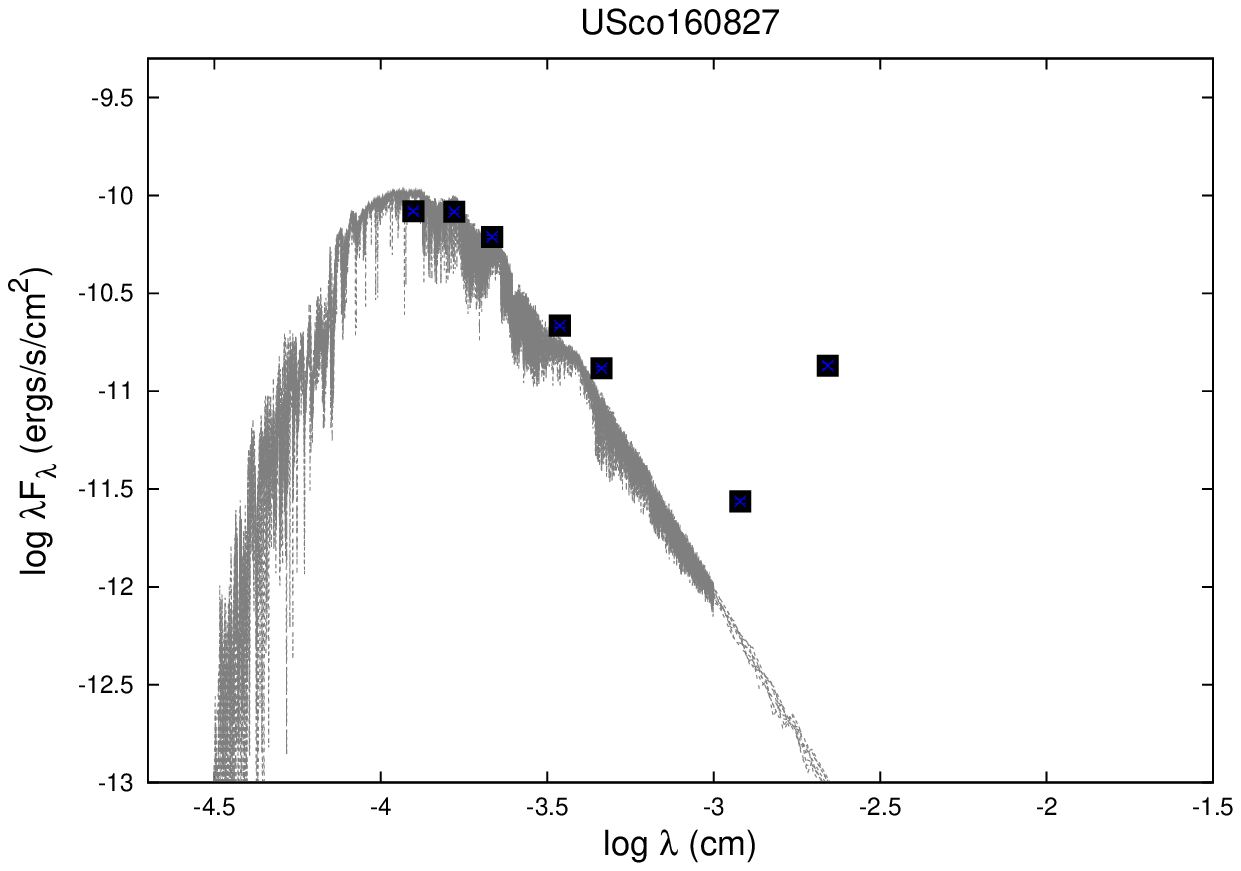} 
     \includegraphics[width=50mm]{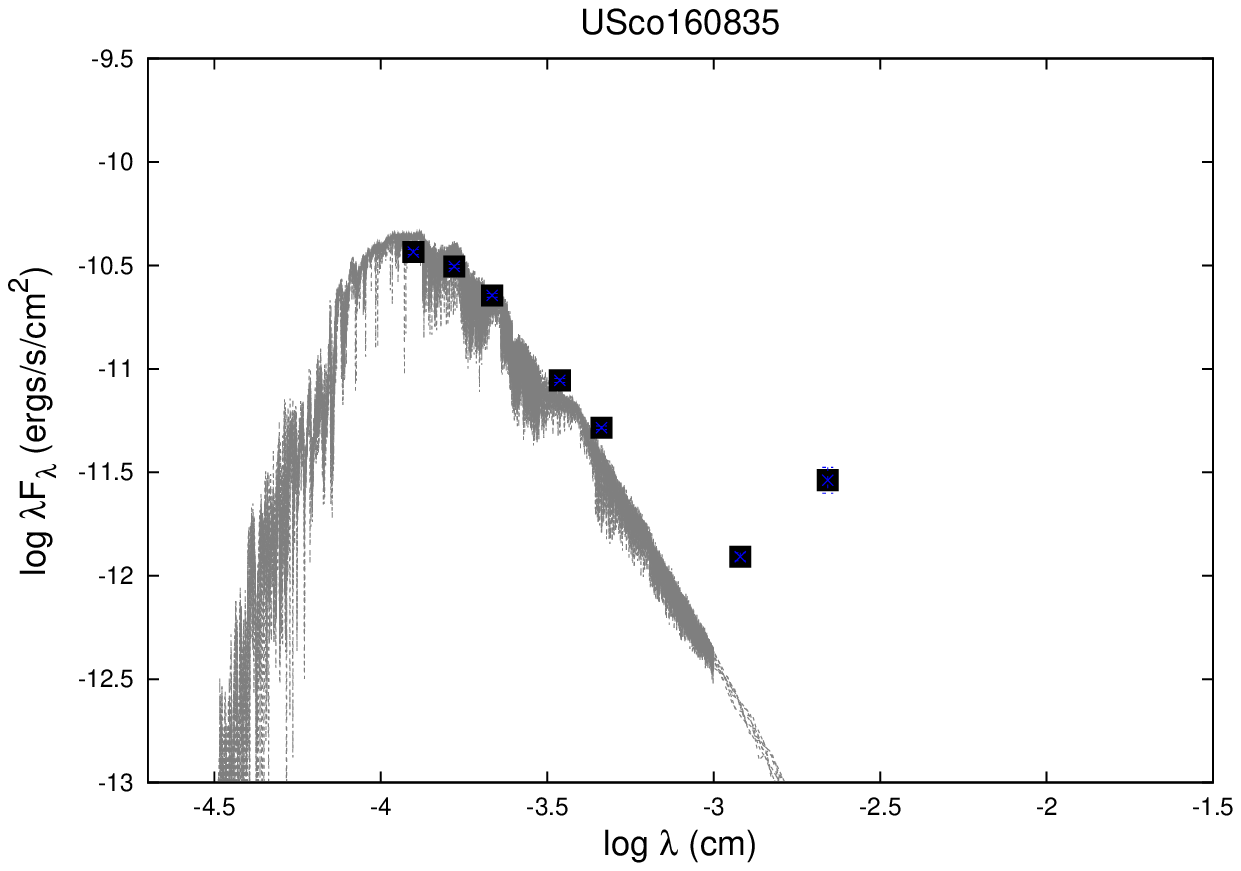} \\
      \includegraphics[width=50mm]{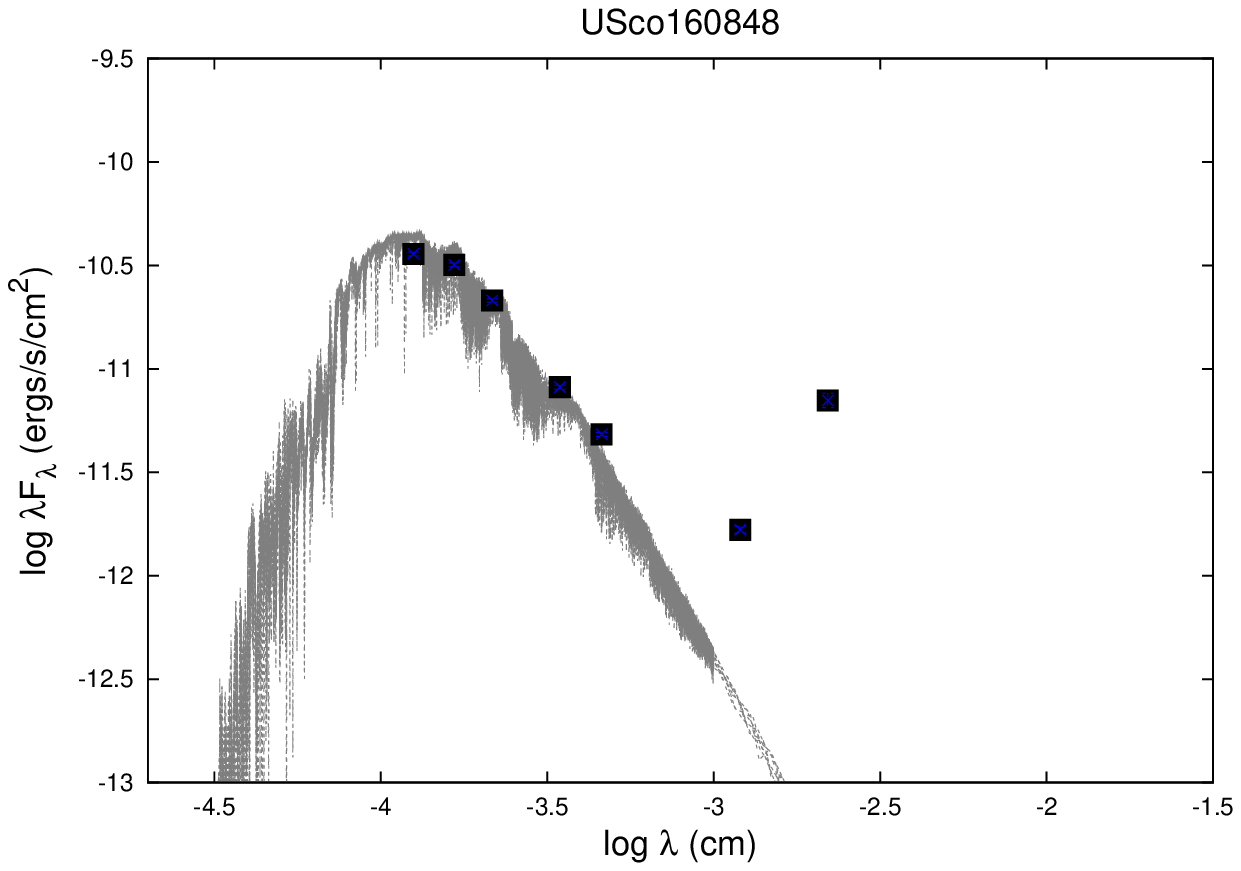} 
       \includegraphics[width=50mm]{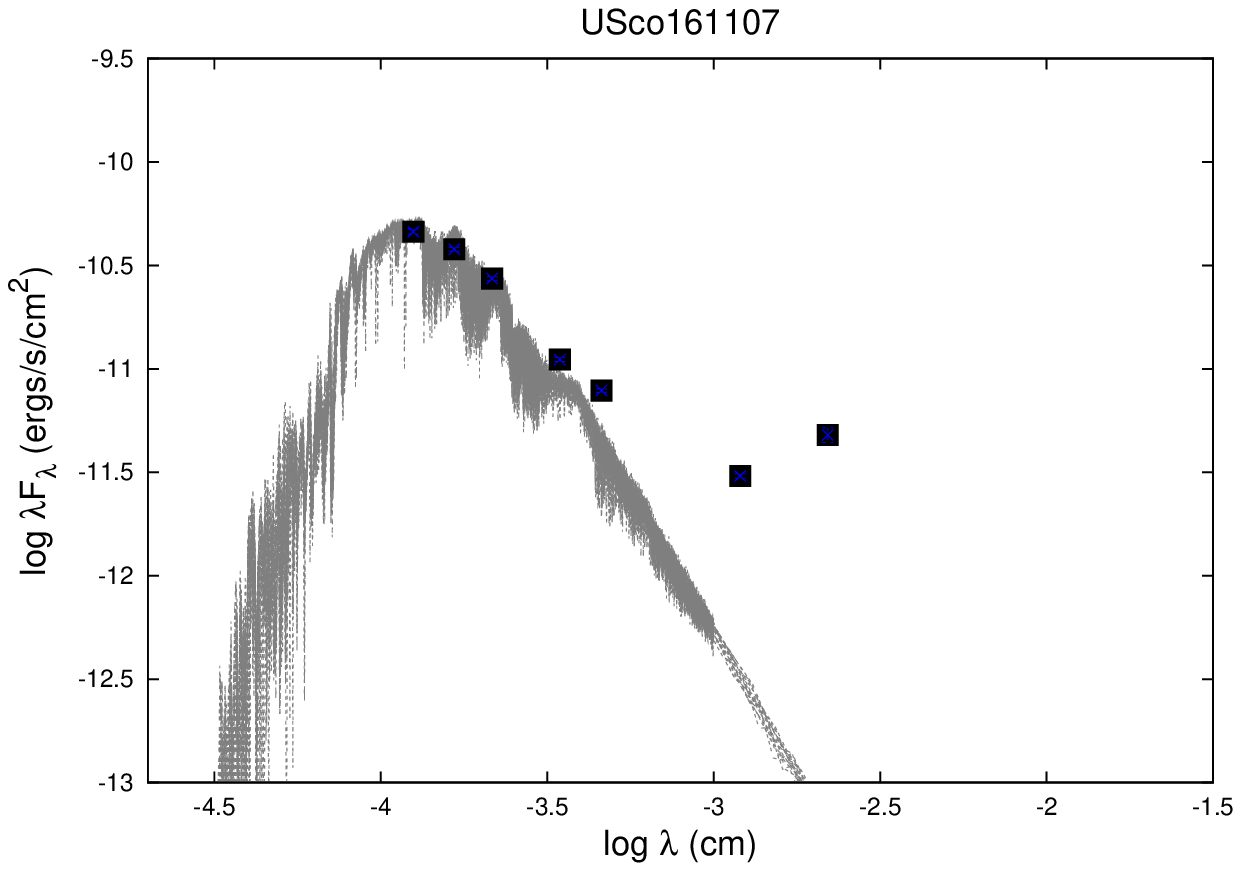} 
        \includegraphics[width=50mm]{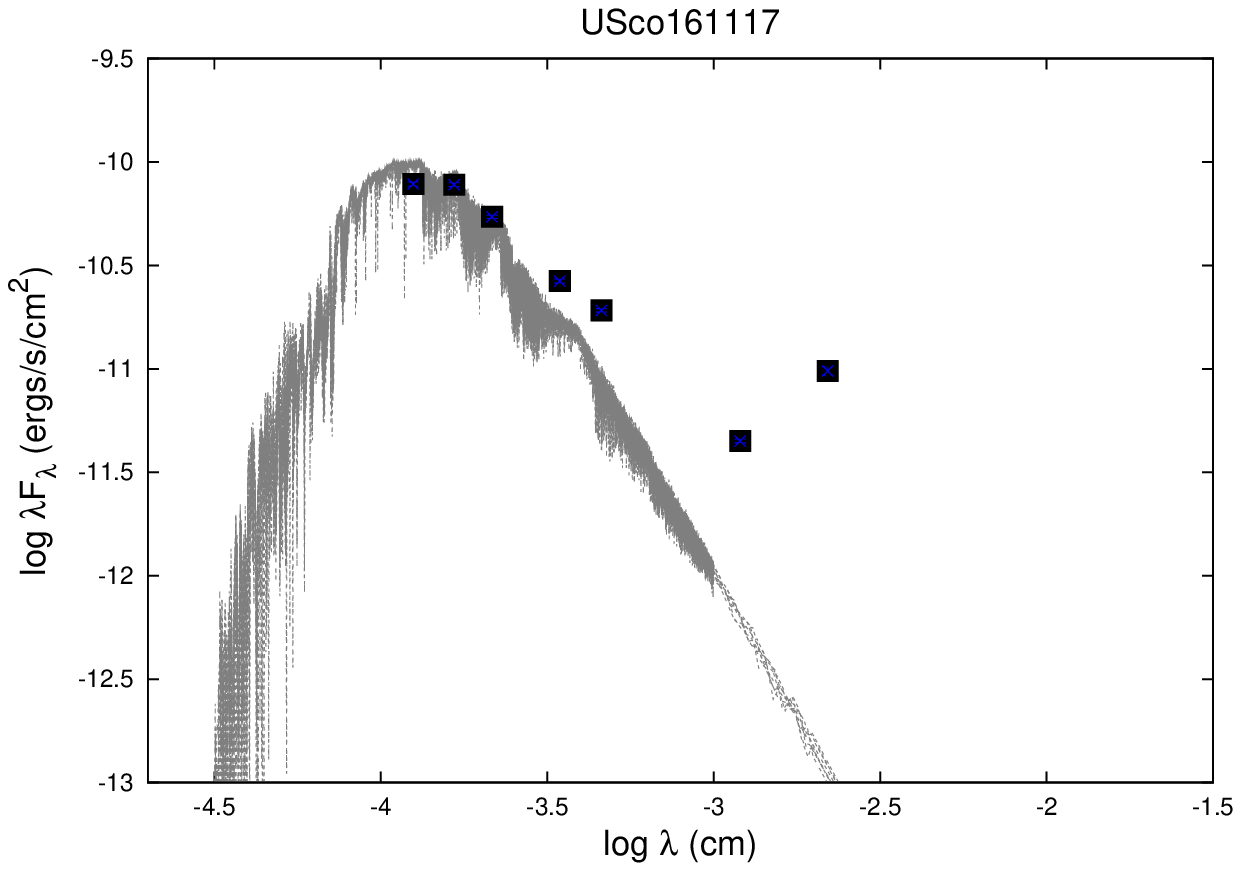} \\
         \includegraphics[width=50mm]{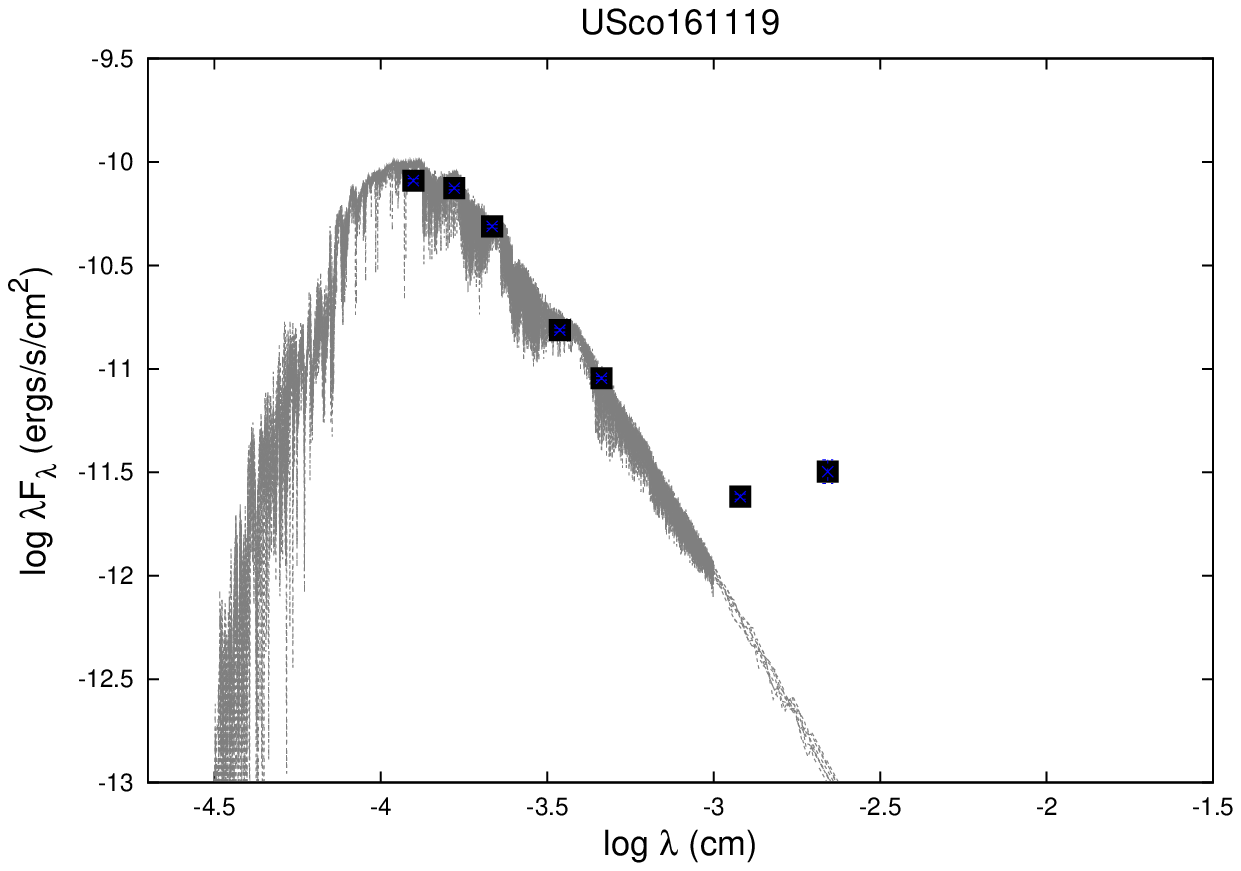} 
          \includegraphics[width=50mm]{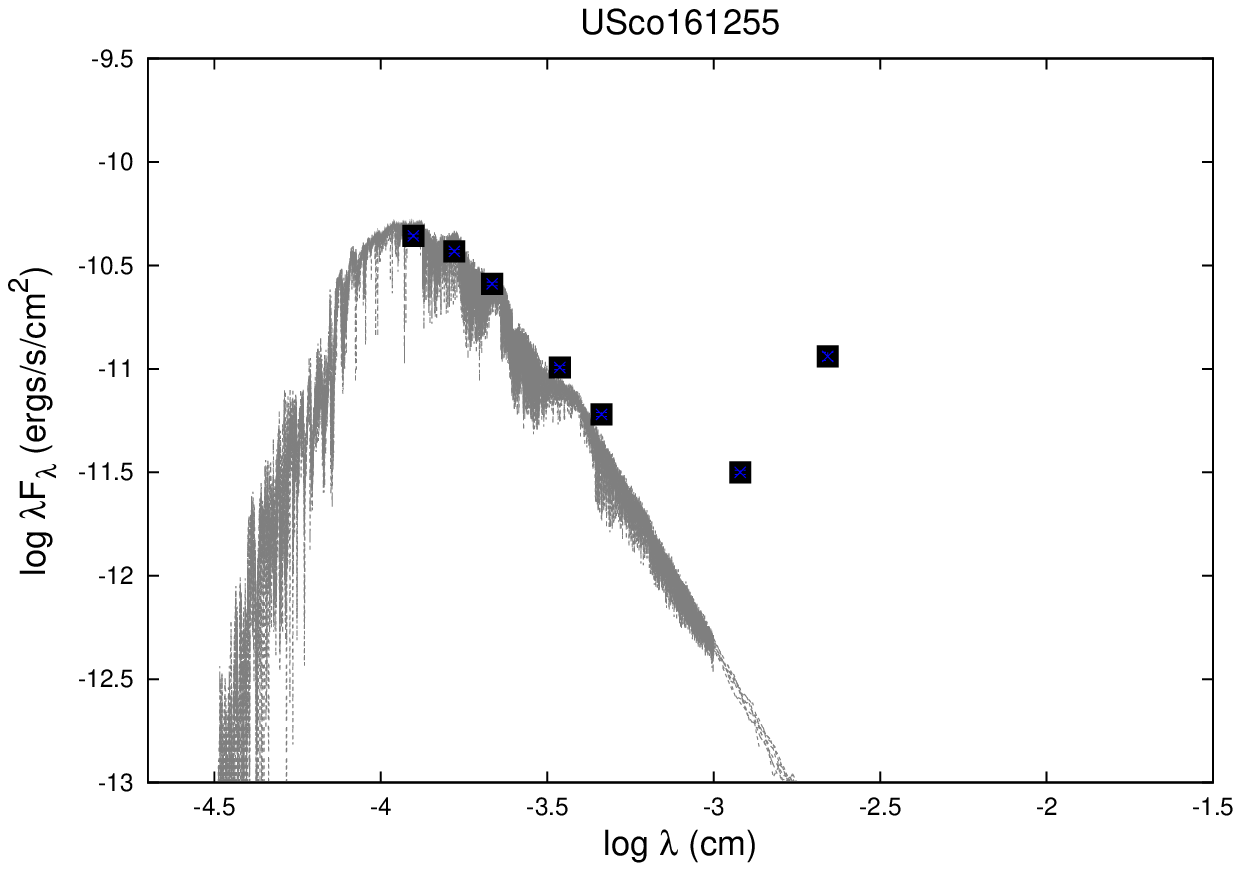} 
           \includegraphics[width=50mm]{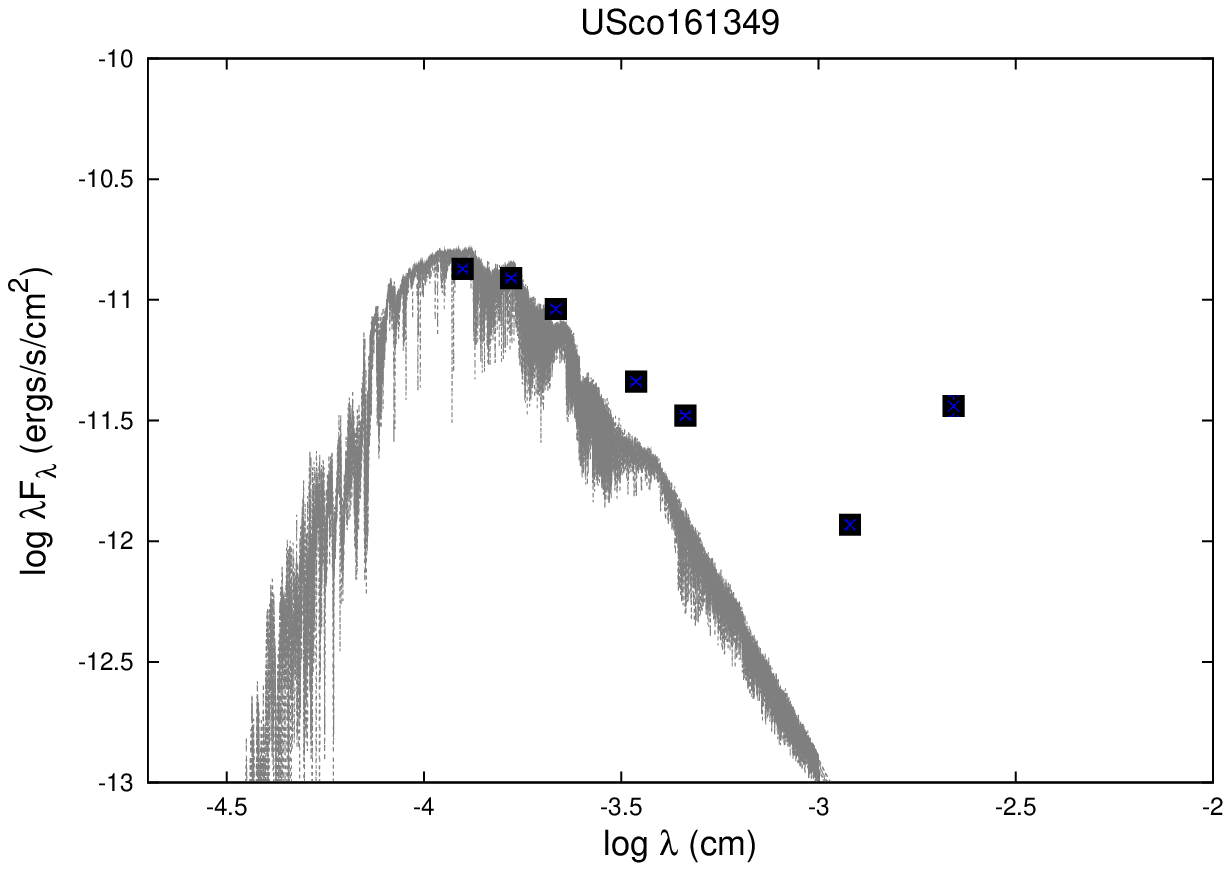} \\
            \includegraphics[width=50mm]{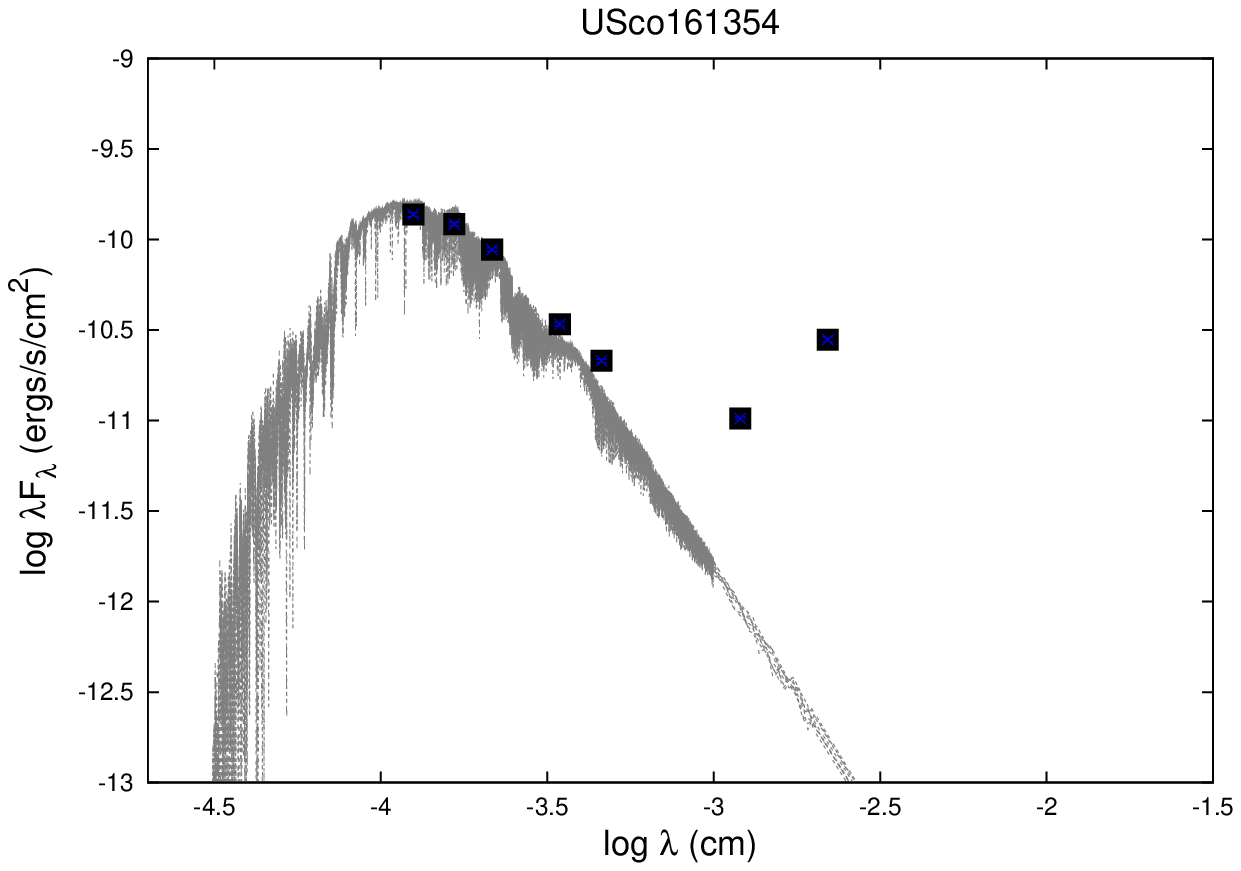} 
             \includegraphics[width=50mm]{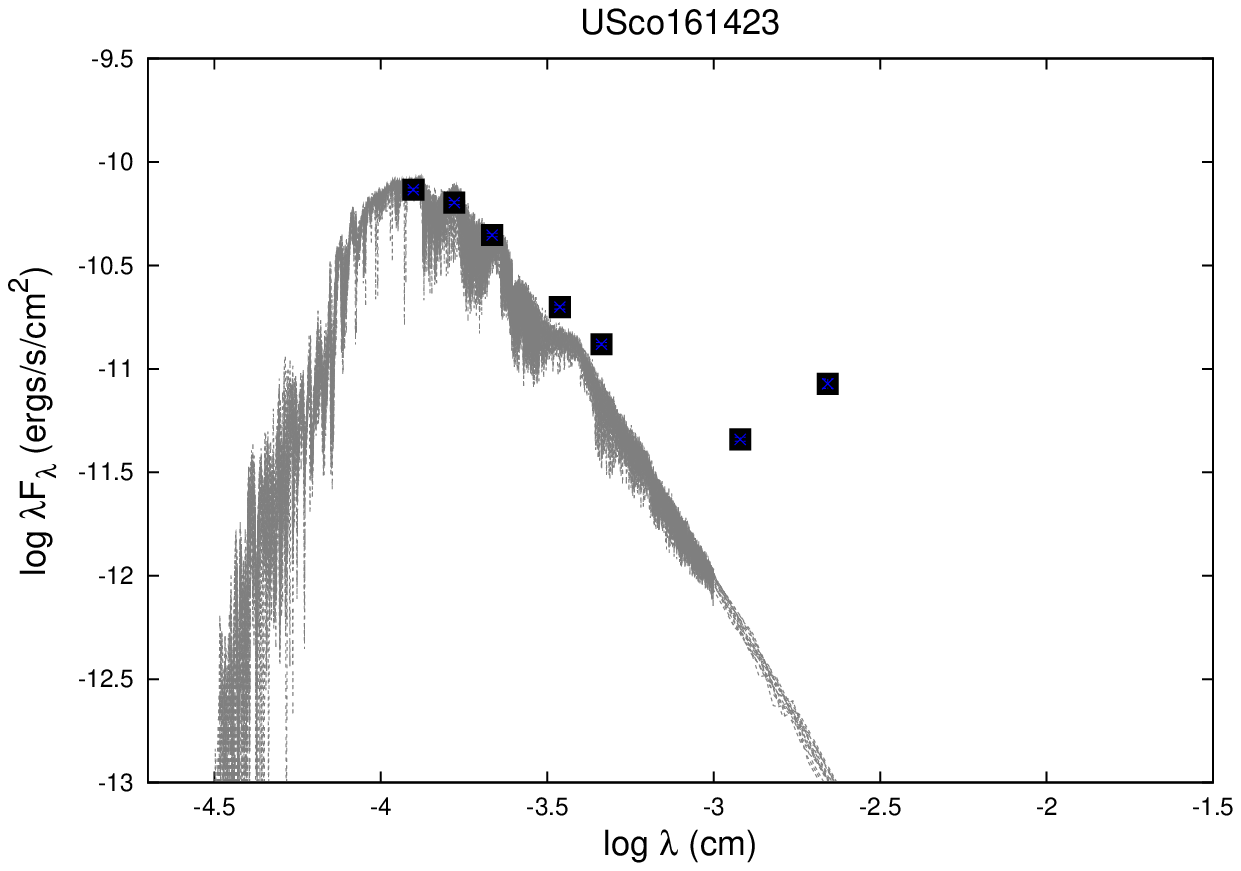} 
              \includegraphics[width=50mm]{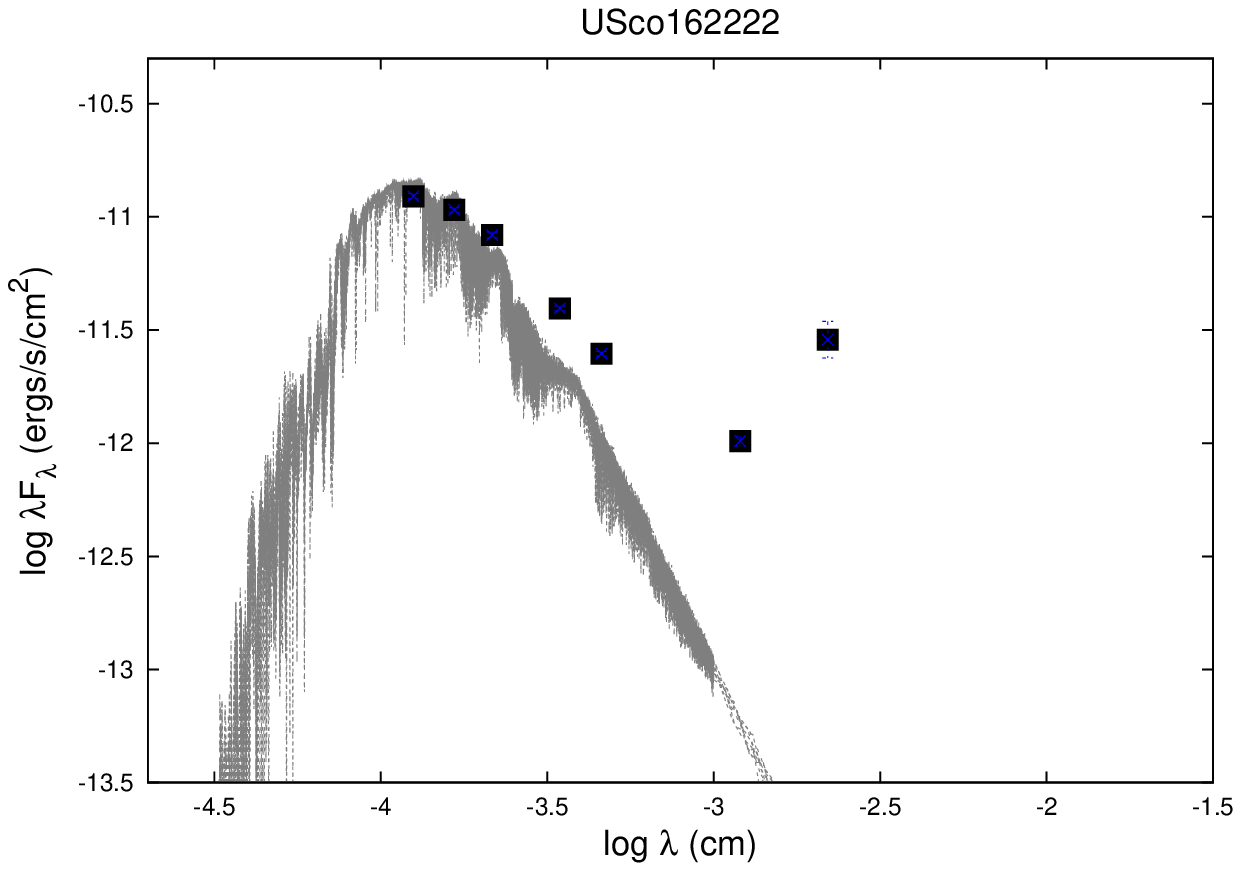} \\
               \includegraphics[width=50mm]{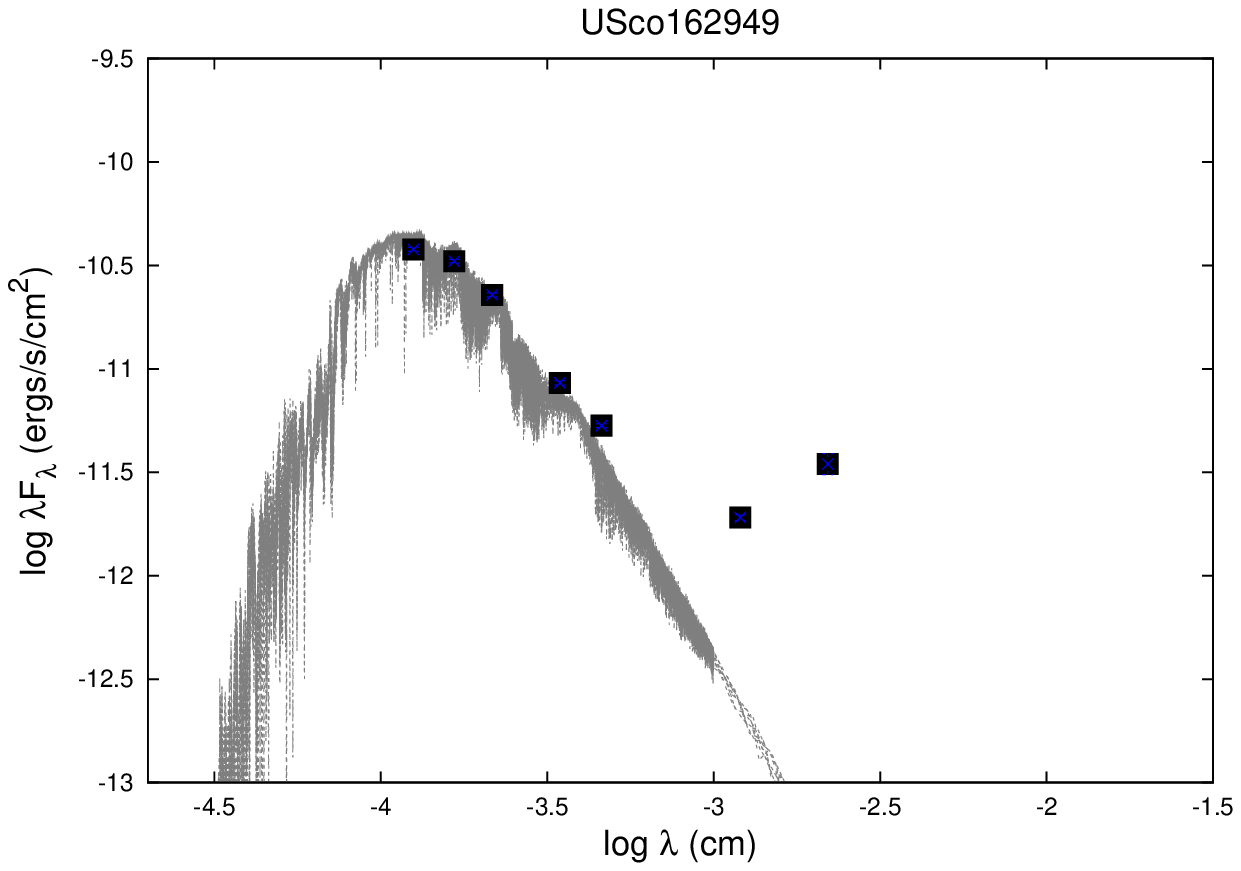} 
              \includegraphics[width=50mm]{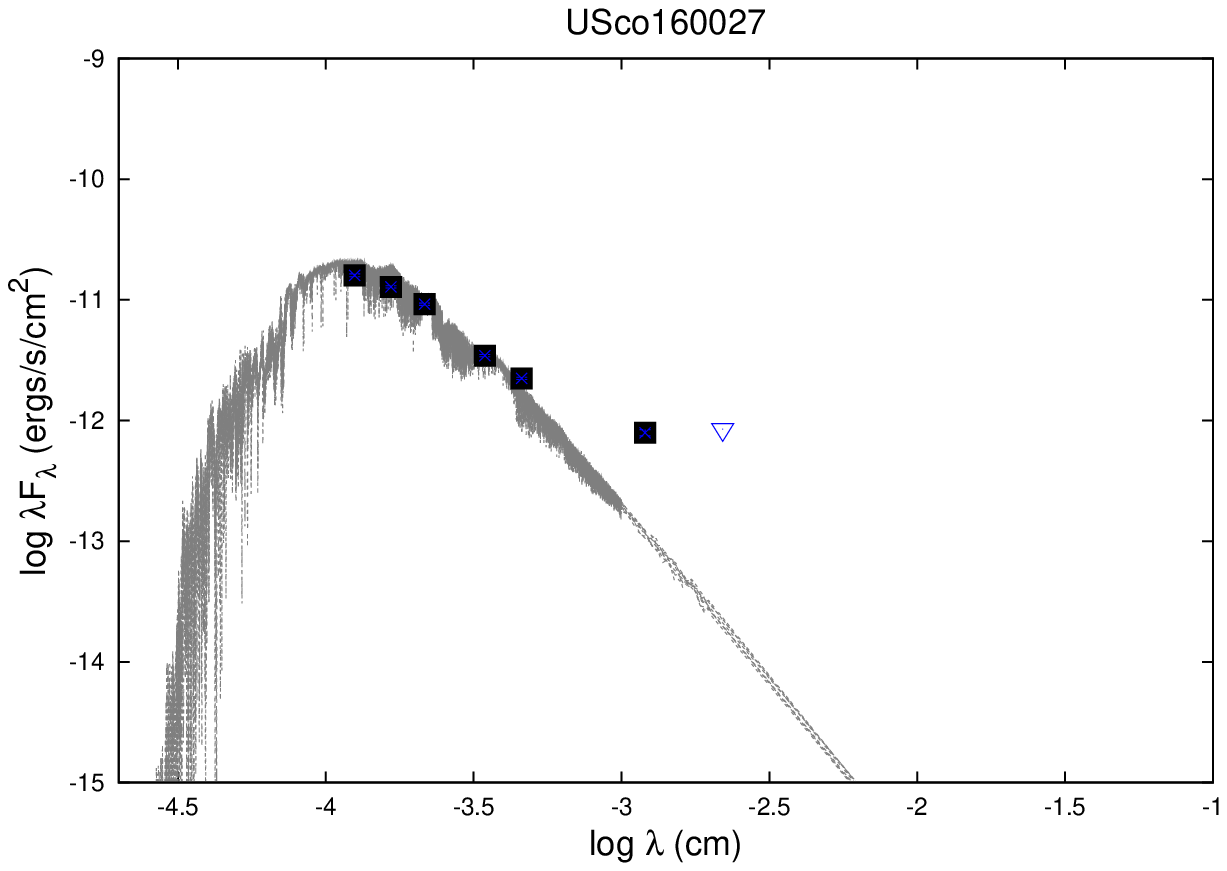} 
               \includegraphics[width=50mm]{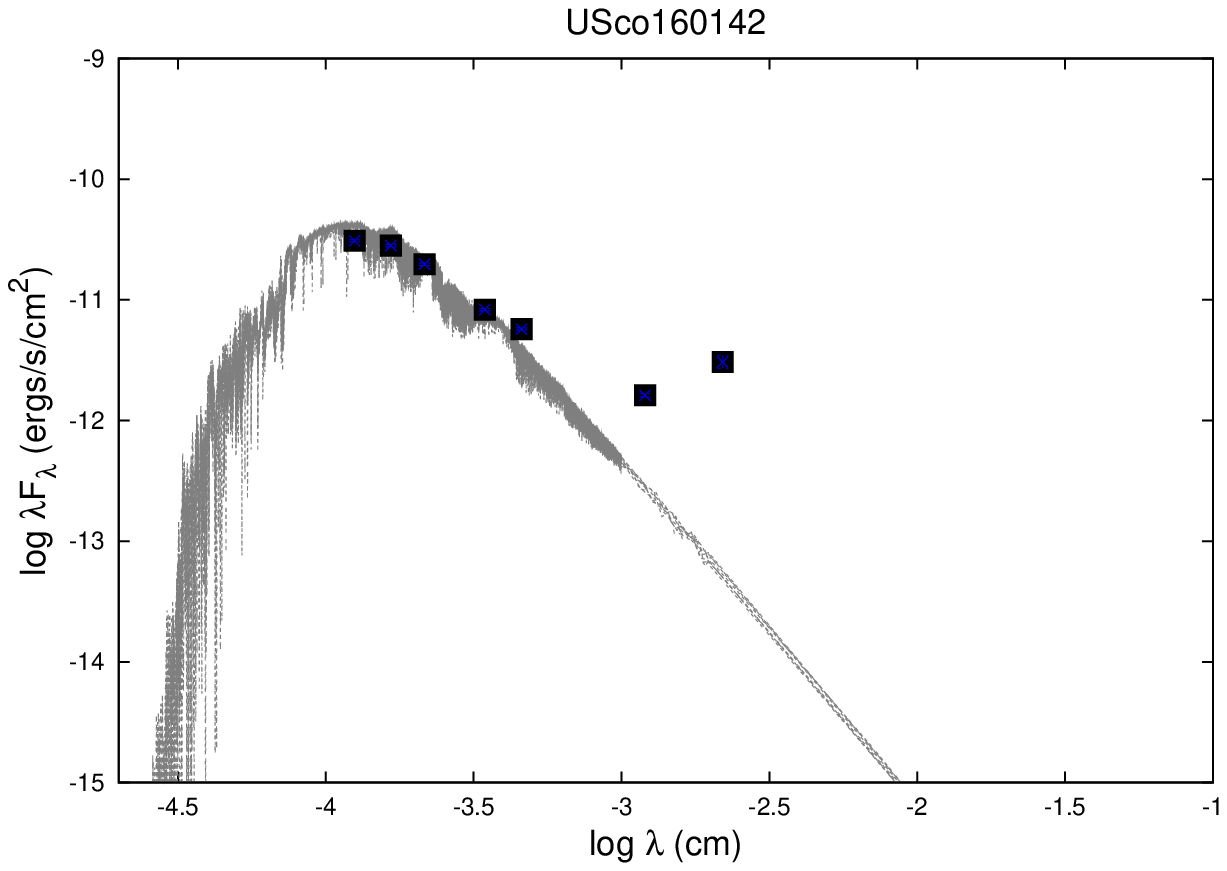} \\                
    \caption{The M dwarf discs in USco.}
   \label{USco-Mdwarf}
 \end{figure*}
 
 \begin{figure*}
 \setcounter{figure}{7}
\includegraphics[width=50mm]{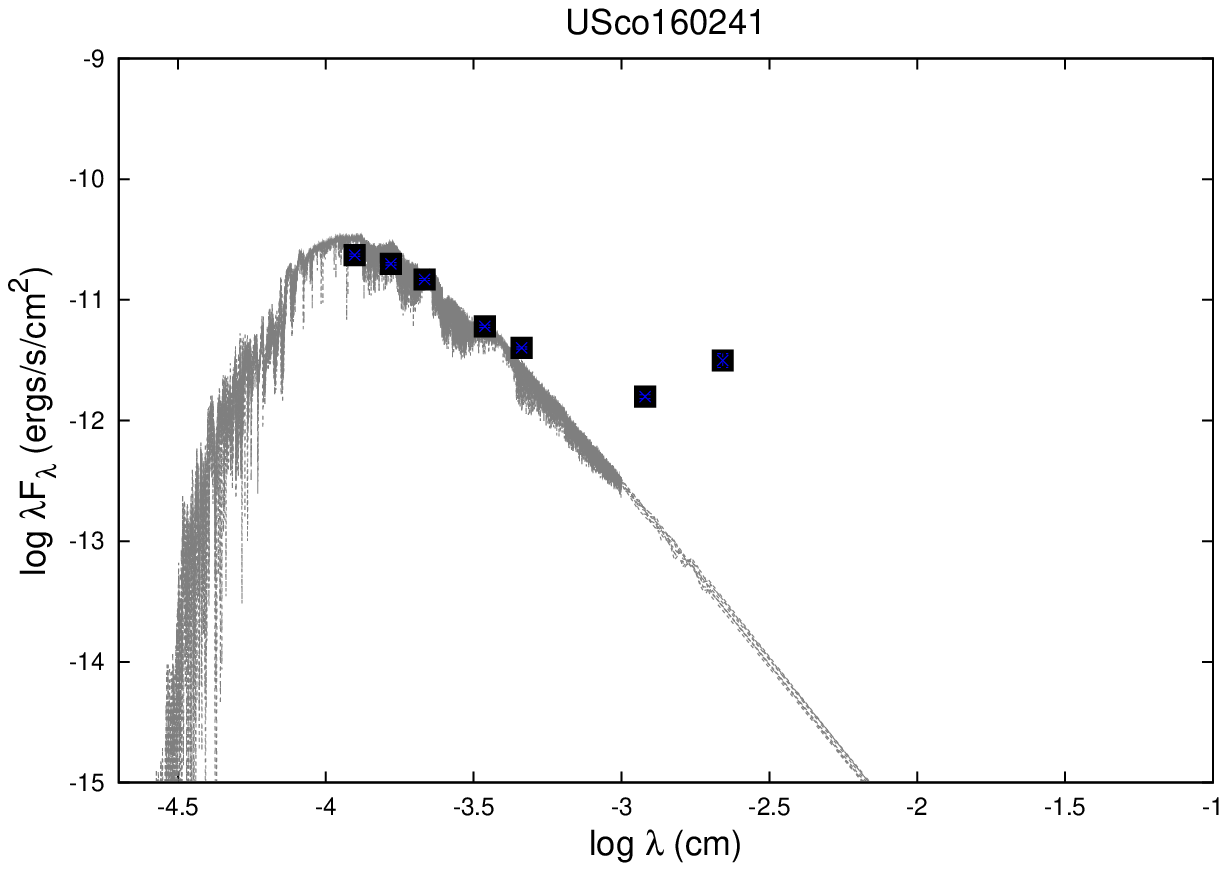} 
 \includegraphics[width=50mm]{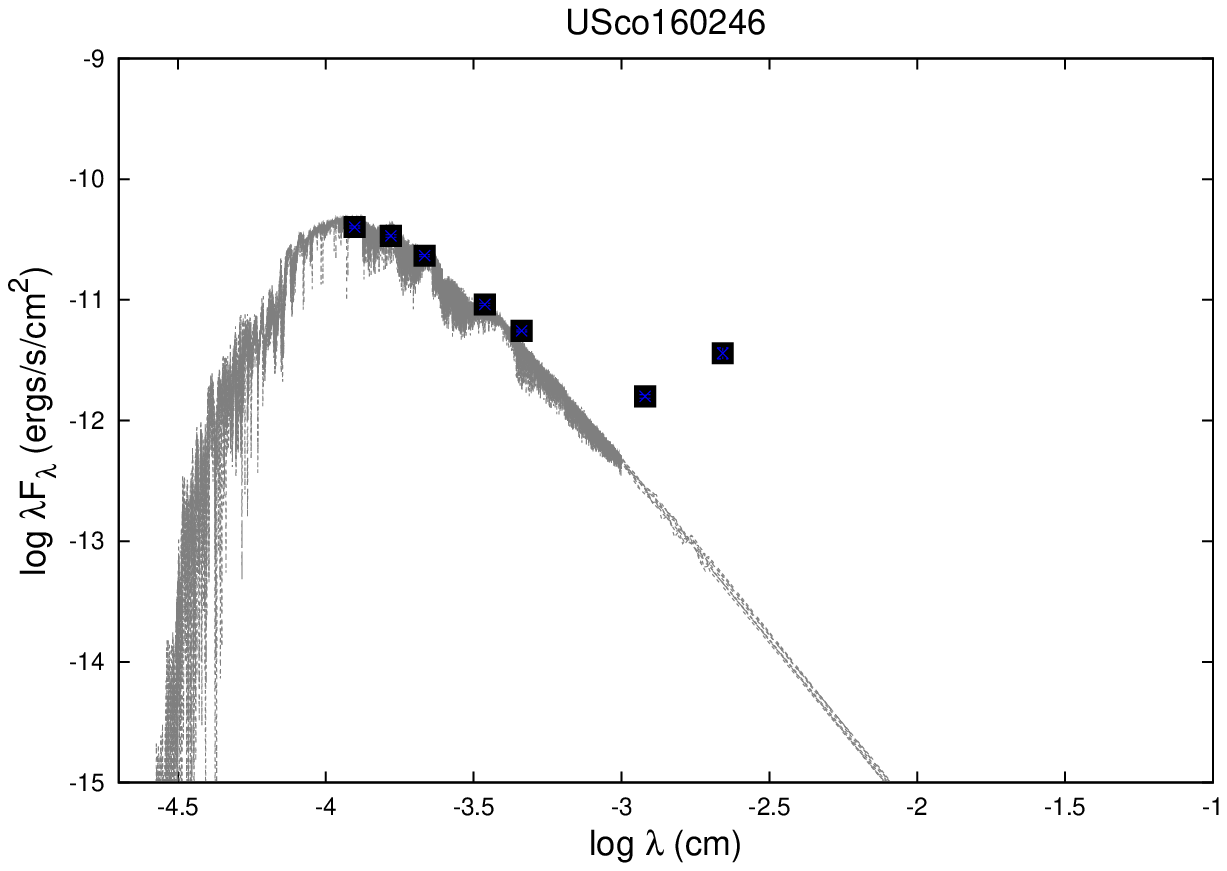} 
  \includegraphics[width=50mm]{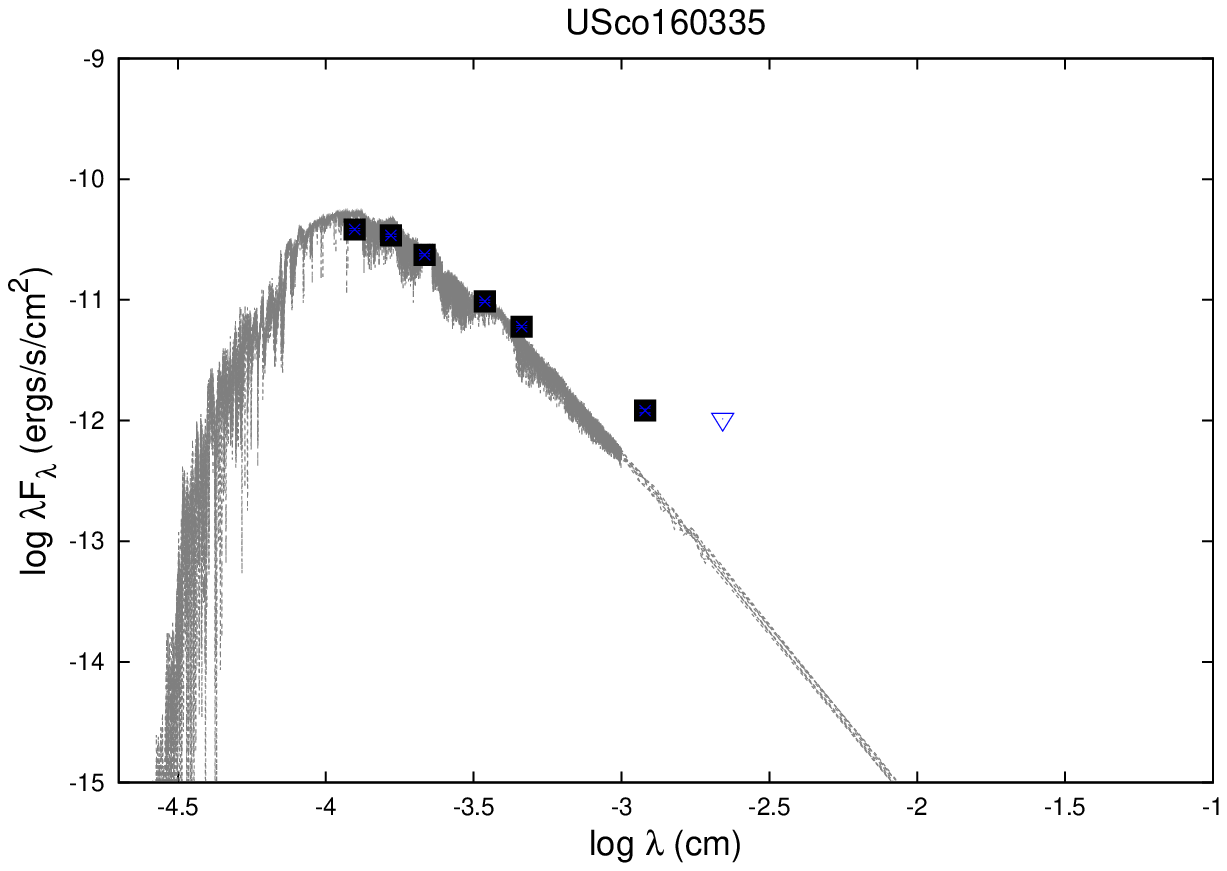} \\
   \includegraphics[width=50mm]{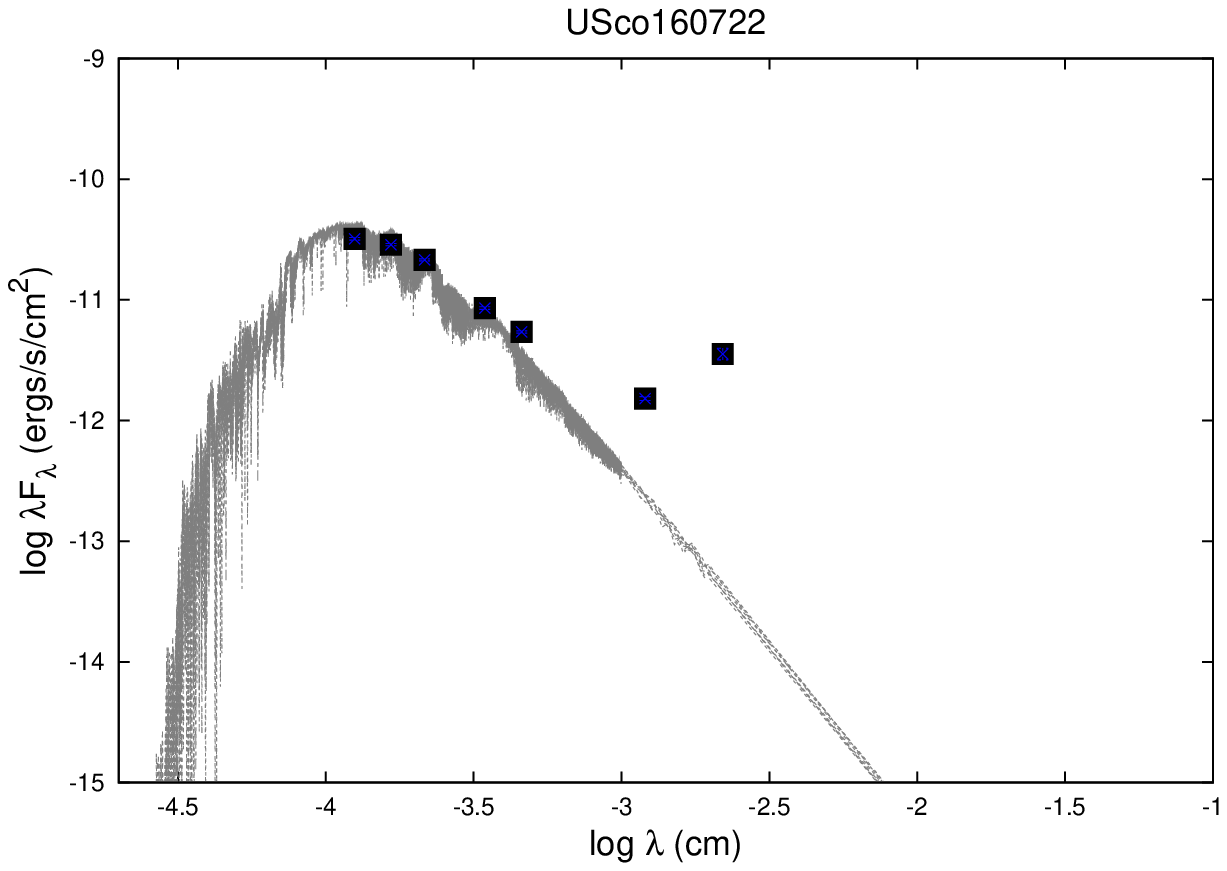} 
    \includegraphics[width=50mm]{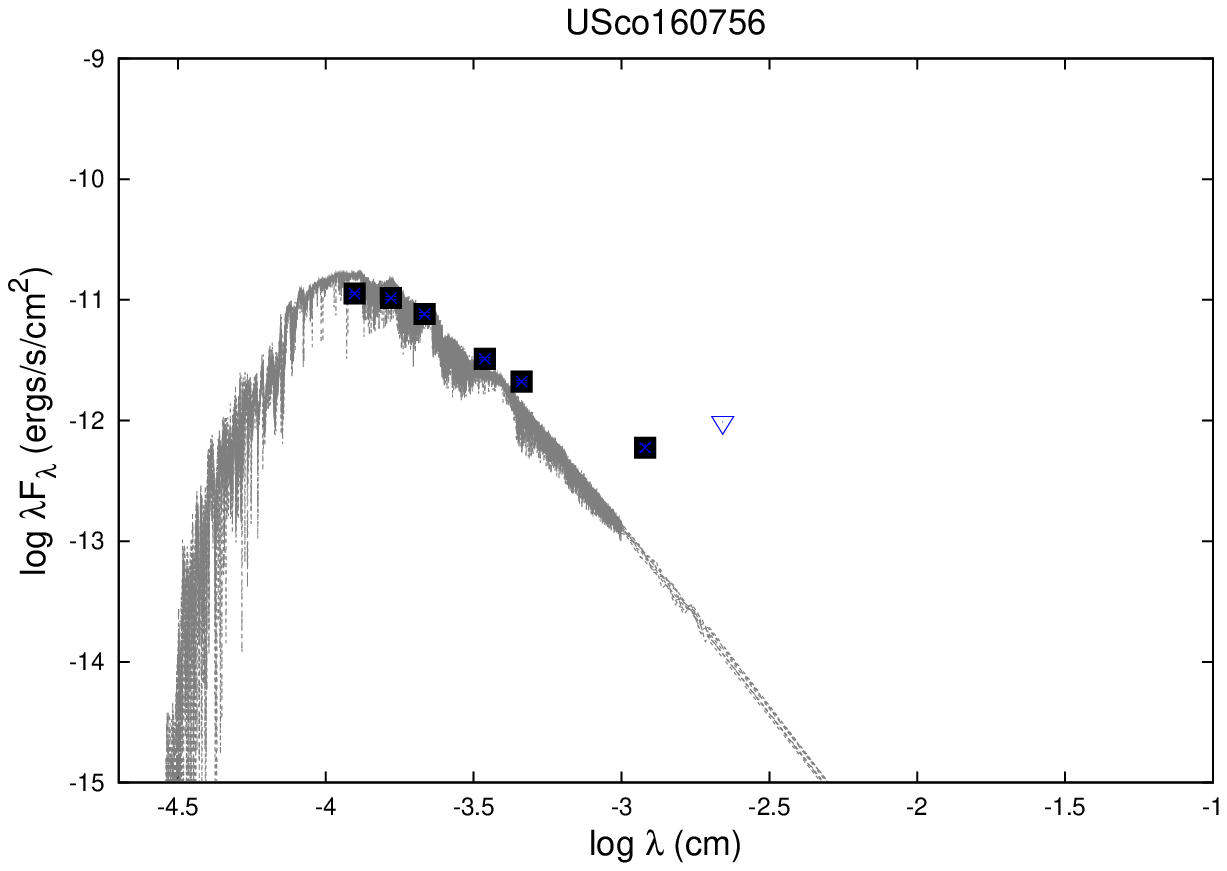} 
     \includegraphics[width=50mm]{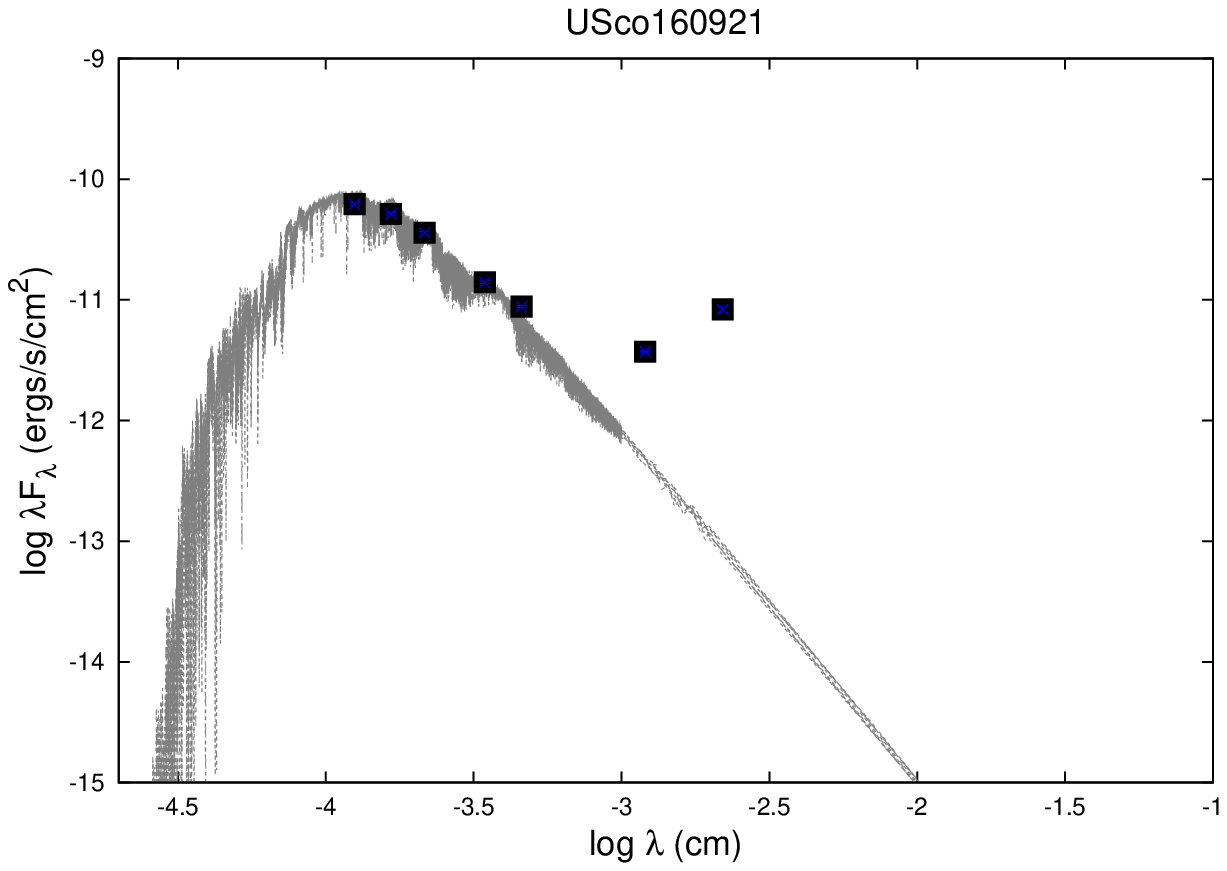} \\
      \includegraphics[width=50mm]{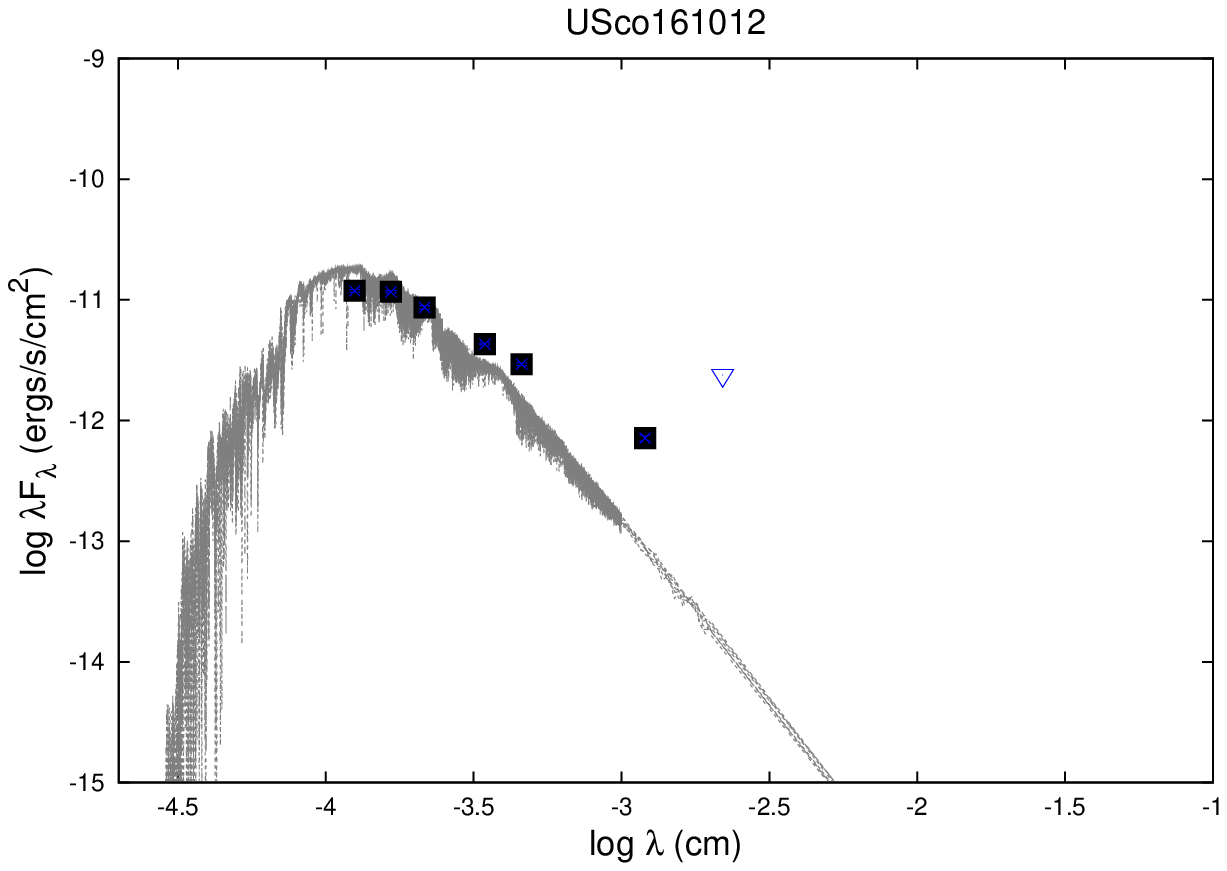} 
       \includegraphics[width=50mm]{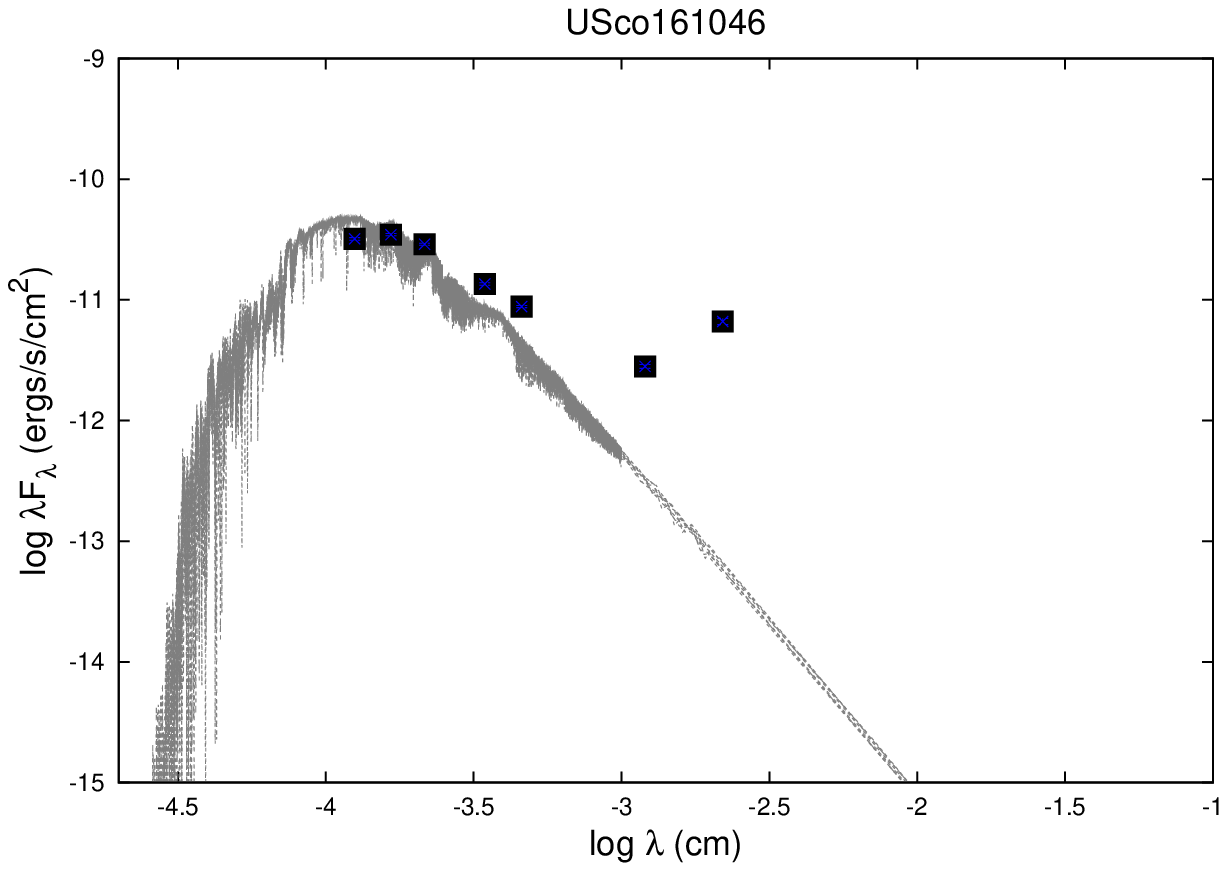} 
        \includegraphics[width=50mm]{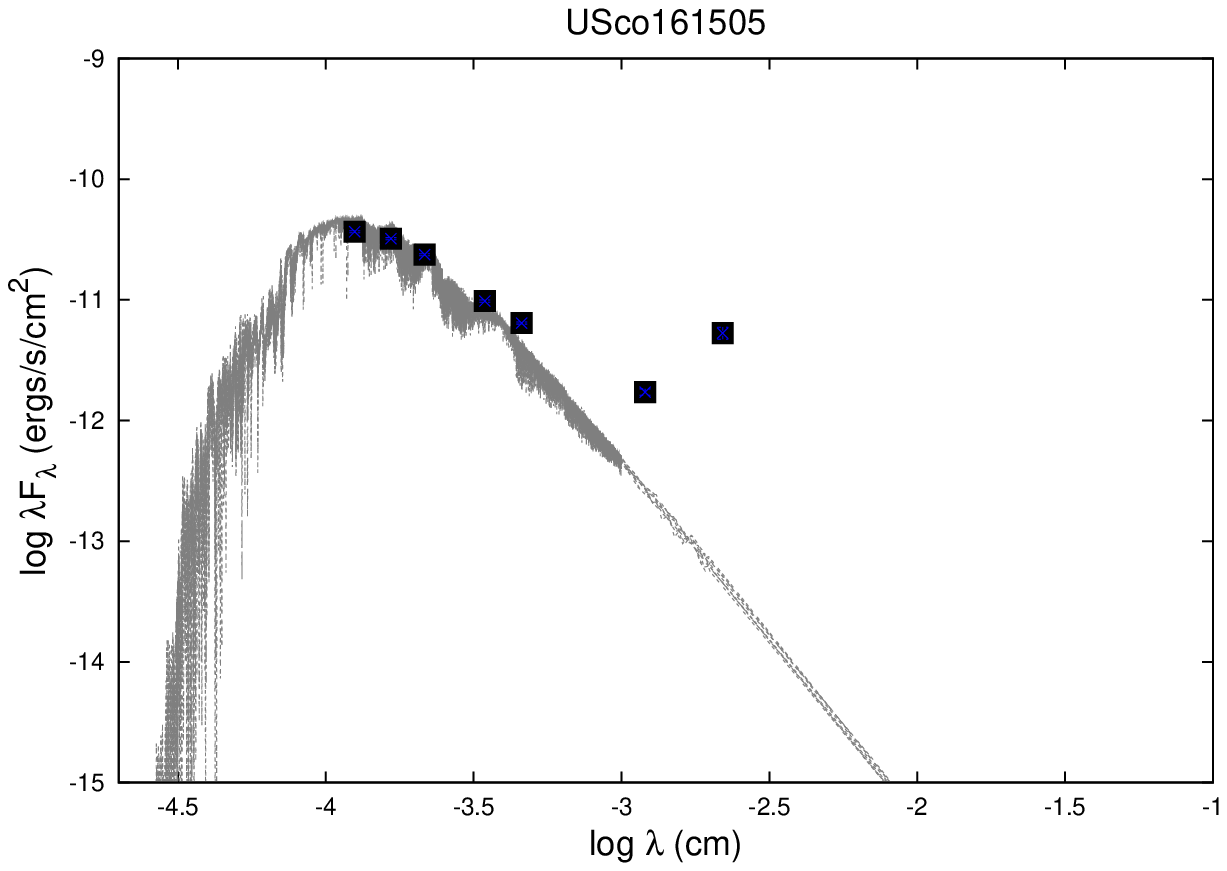} \\
         \includegraphics[width=50mm]{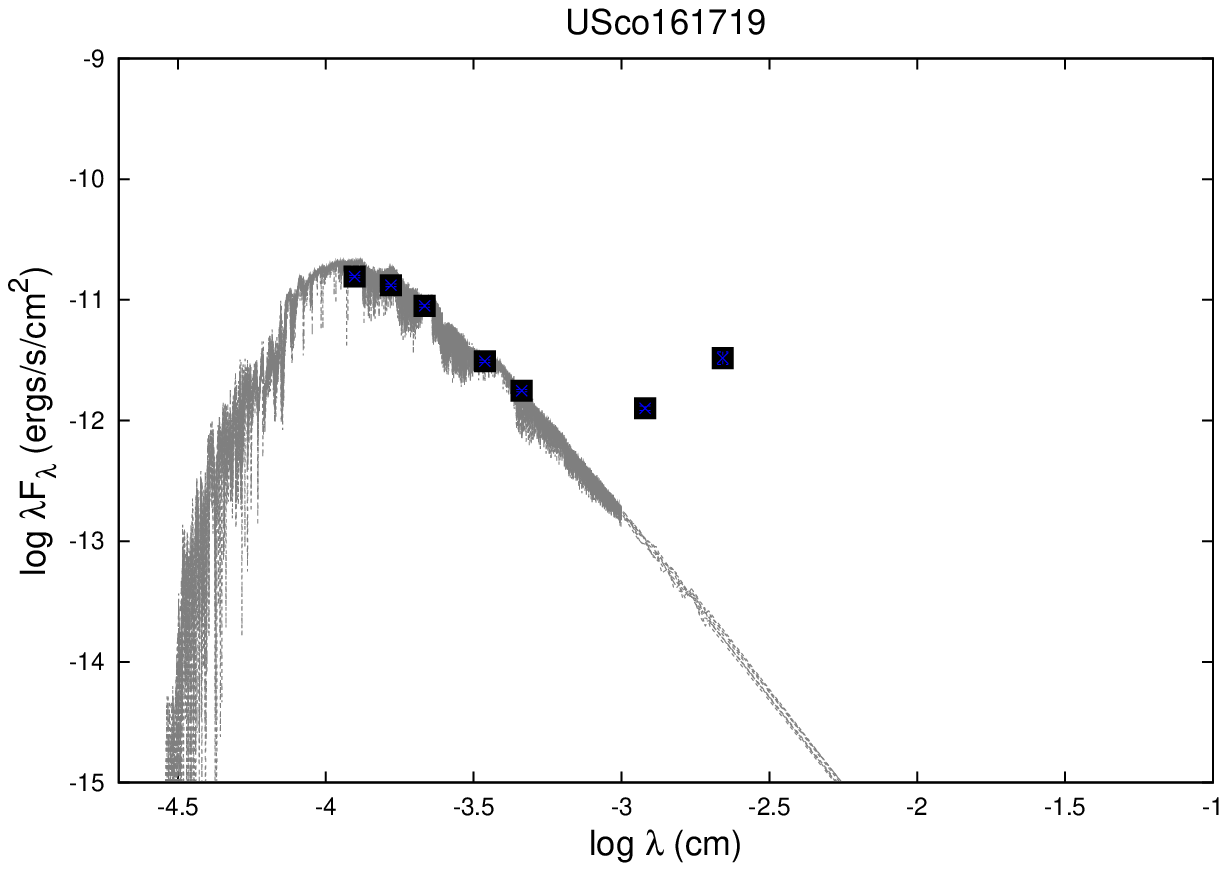} 
          \includegraphics[width=50mm]{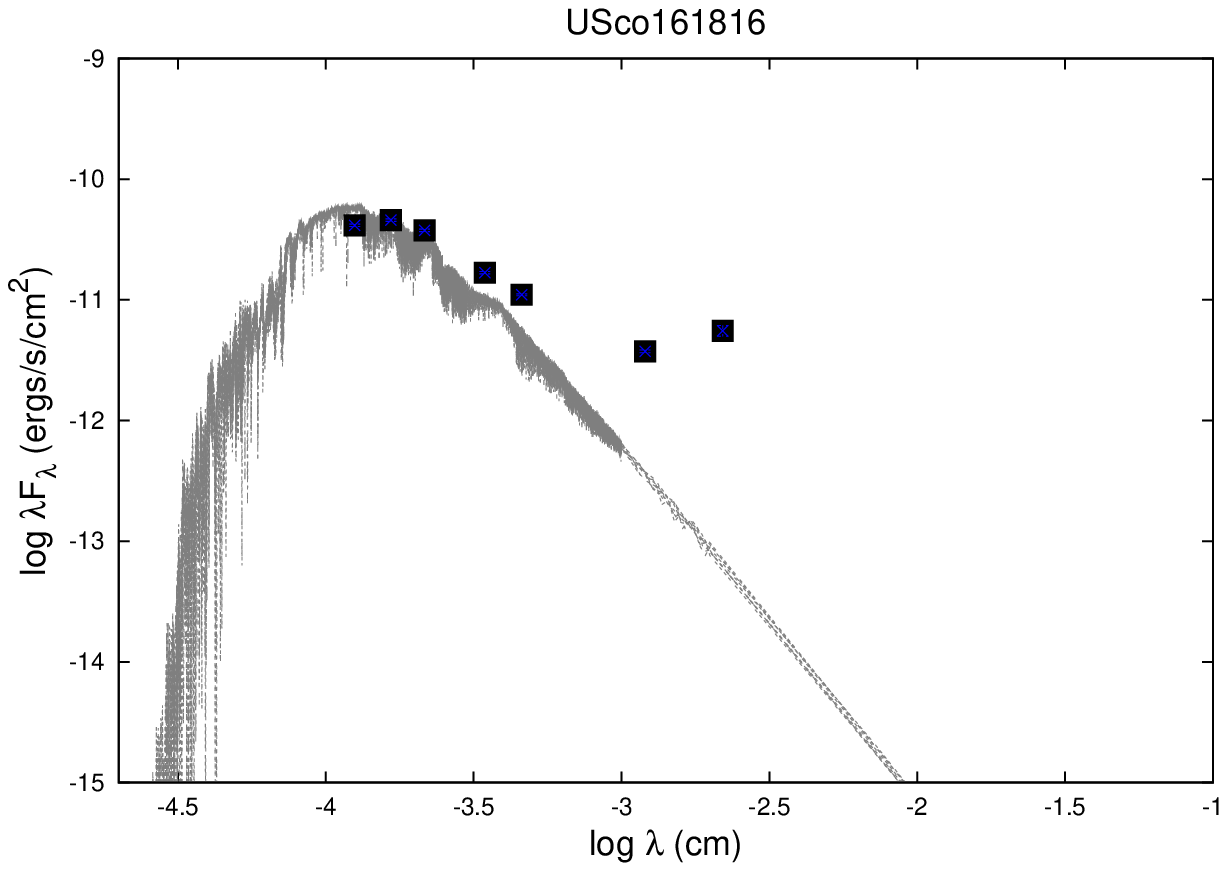} 
           \includegraphics[width=50mm]{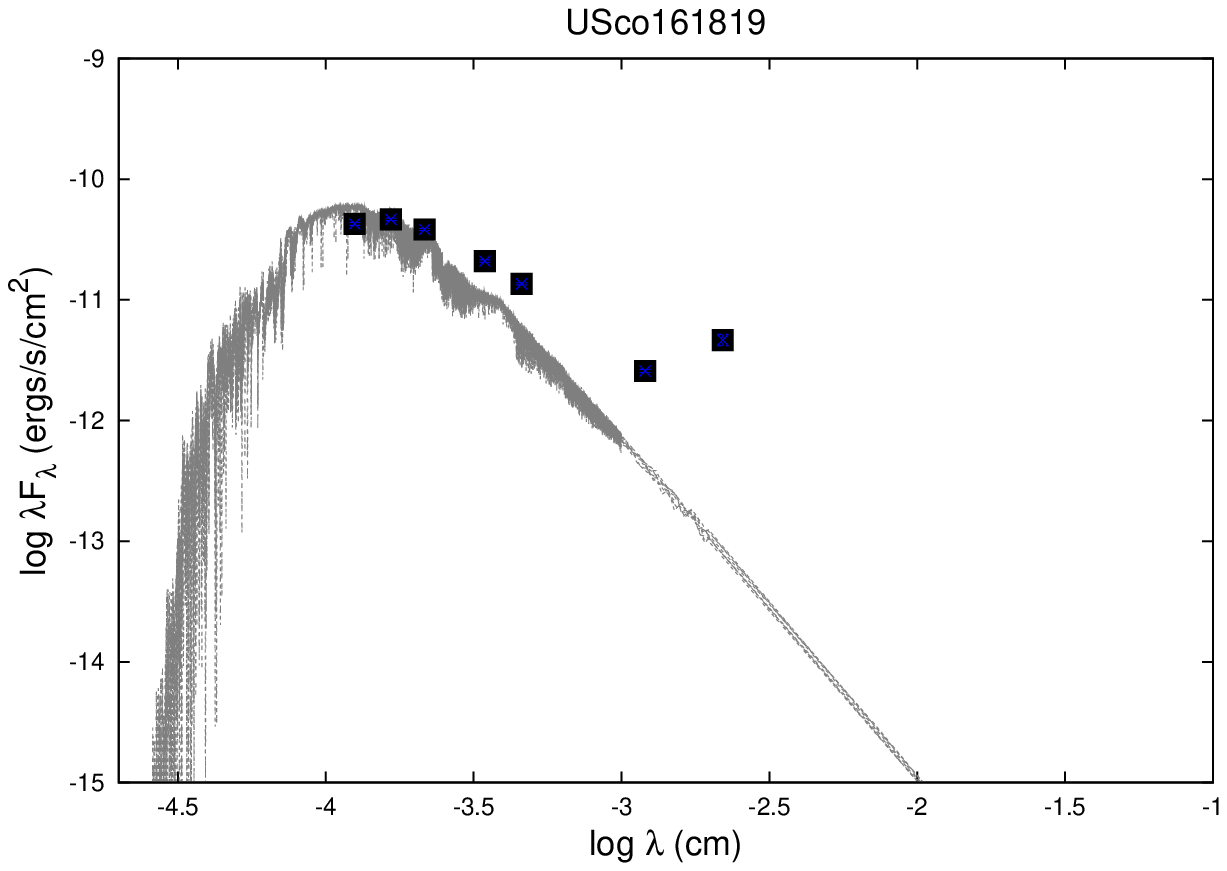} \\
            \includegraphics[width=50mm]{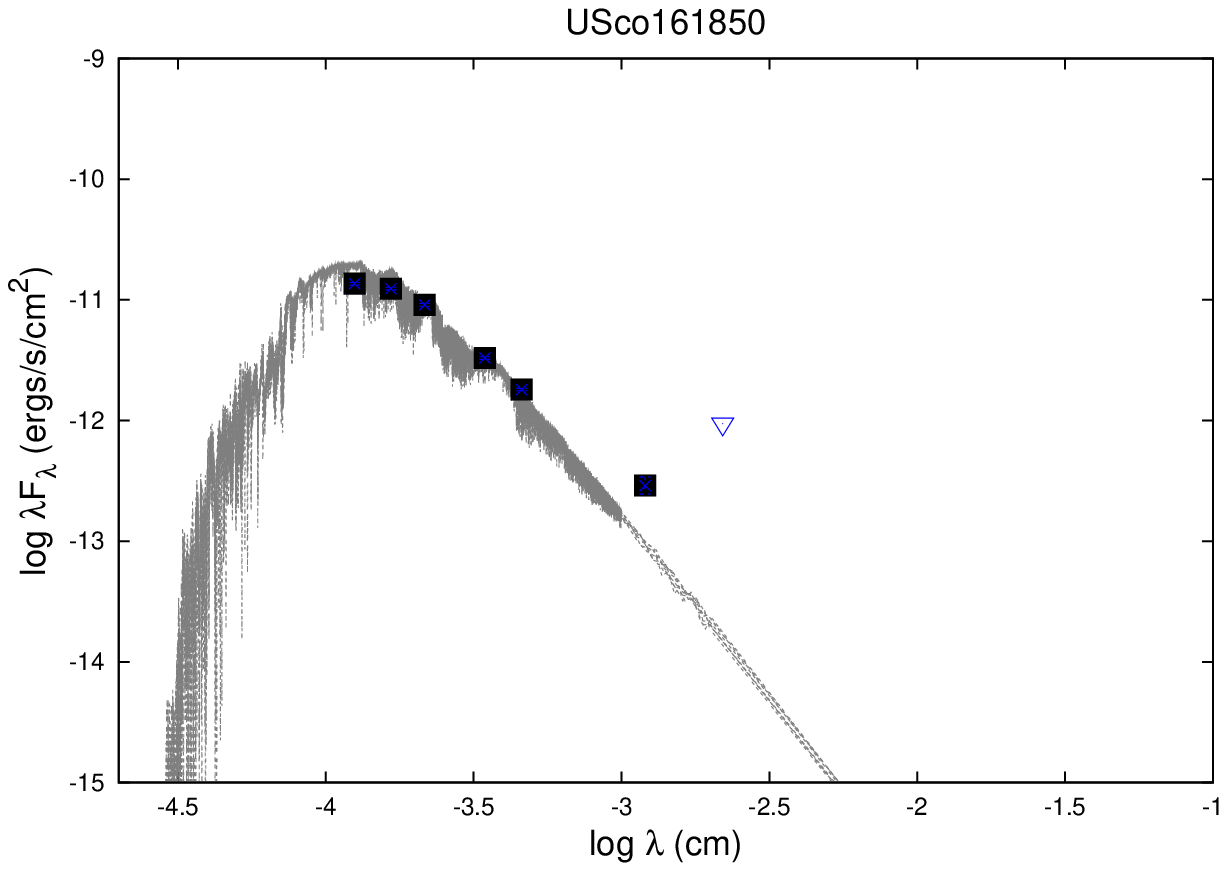} 
             \includegraphics[width=50mm]{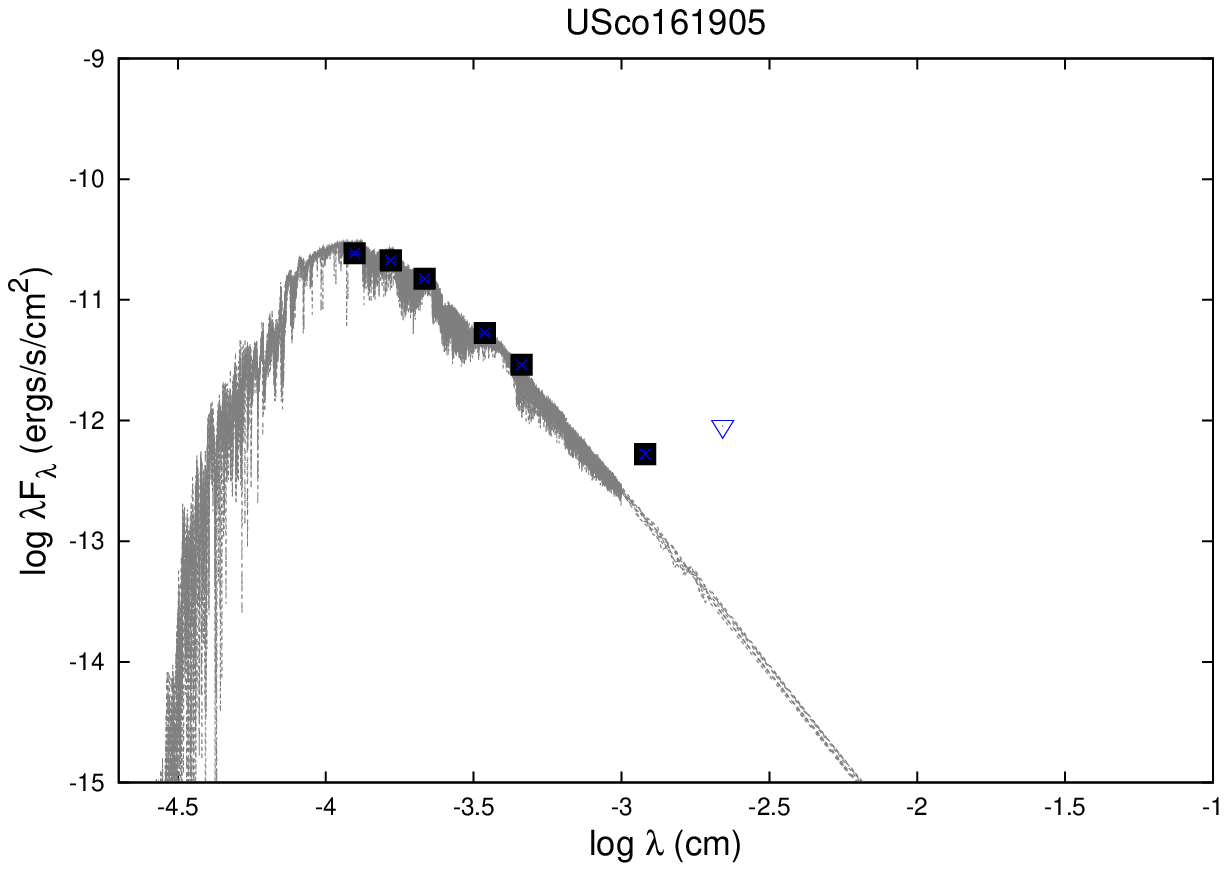} 
              \includegraphics[width=50mm]{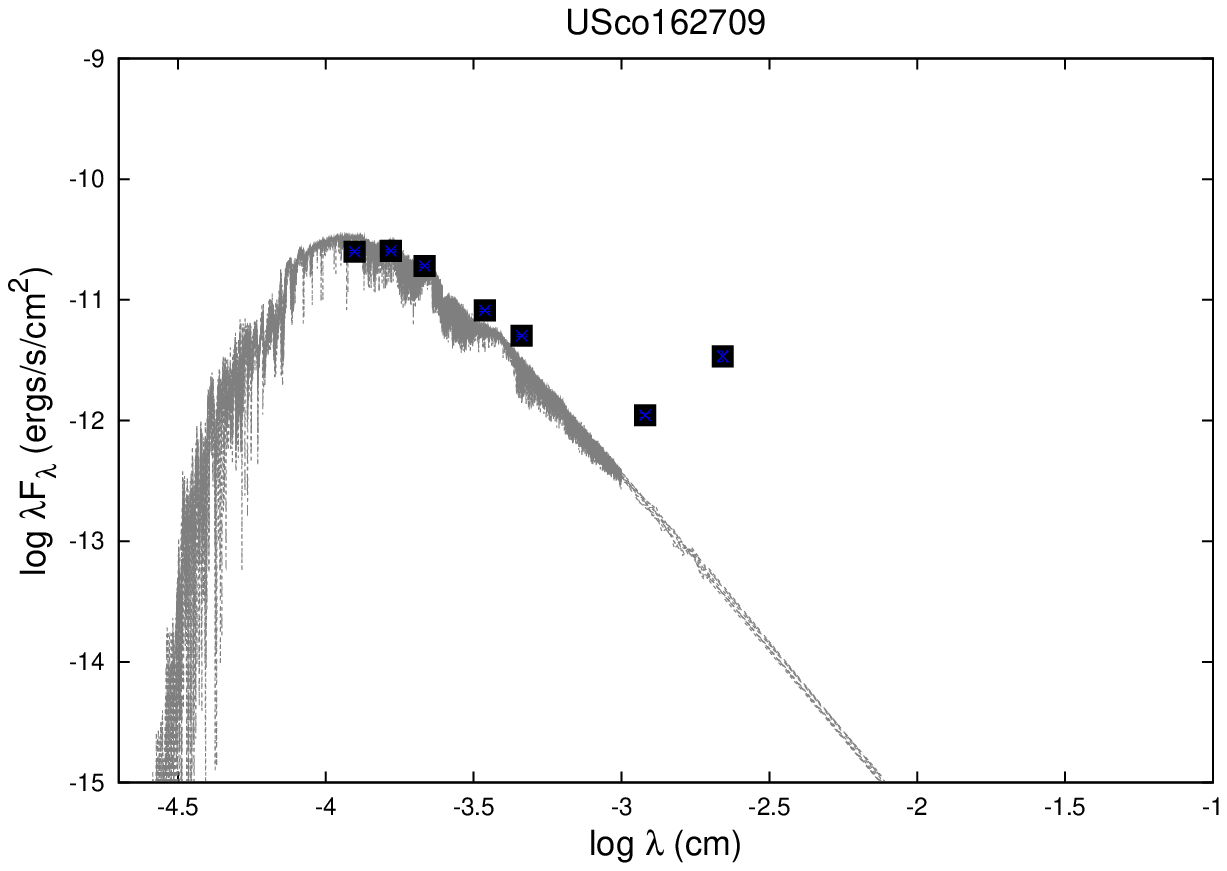} \\
               \includegraphics[width=50mm]{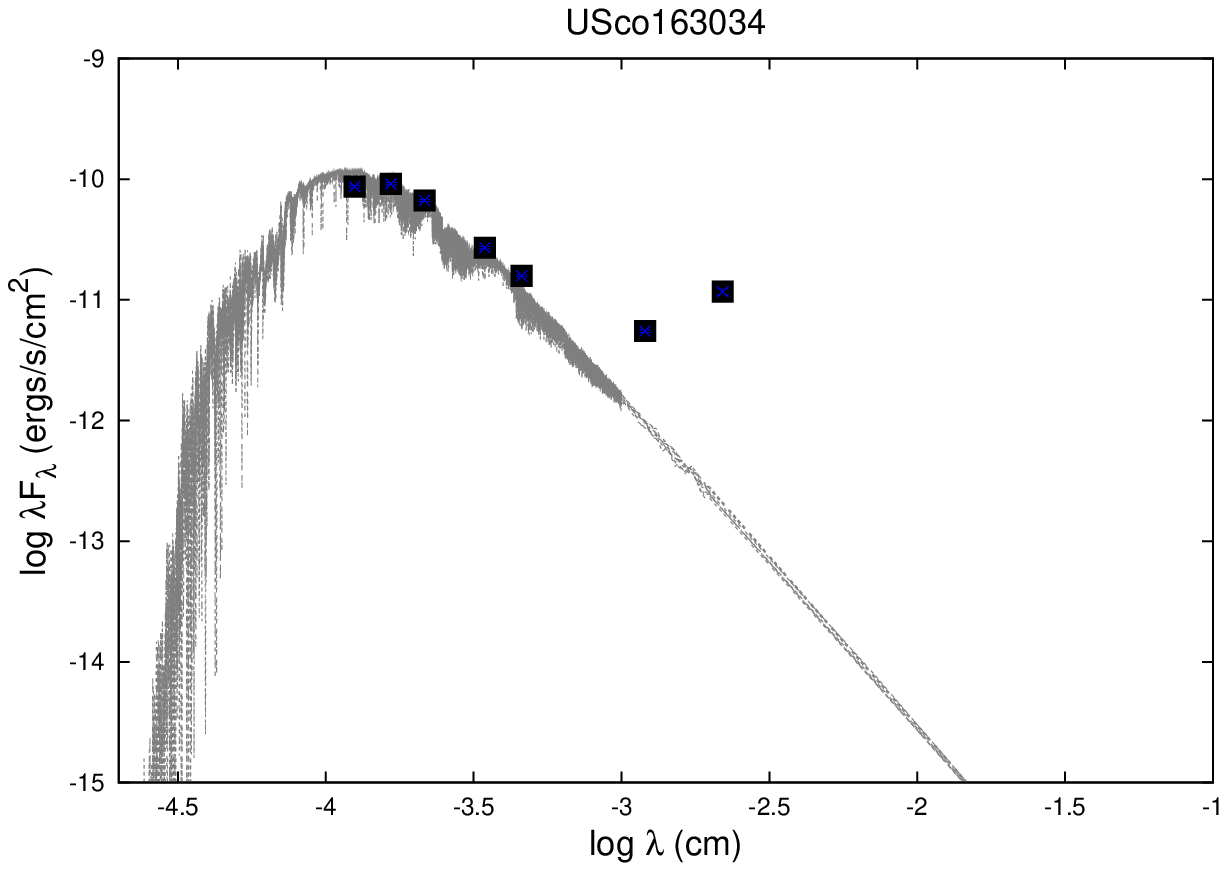} 
              \includegraphics[width=50mm]{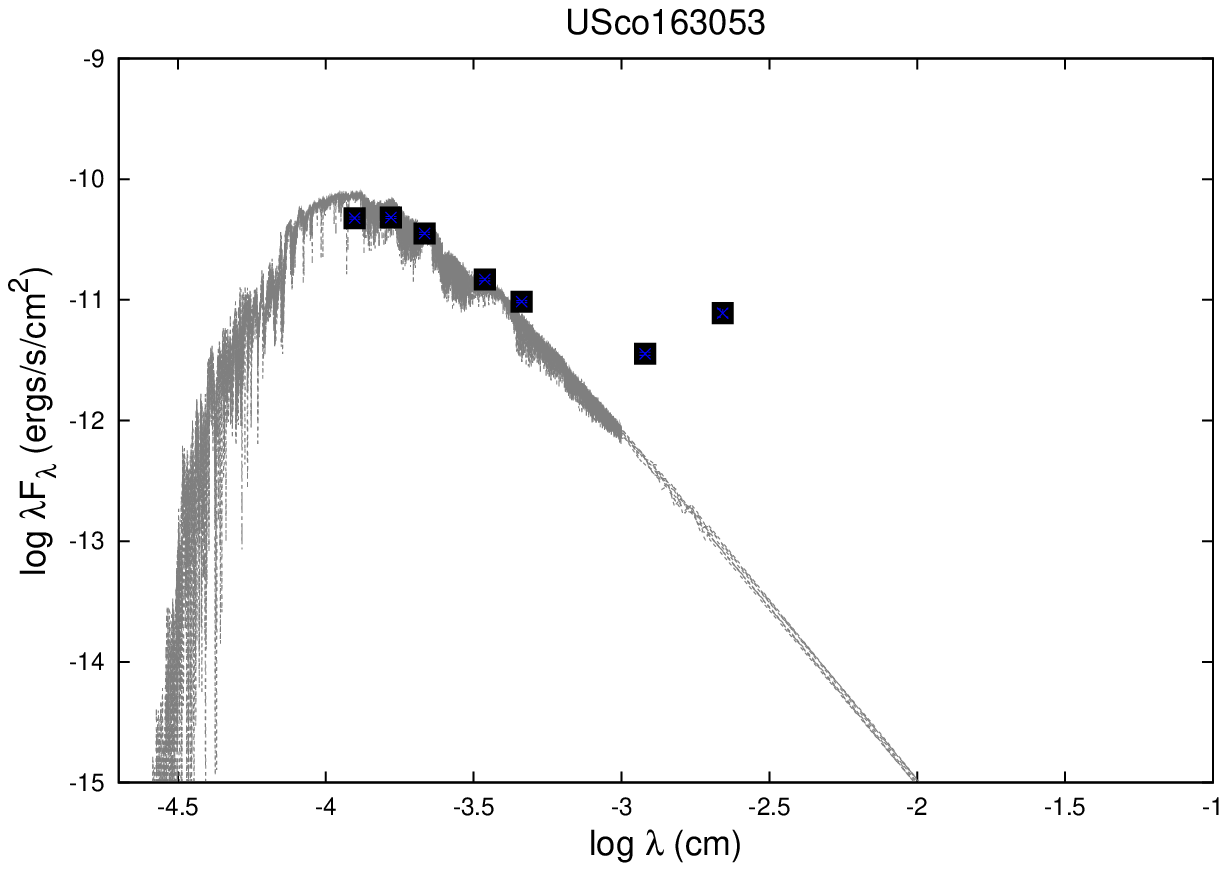} 
               \includegraphics[width=50mm]{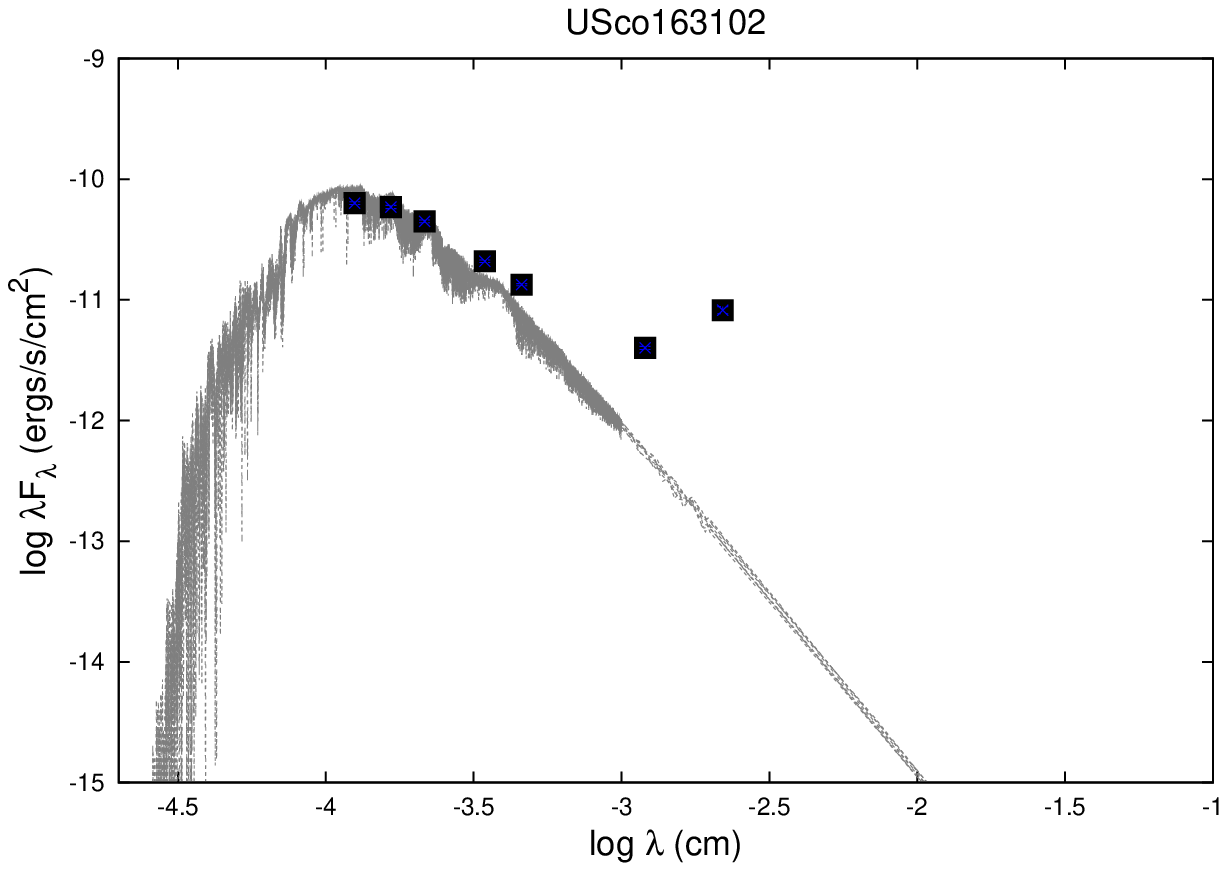} \\                
    \caption{Contd.}
   \label{USco-Mdwarf}
 \end{figure*}

\begin{figure*}
\includegraphics[width=100mm]{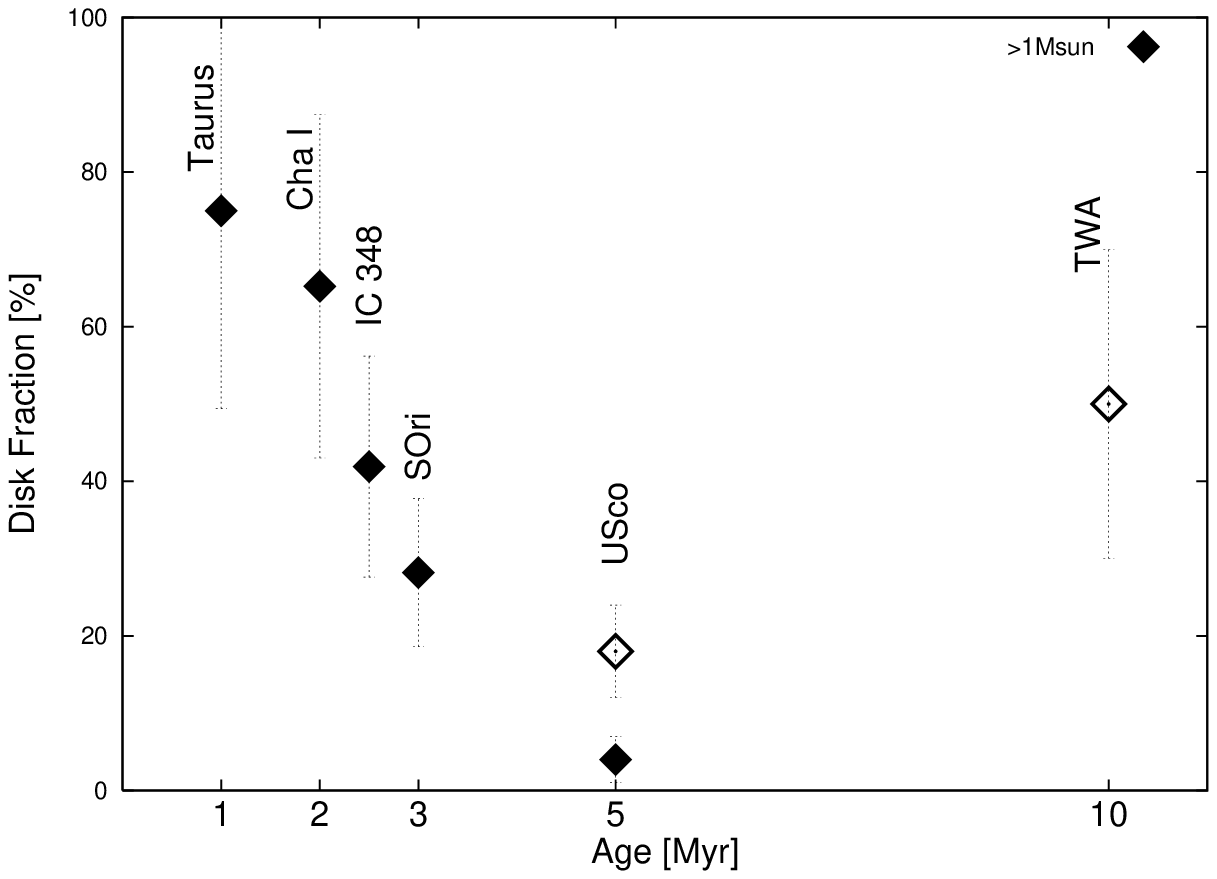}  \\
 \includegraphics[width=100mm]{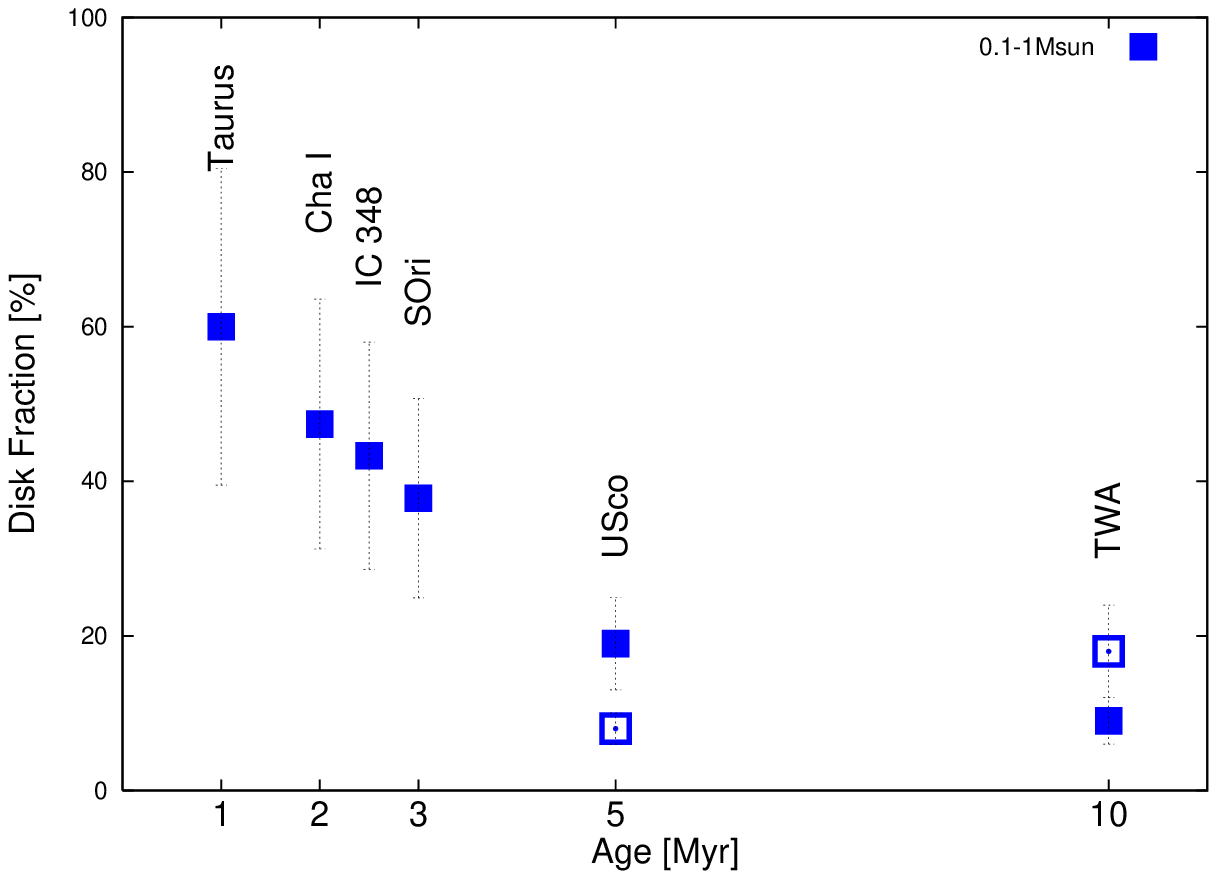}  \\
  \includegraphics[width=100mm]{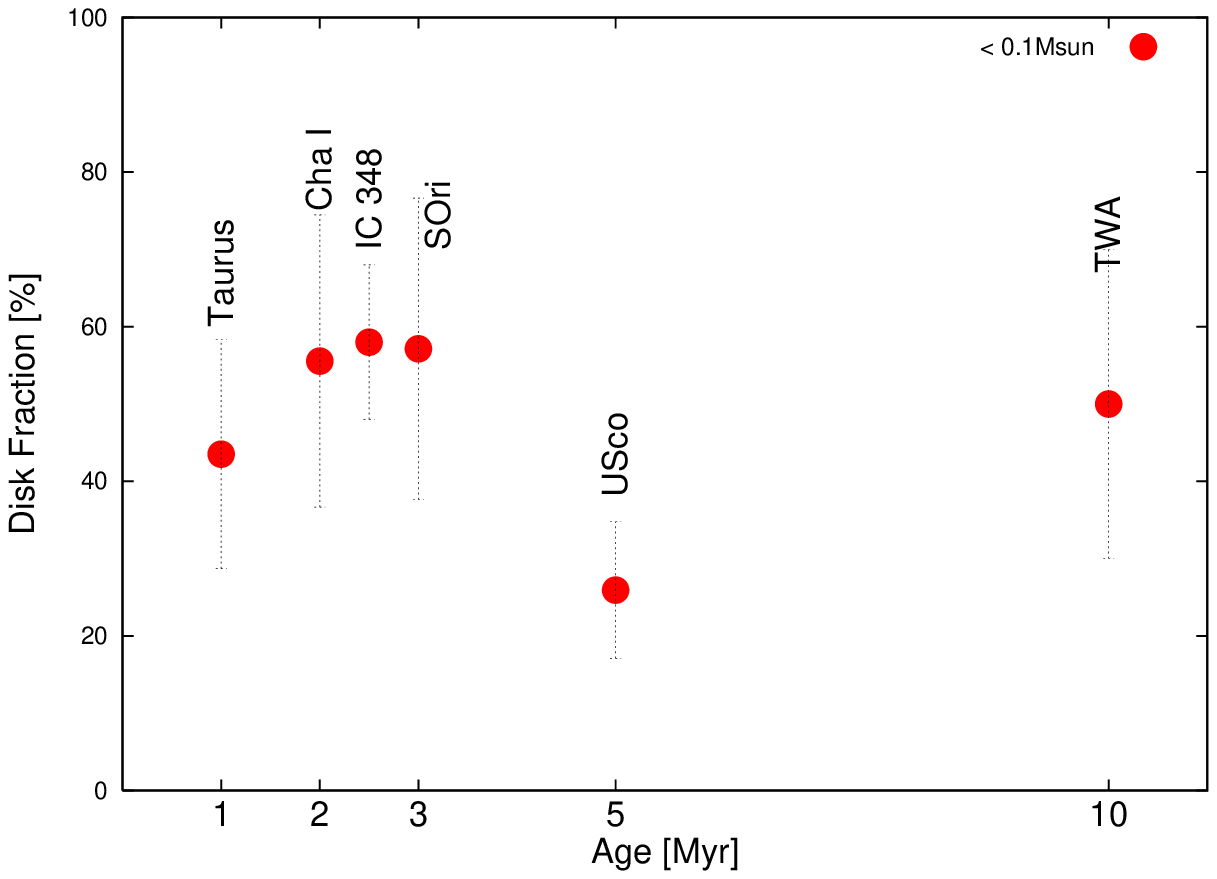}  \\
    \caption{Disc fractions versus stellar age for the 3 mass bins considered: {\it top}--high-mass, {\it middle}--low-mass, {\it bottom}--BDs. Both Cha I and IC 348 have ages of $\sim$2 Myr, we have plotted the IC 348 point at 2.5 Myr for clarification. Primordial disc fractions are denoted by filled symbols; open symbols for the high- and low-mass stars represent the debris disc fractions (Table~\ref{frac-den}). }
   \label{fractions}
 \end{figure*}

 \begin{figure*}
\includegraphics[width=100mm]{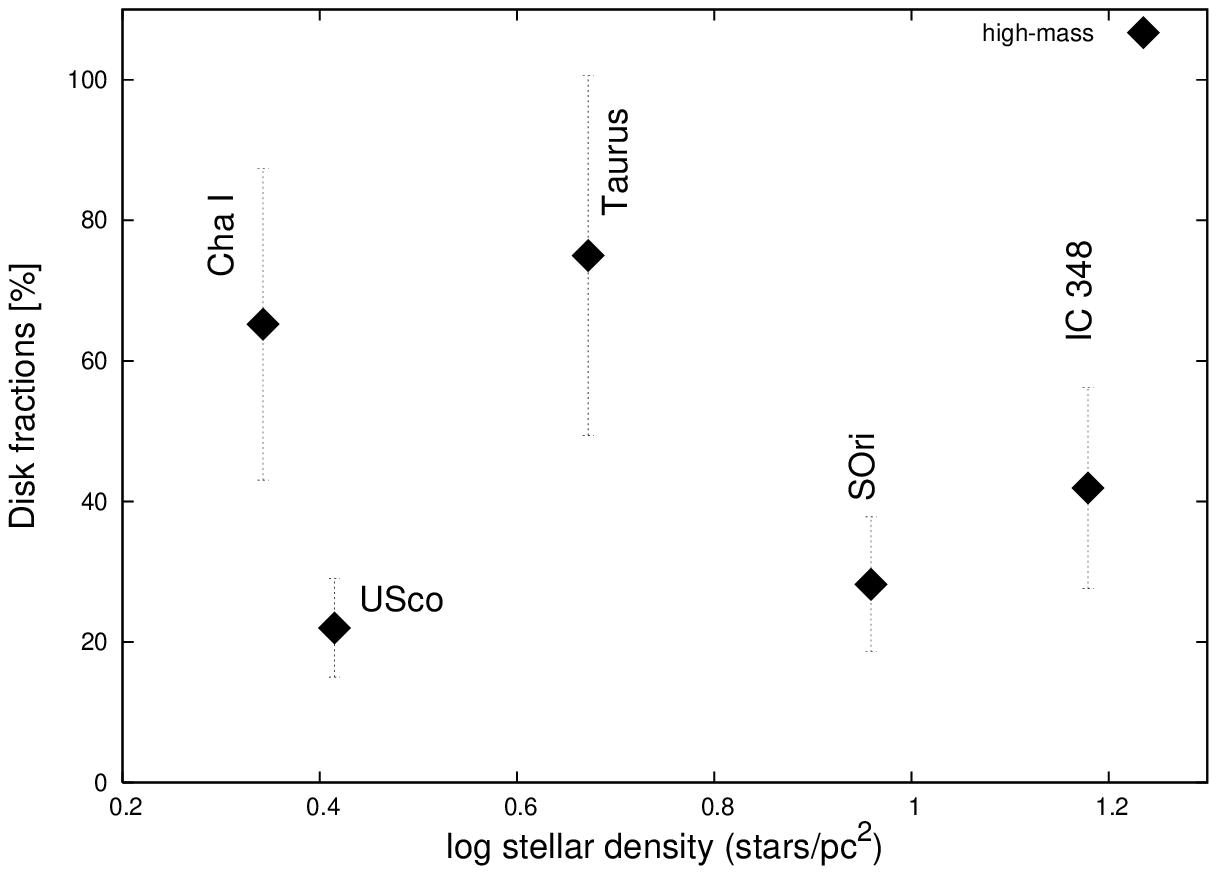}  \\ 
 \includegraphics[width=100mm]{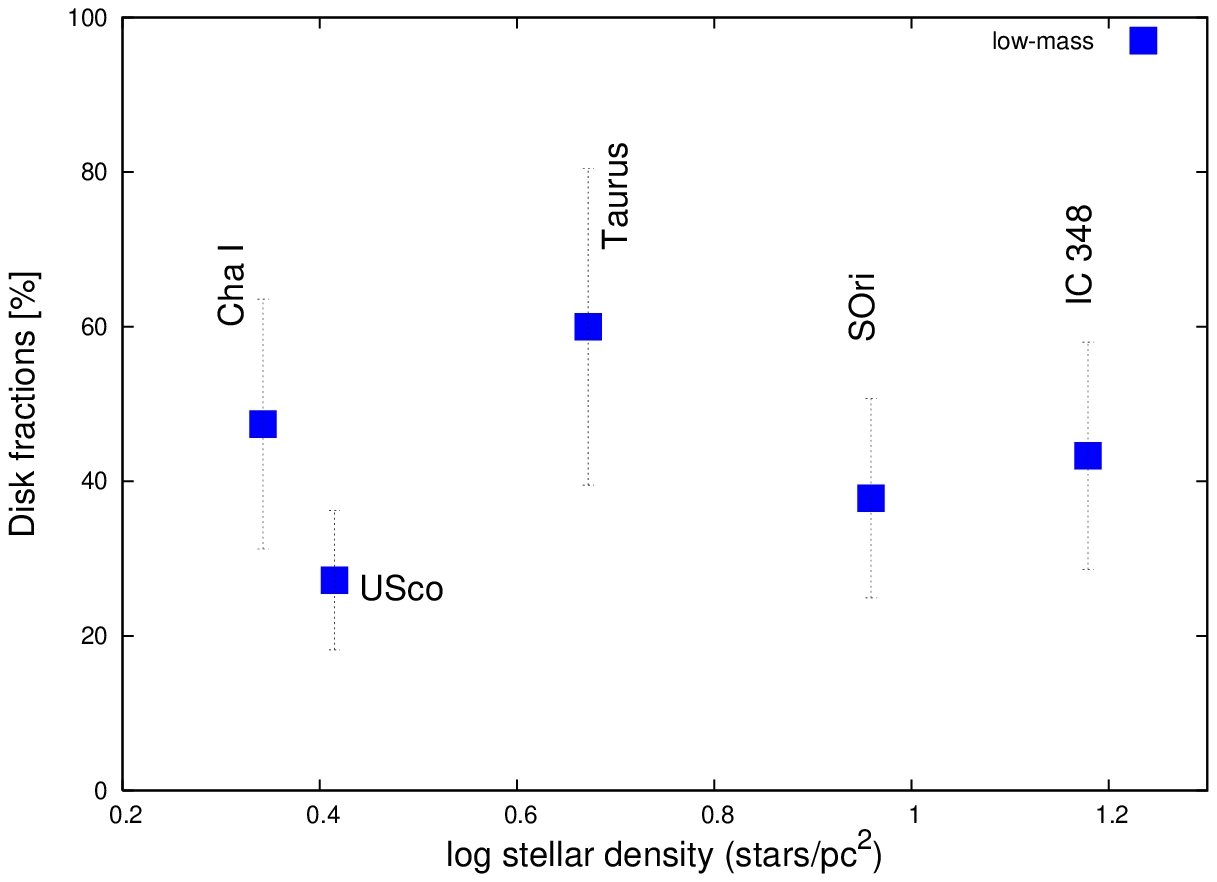}  \\ 
  \includegraphics[width=100mm]{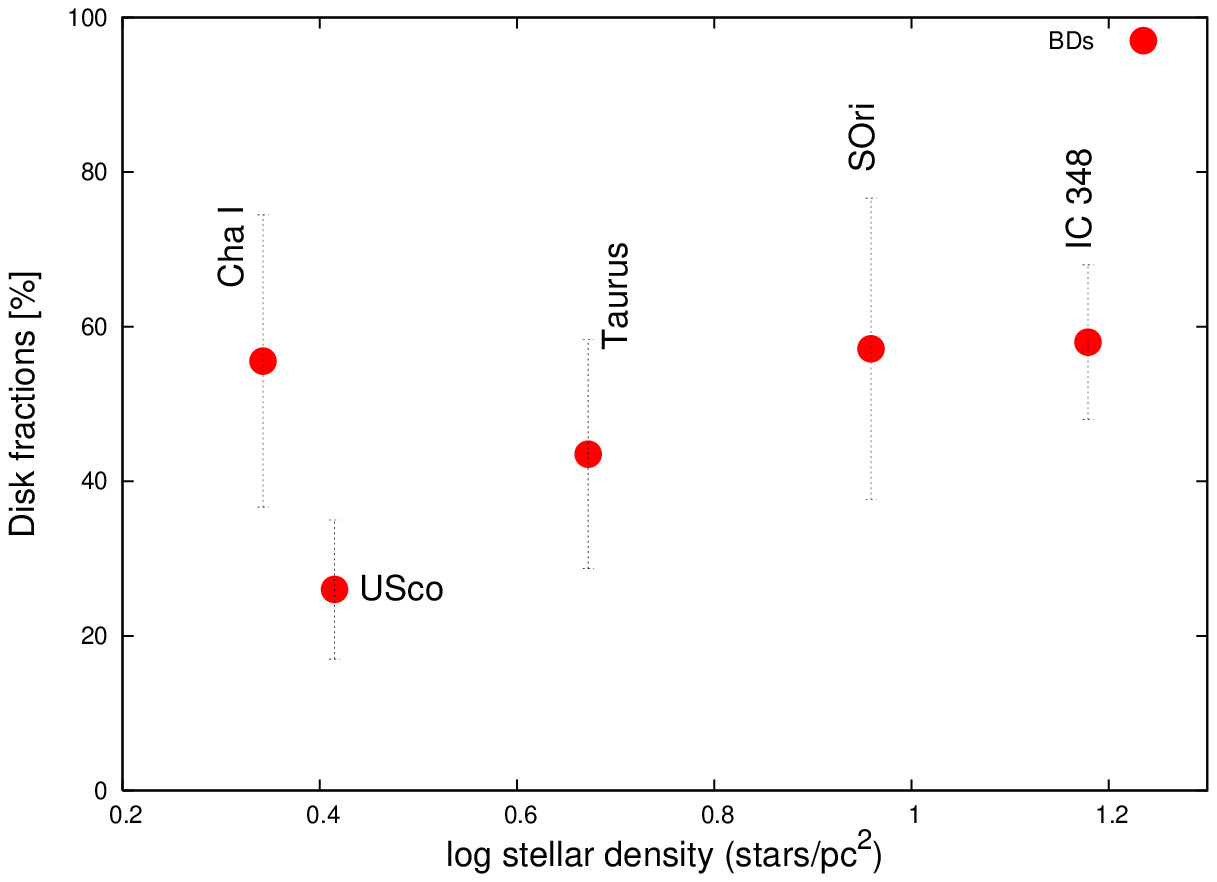}  \\ 
    \caption{Disc fractions versus stellar density for the Taurus ($\sim$1 Myr), Cha I ($\sim$2 Myr), IC 348 ($\sim$2 Myr), SOri ($\sim$3 Myr) and USco ($\sim$5 Myr) regions. Symbols and the figure sequence is the same as in Fig.~\ref{fractions}. For the high- and low-mass stars, the total disc fractions (primordial+debris) have been plotted.  }
   \label{density}
 \end{figure*}

 \begin{figure*}
 \resizebox{160mm}{!}{\includegraphics[angle=0]{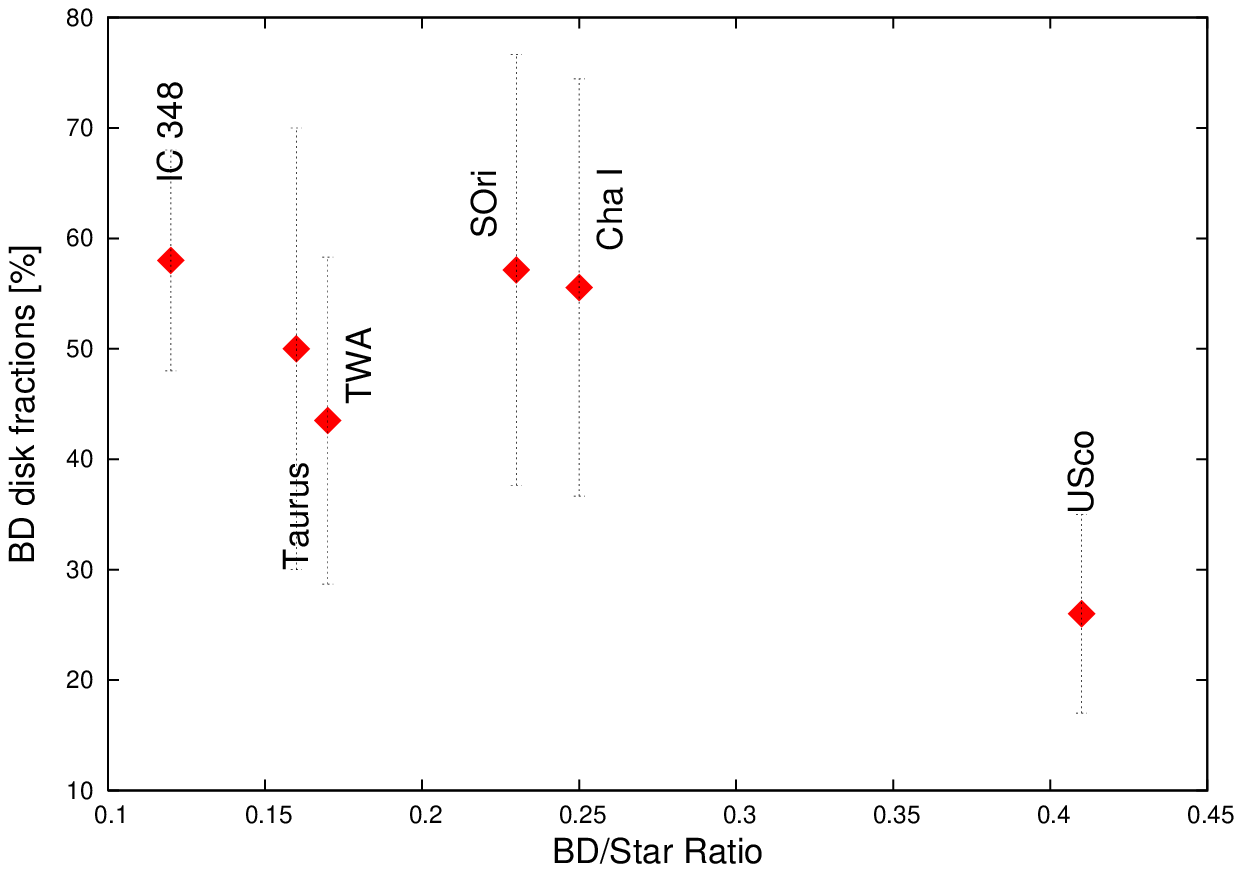}}  \\ 
    \caption{BD disc fractions versus BD/Star number ratio.}
   \label{ratio}
 \end{figure*}
 

\begin{onecolumn}
\begin{landscape}
\begin{table*}
\begin{minipage}{\linewidth}
\tiny
\caption{New Disc Sources in USco}
\label{USco-phot}
\begin{tabular}{lccccccccccccc@{}c@{}c}
\hline

Position (J2000)	&	SpT	&	H$\alpha$ EW 	&	[3.4] &	[3.4] S/N &	[4.6] 	&	[4.6] S/N &	[12]	&	[12] S/N	&	[22]	&	[22] S/N	&	2MASS J &	2MASS H	&	2MASS $K_{s}$	&	$\alpha_{WISE}$	&	SED Class	\\ 
 & &($\AA$) & (mag) & & (mag) & &(mag) & &(mag) & &(mag) & (mag) & (mag) & & \\

\hline

USco15:51:40	-21:46:11	& M4	&	-9	&	10.528 (0.023) &	47	&	10.049 (0.021) &	52.7	&	8.733 (0.029) &	37.1	&	7.41 (0.134) &	8.1	&	12.028 (0.022) &	11.346 (0.023) &	11.004 (0.024) &	-1.55	&	II	\\
USco15:53:01	-21:14:14	& M4	&	-30.7	&	10.739 (0.024) &	44.3	&	10.294 (0.023) &	47.4	&	8.404 (0.027) &	40.5	&	6.686 (0.07) &	15.4	&	12.013 (0.021) &	11.362 (0.023) &	11.025 (0.023) &	-1.32	&	II	\\
USco16:01:42	-21:11:39	& M4	&	-79.2	&	11.296 (0.025) &	43.8	&	10.728 (0.023) &	47.1	&	9.225 (0.035) &	30.8	&	7.496 (0.129) &	8.4	&	12.745 (0.022) &	12.044 (0.025) &	11.68 (0.026) &	-1.43	&	II	\\
USco16:30:34	-24:28:07	& M4	&	-5.3	&	10.01 (0.025) &	43.6	&	9.624 (0.02) &	55.2	&	7.894 (0.028) &	38.2	&	6.04 (0.062) &	17.5	&	11.625 (0.026) &	10.762 (0.024) &	10.36 (0.02) &	-1.42	&	II	\\
USco16:00:27	-20:56:32	& M4.5	&	-16.6	&	12.247 (0.024) &	45.3	&	11.75 (0.024) &	45.9	&	10.001 (0.065) &	16.8	&	8.89 (0.51) &	2.1	&	13.463 (0.029) &	12.898 (0.023) &	12.507 (0.023) &	-1.38	&	II	\\
USco16:08:27	-22:17:29	& M4.5	&	-9.3	&	10.254 (0.024) &	45.2	&	9.83 (0.02) &	54.6	&	8.654 (0.031) &	35	&	5.885 (0.04) &	27.3	&	11.672 (0.026) &	10.876 (0.022) &	10.447 (0.021) &	-1.78	&	II	\\
USco16:10:46	-18:41:00	& M4.5	&	-8.8	&	10.764 (0.025) &	43.2	&	10.261 (0.022) &	48.7	&	8.627 (0.039) &	27.9	&	6.66 (0.065) &	16.7	&	12.708 (0.027) &	11.816 (0.024) &	11.266 (0.022) &	-1.34	&	II	\\
USco16:13:54	-23:20:34	& M4.5	&	-156.6	&	9.761 (0.024) &	44.8	&	9.298 (0.02) &	55.2	&	7.218 (0.017) &	63.9	&	5.094 (0.035) &	30.9	&	11.126 (0.029) &	10.453 (0.022) &	10.056 (0.023) &	-1.20	&	II	\\
USco16:27:09	-21:48:46	& M4.5	&	-11.1	&	11.308 (0.026) &	42.2	&	10.866 (0.022) &	48.9	&	9.636 (0.047) &	23.2	&	7.383 (0.112) &	9.7	&	12.973 (0.024) &	12.152 (0.022) &	11.713 (0.023) &	-1.65	&	II	\\
USco16:18:19	-20:28:48	& M4.75	&	-11.3	&	10.288 (0.024) &	44.8	&	9.792 (0.021) &	51	&	8.719 (0.032) &	33.6	&	7.043 (0.118) &	9.2	&	12.396 (0.029) &	11.501 (0.026) &	10.959 (0.021) &	-1.59	&	II	\\
USco16:02:46	-23:04:51	& M5	&	-9.6	&	11.186 (0.025) &	43.9	&	10.767 (0.021) &	51	&	9.246 (0.037) &	29.3	&	7.318 (0.113) &	9.6	&	12.461 (0.023) &	11.844 (0.027) &	11.502 (0.023) &	-1.53	&	II	\\
USco16:03:35	-18:29:31	& M5	&	-36.6	&	11.121 (0.025) &	44.1	&	10.678 (0.023) &	46.3	&	9.54 (0.046) &	23.6	&	8.678 (0.377) &	2.9	&	12.513 (0.023) &	11.831 (0.025) &	11.484 (0.026) &	-1.72	&	II	\\
USco16:08:35	-22:11:56	& M5	&	-12.2	&	11.231 (0.024) &	44.9	&	10.836 (0.022) &	49.8	&	9.518 (0.048) &	22.5	&	7.557 (0.156) &	7	&	12.559 (0.031) &	11.931 (0.023) &	11.53 (0.023) &	-1.67	&	II	\\
USco16:08:48	-23:41:21	& M5	&	-9.4	&	11.313 (0.025) &	43.6	&	10.915 (0.024) &	44.9	&	9.193 (0.04) &	27	&	6.593 (0.087) &	12.5	&	12.583 (0.024) &	11.911 (0.023) &	11.591 (0.019) &	-1.45	&	II	\\
USco16:11:17	-22:13:09	& M5	&	-8.1	&	10.03 (0.024) &	45.6	&	9.417 (0.021) &	50.8	&	8.118 (0.023) &	47.2	&	6.236 (0.05) &	21.7	&	11.737 (0.026) &	10.943 (0.023) &	10.58 (0.021) &	-1.45	&	II	\\
USco16:11:19	-23:19:20	& M5	&	-3.5	&	10.623 (0.025) &	43.8	&	10.238 (0.023) &	47.4	&	8.791 (0.031) &	35.5	&	7.453 (0.142) &	7.6	&	11.697 (0.026) &	10.983 (0.023) &	10.696 (0.022) &	-1.70	&	II	\\
USco16:13:49	-25:09:01	& M5	&	-25.5	&	11.938 (0.026) &	41.8	&	11.322 (0.023) &	47.2	&	9.576 (0.046) &	23.5	&	7.31 (0.106) &	10.2	&	13.649 (0.022) &	12.942 (0.022) &	12.512 (0.031) &	-1.18	&	II	\\
USco16:15:05	-24:59:35	& M5	&	-38.4	&	11.116 (0.024) &	44.5	&	10.603 (0.021) &	50.5	&	9.158 (0.066) &	16.5	&	6.901 (0.123) &	8.9	&	12.559 (0.022) &	11.901 (0.022) &	11.482 (0.021) &	-1.50	&	II	\\
USco16:18:50	-24:24:32	& M5	&	-6.5	&	12.296 (0.027) &	39.5	&	11.98 (0.025) &	43.1	&	11.092 (0.178) &	6.1	&	8.774 (99.999) &	0.2	&	13.635 (0.027) &	12.934 (0.026) &	12.519 (0.029) &	-2.00	&	II	\\
USco16:22:22	-22:17:31	& M5	&	-60.3	&	12.102 (0.026) &	41.8	&	11.635 (0.025) &	43.1	&	9.724 (0.053) &	20.4	&	7.567 (0.203) &	5.3	&	13.742 (0.03) &	13.09 (0.026) &	12.617 (0.032) &	-1.20	&	II	\\
USco16:29:49	-21:37:09	& M5	&	-30	&	11.261 (0.026) &	41.7	&	10.809 (0.023) &	47.1	&	9.044 (0.034) &	32	&	7.362 (0.127) &	8.5	&	12.527 (0.024) &	11.867 (0.021) &	11.524 (0.024) &	-1.40	&	II	\\
USco16:31:02	-24:08:43	& M5	&	-13.5	&	10.291 (0.024) &	45.9	&	9.81 (0.022) &	49.4	&	8.243 (0.032) &	33.8	&	6.424 (0.074) &	14.7	&	11.964 (0.029) &	11.244 (0.033) &	10.789 (0.029) &	-1.39	&	II	\\
USco16:02:41	-22:48:42	& M5.5	&	-18.4	&	11.641 (0.025) &	44	&	11.119 (0.024) &	44.7	&	9.253 (0.04) &	27.1	&	7.469 (0.144) &	7.5	&	13.046 (0.025) &	12.42 (0.025) &	11.995 (0.023) &	-1.26	&	II	\\
USco16:07:22	-20:11:59	& M5.5	&	-14.2	&	11.262 (0.027) &	40.7	&	10.788 (0.024) &	46.1	&	9.292 (0.04) &	27	&	7.332 (0.12) &	9.1	&	12.705 (0.026) &	12.024 (0.023) &	11.588 (0.021) &	-1.51	&	II	\\
USco16:07:56	-24:43:27	& M5.5	&	-47.3	&	12.313 (0.026) &	41.4	&	11.817 (0.024) &	44.4	&	10.31 (0.076) &	14.4	&	8.741 (99.999) &	0.8	&	13.839 (0.026) &	13.13 (0.033) &	12.71 (0.033) &	-1.46	&	II	\\
USco16:09:21	-21:39:35	& M5.5	&	-24.5	&	10.728 (0.025) &	44.1	&	10.261 (0.023) &	48.3	&	8.319 (0.027) &	40.8	&	6.407 (0.064) &	17.1	&	11.986 (0.028) &	11.389 (0.024) &	11.028 (0.024) &	-1.27	&	II	\\
USco16:10:12	-21:01:55	& M5.5	&	-14.6	&	12.005 (0.025) &	43.1	&	11.457 (0.026) &	42.3	&	10.111 (0.064) &	17	&	7.768 (0.159) &	6.8	&	13.782 (0.027) &	12.999 (0.023) &	12.574 (0.032) &	-1.45	&	II	\\
USco16:12:55	-22:26:54	& M5.5	&	-6.8	&	11.075 (0.025) &	44.3	&	10.674 (0.024) &	44.9	&	8.498 (0.026) &	41.1	&	6.061 (0.049) &	22	&	12.364 (0.024) &	11.747 (0.023) &	11.391 (0.023) &	-1.17	&	II	\\
USco16:17:19	-21:37:13	& M5.5	&	-13.4	&	12.363 (0.027) &	41	&	12.005 (0.024) &	44.5	&	9.494 (0.046) &	23.8	&	7.418 (0.131) &	8.3	&	13.486 (0.026) &	12.864 (0.027) &	12.54 (0.029) &	-1.06	&	II	\\
USco16:18:16	-23:47:08	& M5.5	&	-16.3	&	10.529 (0.024) &	44.6	&	10.019 (0.021) &	51.9	&	8.314 (0.03) &	35.8	&	6.853 (0.112) &	9.7	&	12.425 (0.026) &	11.515 (0.024) &	10.977 (0.024) &	-1.32	&	II	\\
USco16:19:05	-23:07:53	& M5.5	&	-12.1	&	11.776 (0.026) &	42.3	&	11.465 (0.022) &	49.2	&	10.442 (0.087) &	12.5	&	8.824 (0.451) &	2.4	&	13.005 (0.024) &	12.349 (0.024) &	11.98 (0.023) &	-1.93	&	II	\\
USco16:30:53	-24:24:54	& M5.5	&	-34.8	&	10.67 (0.026) &	41.2	&	10.16 (0.02) &	53.4	&	8.363 (0.032) &	33.9	&	6.485 (0.085) &	12.8	&	12.277 (0.023) &	11.461 (0.023) &	11.041 (0.024) &	-1.30	&	II	\\
USco16:07:50	-22:21:02	& M5.8	&	-14.3	&	10.822 (0.025) &	44.2	&	10.238 (0.021) &	51.6	&	8.877 (0.031) &	34.8	&	7.359 (0.129) &	8.4	&	12.349 (0.027) &	11.69 (0.023) &	11.342 (0.027) &	-1.44	&	II	\\
USco16:08:11	-22:29:43	& M5.8	&	-22	&	11.506 (0.025) &	43	&	11.198 (0.024) &	45	&	9.199 (0.036) &	29.9	&	7.2 (0.099) &	10.9	&	12.619 (0.026) &	11.997 (0.026) &	11.679 (0.023) &	-1.39	&	II	\\
USco16:11:07	-22:28:50	& M5.8	&	-64.7	&	10.979 (0.025) &	43.3	&	10.387 (0.02) &	53.7	&	8.542 (0.028) &	38.9	&	7.011 (0.089) &	12.2	&	12.313 (0.03) &	11.723 (0.023) &	11.328 (0.024) &	-1.23	&	II	\\
USco16:14:23	-22:19:34	& M5.8	&	-94.5	&	10.345 (0.025) &	42.9	&	9.825 (0.021) &	51.3	&	8.099 (0.024) &	45.7	&	6.391 (0.056) &	19.3	&	11.805 (0.023) &	11.159 (0.024) &	10.8 (0.021) &	-1.30	&	II	\\
USco16:09:53	-19:48:17	& M6	&	-21.7	&	11.302 (0.024) &	44.5	&	10.717 (0.022) &	49.9	&	9.081 (0.037) &	29.7	&	6.934 (0.083) &	13.1	&	12.805 (0.023) &	12.167 (0.024) &	11.76 (0.021) &	-1.31	&	II	\\
USco16:10:05	-19:19:36	& M6	&	-49.8	&	12.179 (0.024) &	44.3	&	11.624 (0.024) &	44.5	&	10.202 (0.067) &	16.3	&	9.083 (0.531) &	2	&	14.218 (0.029) &	13.431 (0.025) &	12.7 (0.033) &	-1.42	&	II	\\
USco16:19:40	-21:45:35	& M6	&	-36.5	&	11.646 (0.026) &	42.2	&	10.986 (0.021) &	50.9	&	9.054 (0.034) &	31.7	&	7.299 (0.11) &	9.9	&	13.223 (0.029) &	12.537 (0.023) &	12.115 (0.024) &	-1.09	&	II	\\
USco16:28:47	-24:28:14	& M6	&	-182.7	&	11.277 (0.024) &	44.7	&	10.589 (0.023) &	47.5	&	8.779 (0.055) &	19.7	&	7.312 (0.197) &	5.5	&	13.245 (0.026) &	12.533 (0.023) &	11.952 (0.023) &	-1.06	&	II	\\
USco16:30:27	-23:59:09	& M6	&	-34.7	&	11.034 (0.026) &	42.1	&	10.454 (0.022) &	48.6	&	8.466 (0.046) &	23.5	&	6.092 (0.074) &	14.6	&	12.619 (0.022) &	11.938 (0.024) &	11.479 (0.021) &	-1.12	&	II	\\
USco16:09:59	-23:45:19	& M6.5	&	-34.4	&	11.364 (0.025) &	44	&	10.977 (0.022) &	49.1	&	9.994 (0.059) &	18.5	&	8.41 (99.999) &	1.1	&	12.635 (0.024) &	11.996 (0.024) &	11.624 (0.024) &	-1.8	&	II	\\
USco16:14:33	-22:42:13	& M6.5	&	-73.4	&	12.904 (0.027) &	39.8	&	12.286 (0.027) &	39.7	&	11.405 (0.214) &	5.1	&	8.511 (99.999) &	1.1	&	14.232 (0.03) &	13.664 (0.033) &	13.24 (0.032) &	-1.7	&	II	\\
USco16:15:14	-23:04:26	& M6.5	&	-35.9	&	13.487 (0.031) &	34.9	&	12.826 (0.031) &	35.3	&	11.412 (0.184) &	5.9	&	8.701 (99.999) &	0.7	&	14.814 (0.042) &	14.193 (0.032) &	13.915 (0.055) &	-1.40	&	II	\\
USco16:06:04	-20:56:45	& M7	&	-159	&	12.116 (0.027) &	40.2	&	11.609 (0.025) &	43.5	&	9.868 (0.058) &	18.8	&	8.056 (0.211) &	5.1	&	13.528 (0.03) &	12.906 (0.023) &	12.475 (0.027) &	-1.34	&	II	\\
USco16:14:21	-23:39:15	& M7	&	-43	&	13.58 (0.033) &	33.3	&	12.958 (0.032) &	34	&	11.205 (0.147) &	7.4	&	8.704 (0.387) &	2.8	&	15.053 (0.037) &	14.363 (0.033) &	13.999 (0.056) &	-1.24	&	II	\\
USco15:54:20	-21:35:43	& M8	&	-29.4	&	13.265 (0.029) &	37.5	&	12.548 (0.029) &	37.8	&	10.981 (0.126) &	8.6	&	8.538 (0.316) &	3.4	&	15.016 (0.037) &	14.255 (0.051) &	13.772 (0.049) &	-1.25	&	II	\\
USco16:08:30	-23:35:11	& M8.5	&	-97.4	&	13.357 (0.029) &	37.8	&	12.833 (0.031) &	34.9	&	11.41 (0.19) &	5.7	&	8.876 (99.999) &	0.4	&	14.916 (0.032) &	14.214 (0.036) &	13.814 (0.054) &	-1.47	&	II	\\

\hline
\end{tabular}
\end{minipage}
\end{table*}
\end{landscape}
\end{onecolumn}


\begin{table*}
\begin{minipage}{\linewidth}
\caption{Disc Fractions and Stellar Densities}
\label{frac-den}
\begin{tabular}{cccccclcccccc}
\hline

Name & Age (Myr)\footnote{Reference for the median age estimates are: Taurus (Luhman 2004), Cha I (Luhman 2007), IC 348 (Luhman et al. 2003), SOri (Zapatero Osorio et al. 2002), USco (Preibisch \& Zinnecker 1999), TWA (Barrado y Navascu\'{e}s 2006). } & Distance (pc)\footnote{References for distance estimates are: Taurus (Luhman 2004; Luhman et al. 2010), Cha I (Luhman 2007), IC 348 (Luhman et al. 2003), SOri (Perryman et al. 1997), USco (de Bruijne et al. 1997), TWA (Biller \& Close 2007). } & & & Disc Fraction [\%] & & Stellar Density [objs/deg$^{2}$; objs/pc$^{2}$]  & BD/Star Ratio	\\  \hline
                &&&               &  high-mass & low-mass & BDs & & \\

\hline

Taurus\footnote{These discs are classified as primordial/evolved discs. The debris disc fraction for the younger clusters is negligible ($\sim$1-2\%). The disc fractions for these clusters are from Luhman et al. (2008; 2010). } & 1-1.5 & 140 & & 75$\pm$26 & 60$\pm$20 & 43$\pm$15 & 28; 4.7 & 0.17$^{+0.08}_{-0.06}$\\
Cha I$^{c}$ & 2-3 &160-170 & & 65$\pm$22 & 47$\pm$16 & 55$\pm$19 & 17; 2.2 & 0.25$^{+0.28}_{-0.12}$ \\
IC 348$^{c}$ & 2-3 & 315 & & 42$\pm$14 & 43$\pm$15 & 58$\pm$10 & 415; 15.1 & 0.12$^{+0.06}_{-0.04}$ \\
SOri$^{c}$ & 3$\pm$2 & 352$^{+166}_{-85}$ & & 28$\pm$10 & 38$\pm$13 & 57$\pm$19 & 345; 9.1 & 0.23$\pm$0.02 \\
USco\footnote{The first value is for primordial discs, second value for debris discs. The BD discs are all classified as primordial/evolved discs. The disc fractions are from Carpenter et al. (2009), Riaz et al. (2009), Scholz et al. (2007) and the present work. The total disc fraction for high-mass stars is 22$\pm$7\%, and for low-mass stars is 27$\pm$9\%. } & 5$\pm$2 & 145$\pm$2 & & 4$\pm$3, 18$\pm$6 & 19$\pm$6, 8$\pm$2 & 26$\pm$9 & 17; 2.6 & 0.41$\pm$0.03 \\
TWA\footnote{For the high-mass stars, all discs are classified as debris sources. For the low-mass stars, the first value is for primordial discs, second value for debris discs. The BDs are all classified as primordial/evolved discs. The disc fractions are from Low et al. (2005) and Riaz et al. (2009). The total disc fraction for low-mass stars is 21$\pm$7\%. } & 10$^{+5}_{-2}$ & 58.8$\pm$5.5 & & 50$\pm$20 & 9$\pm$3, 18$\pm$6 & 50$\pm$20 & 0.06; 0.05 & 0.16$\pm$0.04 \\

\hline
\end{tabular}
\end{minipage}
\end{table*}

\clearpage 

\begin{twocolumn}
 
\appendix

\section{Compilation of Disc Fractions and Stellar Densities}
\label{appendix}

The disc fractions for Taurus, Chamaeleon I (Cha I), IC 348 and $\sigma$ Orionis (SOri) are from Luhman et al. (2008) and Luhman et al. (2010). The age estimates for Cha I and IC 348 are $\sim$2 Myr (e.g., Luhman et al. 2008), and 3$\pm$2 Myr for SOri (e.g., Zapatero Osorio et al. 2002). Luhman et al. (2008; 2010) have converted the spectral types to stellar masses by combining the temperature scale of Luhman et al. (2003a), and the evolutionary models from Baraffe et al. (2003). They have considered the mass limits of 0.1 $M_{\sun}$ and 1 $M_{\sun}$ at SpT of M6 and K5, respectively. In Fig.~\ref{fractions}, the sub-stellar disc fraction is for sources with SpT between M6 and M9, implying masses between $\sim$0.1 and 0.02$M_{\sun}$, the low-mass disc fraction is for SpT of K5-M5 (0.1-1$M_{\sun}$), and the high-mass disc fraction is for SpT earlier than K5. The earliest type in these clusters is a mid-B star (B5 or B6; Luhman et al. 2003a; 2007). From the models of Seiss et al. (2000), at an age of 2-3 Myr, a B5 SpT corresponds to a mass of $\sim$4 $M_{\sun}$. The high-mass disc fractions plotted are thus approximately for the mass range of 1-4$M_{\sun}$. Since most of the SOri sources do not have spectral classification, Luhman et al. (2008) have used the $M_{J}$ magnitudes as a proxy for stellar masses, and have considered the mass limits of 1$M_{\sun}$ and 0.1$M_{\sun}$ at $M_{J}$ of 3 and 7 mag, respectively. From the $M_{J}$ ranges listed in Luhman et al. (2008; Table 4), we have considered $M_{J}$ between 0 and 3 mag for high-mass stars, $M_{J}$=4-8 mag for low-mass stars, and $M_{J}$=8-10 mag for sub-stellar sources in SOri. These $M_{J}$ ranges correspond to similar mass ranges as considered for the other three clusters. In USco, the BD disc fraction (26$\pm$9\%) is the combined disc fraction from the work of Scholz et al. (2007), Riaz et al. (2009), and the present work. The disc fraction for low-mass stars in USco is for the K5-M5 stars from Carpenter et al. (2009), and includes the 36 M dwarf discs (M3.5-M5) found in the present survey. The disc fraction plotted for high-mass stars in USco is for SpT earlier than K5 obtained from Carpenter et al. (2009). In the $\sim$10 Myr old TWA, the BD disc fraction (M6-M8) is from Riaz et al. (2009), and the disc fraction for the low-mass stars (K5-M5) is from Low et al. (2005). Among the higher mass stars in TWA, the Low et al. sample consists of a G5 star (TWA 19A) and an A0 star (HR 4796A). TWA 19A is found to be photospheric, while HR4796A shows excess emission in the {\it Spitzer}/MIPS bands. This results in a 50$\pm$20\% disc fraction for the high mass stars. We note that the TWA is a sparsely populated association with a small number of 8 discs found to date. The disc fractions for this association thus have large uncertainties. The error bars for the disc fractions are the 1-$\sigma$ Gaussian distribution uncertainties. 

The quoted disc fractions in other clusters are based on deep Spitzer surveys, which have high sensitivities capable of detecting the photospheres for the brown dwarfs. If we consider the Spitzer 8$\micron$ band, which is the closest in wavelength to the WISE 12$\micron$ band, then the predicted photospheric flux at this wavelength for an M6 dwarf is $\sim$0.3 mJy, and for an M8 dwarf the flux is $\sim$0.13 mJy. The deep surveys conducted with Spitzer in SOri, Taurus, Cha I and IC 348 have a 1-$\sigma$ sensitivity of $\sim$0.01-0.02mJy at 8$\micron$. These surveys were thus capable of detecting the photospheres and therefore the smallest excesses possible. The smallest 12$\micron$ excess that we can detect with WISE is Fobs/Fphot $\sim$ 3-5 for M6-M8 objects. We have checked in the WISE catalog for detection of all of the Spitzer known discs, in order to determine the fraction of discs that WISE might have missed around M6-M8.5 type objects due to limited sensitivity. In SOri, all of the Spitzer brown dwarf discs at masses of $\sim$0.08-0.04 M$_{\sun}$, corresponding to spectral type of $\sim$M6-M8, were recovered at S/N $\geq$ 3 in WISE 1, 2, 3 bands. The S/N is $\sim$1-2 in the 22$\micron$ band. For lower masses of $\sim$0.04-0.02 M$_{\sun}$ (M8-M9), none of the sources were detected in WISE 12$\micron$ or 22$\micron$ bands (S/N $<$ 1), and detections were only obtained in the 1 and 2 bands at S/N$\sim$10. From the number of missed discs, we estimate that the WISE BD disc fraction in SOri would be $\sim$10\% lower than the Spitzer fraction. SOri is also the farthest cluster ($\sim$352 pc) considered in our study, which explains why the later type discs were not detected. IC 348 is another distant cluster ($\sim$315pc) and WISE has missed the M8.25-M8.75 discs in this cluster, which results in a BD disc fraction which is $\sim$8\% lower than that estimated by the Spitzer survey. We note that due to the 3-$\sigma$ quality criteria, the full sample detected in the WISE 1, 2, and 3 bands at S/N$\geq$3 would actually be smaller in size than the original sample, resulting in a higher disc fraction. Combining these limitations, we estimate that the WISE-recovered disc fractions in SOri and IC 348 will be $\sim$52\% and $\sim$53\%, respectively, which are similar to the Spitzer fractions of 58\% and 57\% (Table~\ref{frac-den}). In the Taurus and Cha I regions ($\sim$140-160pc), all of the brown dwarf discs as well as the full samples down to a spectral type of M9 were recovered by WISE, at a $\geq$3 sigma level at 12$\micron$, as well as in WISE bands 1 and 2. Therefore the WISE-recovered disc fractions in these regions are similar to the Spitzer ones. The WISE data on TWA ($\sim$60pc) has not been released yet, but from the Spitzer/IRS data discussed in Riaz et al. (2008), the observed flux density at 12$\micron$ is $\sim$5 mJy for the TWA brown dwarf discs, which is higher than the 5-$\sigma$ sensitivity at this wavelength ($\sim$0.7 mJy). These discs thus should be recovered with WISE. We therefore do not expect any significant changes in the overall trend with age observed for the BD disc fractions shown in Figs.~\ref{fractions}, and the USco disc fraction would still be a factor of $\sim$2 lower then the SOri fraction. In Figs.~\ref{fractions}, ~\ref{density} and Table~\ref{frac-den}, we have used the Spitzer fractions quoted in the literature, and not the WISE-recovered disc fractions.

It is important to determine the wavelengths at which these disc sources show excess emission. The disc fractions in Taurus, Cha I, IC 348 and SOri have been determined based on observations made in the {\it Spitzer}/IRAC bands of 3.6-8$\micron$ and MIPS 24$\micron$ band. The intermediate wavelengths of $\sim$8-24$\micron$ are important in making a distinction between a primordial and a debris disc. Luhman et al. (2010) have classified sources that show excess emission at 24$\micron$ but photospheric emission at shorter wavelengths as debris discs, while primordial discs are sources that show excess emission at 5.8, 8 and 24$\micron$. Intermediate between the primordial and debris discs are the `evolved' discs that show weaker excess emission in these bands compared to the primordial discs. These evolved discs are similar to the primordial transition discs discussed in Section \S\ref{USco}. Based on this definition, there are just 4 debris discs in Taurus, 2 in IC 348, 1 in Cha I, and 2 in SOri (Luhman et al. 2010). We can thus consider that the fraction of such debris systems with excesses at $\geq$24$\micron$ are negligible in these clusters ($\sim$1-2\%). A small $\sim$12\% of the discs in Taurus are classified as evolved discs, while a large fraction is found to be primordial. Similar is the case in the other young clusters of IC 348, Cha I and SOri, where nearly 85\% of the disc population is found to be primordial, and $\sim$10-15\% are evolved sources (Luhman et al. 2010). 


In USco, 19 of the 23 earlier type discs (SpT$<$K5) show excess emission at 24$\micron$ only, and have been classified as debris discs by Carpenter et al. (2009). The rest of the 4 discs have excesses at 8 and 24$\micron$ and have been classified as primordial discs. Among the K5-M5 group, 58 are primordial discs with both 8 and 24$\micron$ excesses, while 10 are debris with 24$\micron$ excess but  weak or no detectable 8$\micron$ excesses. The Carpenter et al. criteria is thus consistent with Luhman et al. (2010), wherein a debris disc is classified based on 24$\micron$ excess {\it only}, while a primordial disc will show excess emission at {\it both} 8 and 24$\micron$. Carpenter et al. have not considered the intermediate evolved disc category, but the primordial discs in their survey are found to show weak or no excess shortward of 8$\micron$, so these are likely to be evolved or primordial transition discs. In our present USco work based on the WISE bands, it is only the 12$\micron$ point that can make such a distinction. All of the discs from this survey have an excess at 12$\micron$, and we do not have any discs that show an excess at 22$\micron$ {\it only}. Therefore none of these have been classified as debris discs. As noted in Section \S\ref{USco}, the colors for these discs are also similar to the primordial discs classified by Carpenter et al. (2009). Only six of these show small excess at 3.4 and 4.6$\micron$, so a majority are primordial transition discs. For the other BD surveys in USco, all of the discs from Scholz et al. (2007) show excess emission at 8$\micron$ and at longer wavelengths. The discs from Riaz et al. (2009) also have excess at 8$\micron$ and at longer wavelengths, with one disc showing excess in the IRAC shorter wavelengths bands of $<$8$\micron$. Thus all of the USco BD discs would fall into the primordial/primordial transition disc category. 

In the TWA, the BD discs have excesses at $\sim$5-24$\micron$, and can be classified as primordial (Riaz \& Gizis 2008). These discs show weak signs of inner disc clearing. Among the low-mass stars, TW Hya and Hen 3-600 can be classified as primordial transition discs, since both show strong excesses at 10$\micron$ and longer wavelengths, but photospheric emission at shorter wavelengths. The sources TWA 7 and TWA 13 show excesses only at 70$\micron$, while two other discs show weak or no excess shortward of 24$\micron$. These 4 discs can be classified as debris systems. In Fig.~\ref{fractions}, we have plotted the total (primordial/evolved + debris) disc fraction. It is difficult to clearly separate the primordial and evolved disc sources since the definitions vary from survey to survey. As noted earlier, the debris disc fraction for the younger clusters is negligible ($\sim$1-2\%), therefore a majority of these can be considered as primordial discs. In comparison, a majority of the high- and low-mass discs in USco and TWA are in the debris phase. In Table~\ref{frac-den}, we have listed the separate primordial and debris disc fractions for these clusters.

Figure~\ref{density} plots the disc fractions versus the stellar density in a cluster. By stellar density, we imply the total number of objects (stars + BDs) detected in a given survey area. The stellar density in Taurus is from the work of Luhman (2004), where 112 objects (96 stars and 16 BDs) were confirmed as members from imaging of a 4 deg$^{2}$ region, implying a stellar density of 28 objects per square degree. In Cha I, Luhman (2007) detected 85 objects (70 stars and 15 BDs) in his IK2 survey covering a 1.25$\degr$ radius (area=4.9 deg$^{2}$), implying a stellar density of $\sim$17 objects per square degree. In IC 348, Luhman (1999) found 288 members (including 23 BDs) in a 50$\arcmin$$\times$50$\arcmin$ central region of the cluster, implying a stellar density $\sim$415 objects per square degree. In SOri, Lodieu et al. (2009) found 271 objects (198 stars and 73 BDs) in a 30$\arcmin$ radius from the cluster center, implying a stellar density of $\sim$345 objects per square degree. In USco, Lodieu et al. (2011) found 81 stars and 31 BDs in a 6.5 deg$^{2}$ region surveyed, resulting in a stellar density of $\sim$17 objects per square degree. The TWA, in comparison, is a loose association of about 30 confirmed members. From Webb et al. (1999), the TWA searches were mainly conducted in a 12$\degr$ radius of the approximate cluster center. This implies a stellar density of $\sim$0.06 objects per square degree. We note that there are several other surveys conducted for these clusters/associations, which would result in slightly different stellar densities. For example, in Taurus, Luhman et al. (2003b) found 92 members in a 8.4 deg$^{2}$ area, implying a stellar density of $\sim$11 per square degree. In USco, Slesnick et al. (2008) detected 1000 objects in a 62 deg$^{2}$ area, implying a density of 16 per square degree. In IC 348, Luhman et al. (2003a) found 122 members (including 15 BDs) in a 42$\arcmin$x28$\arcmin$ region, resulting in a density of $\sim$373 sources per square degree. Our results for stellar densities indicate SOri and Cha I to be denser than Taurus and USco, which is in agreement with what has been previously noted by several surveys. In Table~\ref{frac-den}, we have listed the stellar densities plotted in Fig.~\ref{density}. The stellar densities are listed in both stars/deg$^{2}$ and stars/pc$^{2}$ units. 

In Fig.~\ref{ratio}, we have plotted the disc fractions versus the ratio of number of BDs to the number of stars in these clusters.  The BD/Star number ratios are listed in Table~\ref{frac-den}. The ratios for Taurus, Cha I and IC 348 have been taken from Andersen et al. (2008), the ratios for SOri and USco are from Lodieu et al. (2009; 2011), and the ratio for TWA has been calculated using data from Low et al. (2005) and Riaz \& Gizis (2008). Note that the mass range in Andersen et al. (2008) are different than Lodieu et al. (2009; 2011). Andersen et al. have used 0.08-1.0 $M_{\sun}$ for "stars", while the range is 0.08-0.35 $M_{\sun}$ in Lodieu et al. (2009; 2011). For BDs, the mass range is 0.03 - 0.08 $M_{\sun}$ in Andersen et al., and 0.02 - 0.08 $M_{\sun}$ in Lodieu et al. However, the ratios are consistent within the uncertainties from these surveys. The values provided in Andersen et al. (2008) have been calculated using data from Luhman et al. (2003) and Luhman (2004; 2007) surveys. The mass ranges are slightly different but the ratios are consistent within the uncertainties. The IC348 sample from Luhman et al. (2003) is unbiased down to 0.03 $M_{\sun}$ and not 0.02 $M_{\sun}$ as used by Andersen et al.. However, BD/star ratios in Luhman et al. (2003) work are defined down to 0.02 $M_{\sun}$, which is also the limit Lodieu et al. (2011) have used. The Taurus sample form Luhman (2004) work is unbiased down to 0.02 $M_{\sun}$. In IC 348, Andersen et al. have listed a ratio of 0.12 (0.08-0.18), Luhman et al. (2003) give a ratio 0.18$\pm$0.04, and Lodieu et al. (2011) list a ratio of 0.20$\pm$0.03. These ratios are all consistent within the uncertainties.

Looking at other BD/Star ratio values for the regions listed in Table~\ref{frac-den}, we have found the following:

\noindent 1) Taurus: 0.17 (0.11-0.25) from Andersen et al. (2008) and 0.14$\pm$0.04 from Luhman (2004) \\
2) SOri: 0.23$\pm$0.02 from Lodieu et al. (2009) \\
3) Cha I: 0.25 (0.13-0.53) from Andersen et al. (2008) and 0.26$\pm$0.06 from Luhman (2007) \\
4) IC348: 0.12 (0.08-0.18) from Andersen et al. (2008) and 0.12$\pm$0.04 from Luhman et al. (2003) \\
5) USco: 0.41$\pm$0.03 from Lodieu et al. (2011) \\

In Table~\ref{frac-den} and Fig.~\ref{ratio}, we have used the values from Andersen et al. (2008) for Taurus, Cha I and IC 348, and from Lodieu et al. (2009; 2011) for SOri and USco.

\end{twocolumn}

\label{lastpage}


\begin{thebibliography}{}

\bibitem[\protect\citeauthoryear{Alexander \& Armitage}{ 2009}]{aa09}Alexander \& Armitage 2009, ApJ, 704, 989
\bibitem[\protect\citeauthoryear{Andersen et al.}{ 2008}]{ander}Andersen et al. 2008, ApJ, 683, L183
\bibitem[\protect\citeauthoryear{Ardila et al.}{ 2000}]{ardila}Ardila et al. 2000, AJ, 120, 479
\bibitem[\protect\citeauthoryear{Baraffe et al.}{ 2003}]{barafe}Baraffe et al. 2003, A\&A, 402, 701
\bibitem[\protect\citeauthoryear{Barrado y Navascu\'{e}s \& Mart\'{i}n }{2003}]{bm03}Barrado y Navascu\'{e}s \& Mart\'{i}n 2003, AJ, 126, 2997
\bibitem[\protect\citeauthoryear{Barrado y Navascu\'{e}s et al. }{2006}]{b06}Barrado y Navascu\'{e}s et al. 2006, A\&A, 459, 511
\bibitem[\protect\citeauthoryear{Bate }{2009}]{bate}Bate 2009, MNRAS, 392, 590
\bibitem[\protect\citeauthoryear{Biller \& Close }{2007}]{biller}Biller \& Close 2007, ApJ, 669, L41
\bibitem[\protect\citeauthoryear{Bonnell et al. }{2008}]{bonnell08}Bonnell et al. 2008, MNRAS, 389, 1556
\bibitem[\protect\citeauthoryear{Carpenter et al. }{2006}]{car06}Carpenter et al. 2006, ApJ, 651, 49 
\bibitem[\protect\citeauthoryear{Carpenter et al. }{2009}]{car09}Carpenter et al. 2009, ApJ, 705, 1646
\bibitem[\protect\citeauthoryear{Clarke et al. }{2008}]{clarke}Clarke et al. 2008, MNRAS, 388, 1171
\bibitem[\protect\citeauthoryear{Dullemond \& Dominik }{2004}]{dd04}Dullemond \& Dominik 2004, A\&A, 421, 1075
\bibitem[\protect\citeauthoryear{Dye et al. }{2006}]{dye}Dye et al. 2006, MNRAS, 372, 1227
\bibitem[\protect\citeauthoryear{de Bruijne et al.\ }{1997}]{debr}de Bruijne, J. H. J. et al.\ 1997, ESA SP-402: Hipparcos, Venice 1997, p575
\bibitem[\protect\citeauthoryear{de Geus et al.\ }{1989}]{degeus}de Geus, E. J. et al.\ 1989, A\&A, 216, 44
\bibitem[\protect\citeauthoryear{de Zeeuw et al. }{1999}]{zeeuw}de Zeeuw et al.\ 1999, AJ, 117, 354
\bibitem[\protect\citeauthoryear{Goodwin et al. }{2004}]{good04}Goodwin et al. 2004, A\&A, 414, 633
\bibitem[\protect\citeauthoryear{Goodwin \& Whitworth }{2007}]{good07}Goodwin \& Whitworth 2007, A\&A, 466, 943
\bibitem[\protect\citeauthoryear{Haisch et al. }{2001}]{haisch}Haisch et al. 2001, ApJ, 553, L153
\bibitem[\protect\citeauthoryear{Hauschildt et. al. }{1999}]{hau99} Hauschildt, P. H., Allard, F., \& Baron, E. 1999, ApJ, 512, 377 
\bibitem[\protect\citeauthoryear{Hennebelle \& Chabrier }{2008}]{hc08}Hennebelle \& Chabrier 2008, ApJ, 684, 395
\bibitem[\protect\citeauthoryear{Herczeg et al. }{2008}]{h08}Herczeg et al. 2008, ApJ, 681, 594
\bibitem[\protect\citeauthoryear{Hillenbrand }{2005}]{hillen}Hillenbrand et al. 2005, STScI Symposium Series 19, A Decade of Discovery: Planets Around Other Stars, ed. M. Livio (Cambridge: Cambridge Univ. Press), arXiv:astro-ph/0511083 
\bibitem[\protect\citeauthoryear{Hernandez et al. }{2007}]{hernan}Hernandez et al. 2007, ApJ, 662, 1067
\bibitem[\protect\citeauthoryear{Lada et al. }{2006}]{lada}Lada, C. et al. 2006, AJ, 131, 1574
\bibitem[\protect\citeauthoryear{Lodieu et al. }{2007}]{lod07}Lodieu, N. et al.\ 2007, MNRAS, 374, 372
\bibitem[\protect\citeauthoryear{Lodieu et al. }{2008}]{lod08}Lodieu, N. et al.\ 2008, MNRAS, 383, 1385
\bibitem[\protect\citeauthoryear{Lodieu et al. }{2011}]{l11}Lodieu et al. 2011, A\&A, 527, 24
\bibitem[\protect\citeauthoryear{Lodieu et al. }{2009}]{l09}Lodieu et al. 2009, A\&A, 505, 1115
\bibitem[\protect\citeauthoryear{Low et. al. }{2005}]{low05} Low, F. J.; Smith, P. S.; Werner, M., Chen, C., Krause, V., Jura, M., \& Hines, D., C. 2005, ApJ, 631, 1170
\bibitem[\protect\citeauthoryear{Luhman }{1999}]{luh99}Luhman 1999, ApJ, 525, 466 
\bibitem[\protect\citeauthoryear{Luhman et al. }{2003a}]{luh03a} Luhman, K. L. et al. 2003a, ApJ, 593, 1093
\bibitem[\protect\citeauthoryear{Luhman et al. }{2003b}]{luh03b} Luhman, K. L. et al. 2003b, ApJ, 590, 348
\bibitem[\protect\citeauthoryear{Luhman }{2004}]{luh04}Luhman, K. L. 2004, ApJ, 617, 1216
\bibitem[\protect\citeauthoryear{Luhman }{2007}]{luh07}Luhman 2007, ApJS, 173, 104
\bibitem[\protect\citeauthoryear{Luhman et al. }{2008}]{luh08}Luhman et al. 2008, ApJ, 688, 362
\bibitem[\protect\citeauthoryear{Luhman et al. }{2010}]{luh10}Luhman et al. 2010, ApJS, 186, 111
\bibitem[\protect\citeauthoryear{Luhman \& Mamajek }{2010}]{lm}Luhman \& Mamajek 2010, ApJ, 716, L120
\bibitem[\protect\citeauthoryear{Machida et al. }{2009}]{mach09}Machida et al. 2009, ApJ, 699, L157
\bibitem[\protect\citeauthoryear{Martin et al. }{2001}]{mar01}Mart\'{i}n et al. 2001, ApJ, 558, 117
\bibitem[\protect\citeauthoryear{Mart\'{i}n et al. }{2004}]{mar04}Mart\'{i}n, E. L. et al. 2004, AJ, 127, 449
\bibitem[\protect\citeauthoryear{Muzerolle et al. }{2005}]{muz05} Muzerolle, J. et al. 2005, ApJ, 625, 906
\bibitem[\protect\citeauthoryear{Muzerolle et al. }{2006}]{muz06} Muzerolle, J. et al. 2006, ApJ, 643, 1003
\bibitem[\protect\citeauthoryear{Oliveira et al. }{2002}]{oliv}Oliveira et al. 2002, A\&A, 382, L22
\bibitem[\protect\citeauthoryear{Padoan \& Nordlund }{2004}]{pn04}Padoan \& Nordlund 2004, ApJ, 617, 559
\bibitem[\protect\citeauthoryear{Payne \& Lodato }{2007}]{pl07}Payne \& Lodato 2007, MNRAS, 381, 1597
\bibitem[\protect\citeauthoryear{Perryman et al. }{1997}]{perry97}Perryman et al. 1997, A\&A, 323, L49
\bibitem[\protect\citeauthoryear{Pfalzner et al. }{2006}]{pfalz}Pfalzner et al. 2006, A\&A, 454, 811
\bibitem[\protect\citeauthoryear{Preibisch et al. }{1998}]{pre98}Preibisch et al.\ 1998, A\&A, 333, 619
\bibitem[\protect\citeauthoryear{Preibisch \& Zinnecker }{1999}]{pz99}Preibisch, T. \& Zinnecker 1999, AJ, 117, 2381
\bibitem[\protect\citeauthoryear{Preibisch et al.\  }{2001}]{preb01}Preibisch, T. et al.\ 2001, AJ, 121, 1040
\bibitem[\protect\citeauthoryear{Preibisch et al. }{2002}]{pre02}Preibisch et al.\ 2002, AJ, 123, 1613
\bibitem[\protect\citeauthoryear{Reipurth \& Clarke }{2001}]{rc01}Reipurth \& Clarke 2001, AJ, 122, 432
\bibitem[\protect\citeauthoryear{Riaz et al. }{2006}]{r06}Riaz et al. 2006, AJ, 132, 866
\bibitem[\protect\citeauthoryear{Riaz \& Gizis }{2008}]{riaz}Riaz, B. \& Gizis 2008, ApJ, 681, 1584
\bibitem[\protect\citeauthoryear{Riaz et al. }{2009}]{r09}Riaz et al. 2009, ApJ, 705, 1173
\bibitem[\protect\citeauthoryear{Rieke et al. }{2005}]{reike}Reike et al. 2005, ApJ, 620, 1010
\bibitem[\protect\citeauthoryear{Scholz et al. }{2007}]{sch07}Scholz, A. et al. 2007, ApJ, 660, 1517 
\bibitem[\protect\citeauthoryear{Seiss et al. }{2000}]{seiss}Seiss et al. 2000, A\&A, 358, 593
\bibitem[\protect\citeauthoryear{Sherry et al. }{2004}]{sherry}Sherry et al. 2004, AJ, 128, 2316
\bibitem[\protect\citeauthoryear{Siegler et al. }{2007}]{sieg}Siegler et al. 2007, ApJ, 654, 580
\bibitem[\protect\citeauthoryear{Slesnick et al.\ }{2006}]{sles}Slesnick, C. et al.\ 2006, AJ, 131, 3016 
\bibitem[\protect\citeauthoryear{Slesnick et al.\ }{2008}]{sles}Slesnick, C. et al.\ 2008, ApJ, 688, 377
\bibitem[\protect\citeauthoryear{Stamatellos \& Whitworth }{2009a}]{stama09a}Stamatellos \& Whitworth 2009a, MNRAS, 392, 413
\bibitem[\protect\citeauthoryear{Stamatellos \& Whitworth}{2009b}]{stama09b}Stamatellos \& Whitworth 2009b, MNRAS, 400, 1563
\bibitem[\protect\citeauthoryear{Thies \& Kroupa}{2007}]{tk07}Thies \& Kroupa 2007, ApJ, 671, 767
\bibitem[\protect\citeauthoryear{Walter et al. }{1994}]{walter}Walter et al. 1994, AJ, 107, 692
\bibitem[\protect\citeauthoryear{Webb et al. }{1999}]{webb}Webb et al. 1999, ApJ, 512, L63
\bibitem[\protect\citeauthoryear{Wright et al. }{2010}]{w10}Wright et al. 2010, AJ, 140, 1868
\bibitem[\protect\citeauthoryear{Young et al.}{2004}]{y04}Young et al. 2004, ApJS, 154, 428
\bibitem[\protect\citeauthoryear{Zapatero et al. }{2000}]{z00}Zapatero Osorio et al. 2000, Science, 290, 103
\bibitem[\protect\citeauthoryear{Zapatero et al. }{2002}]{z02}Zapatero Osorio et al. 2002, A\&A, 384, 937
\bibitem[\protect\citeauthoryear{Zapatero et al. }{2007}]{z07}Zapatero Osorio et al. 2007, A\&A, 472, L9

\end{thebibliography}
\end{document}